\documentclass{aa}

\usepackage[varg]{txfonts}
\usepackage{epstopdf}
\usepackage{lscape}
\usepackage{color}
\usepackage[online]{threeparttable}

\usepackage{natbib,twoopt}

\usepackage[breaklinks=true]{hyperref} 
\bibpunct{(}{)}{;}{a}{}{,}             
\makeatletter
  \newcommandtwoopt{\citeads}[3][][]{\href{http://adsabs.harvard.edu/abs/#3}%
    {\def\hyper@linkstart##1##2{}%
     \let\hyper@linkend\@empty\citealp[#1][#2]{#3}}}
  \newcommandtwoopt{\citepads}[3][][]{\href{http://adsabs.harvard.edu/abs/#3}%
    {\def\hyper@linkstart##1##2{}%
     \let\hyper@linkend\@empty\citep[#1][#2]{#3}}}
  \newcommandtwoopt{\citetads}[3][][]{\href{http://adsabs.harvard.edu/abs/#3}%
    {\def\hyper@linkstart##1##2{}%
     \let\hyper@linkend\@empty\citet[#1][#2]{#3}}}
  \newcommandtwoopt{\citeyearads}[3][][]%
    {\href{http://adsabs.harvard.edu/abs/#3}
{\def\hyper@linkstart##1##2{}%
\let\hyper@linkend\@empty\citeyear[#1][#2]{#3}}}
\makeatother

\begin{document}

\title{Physical parameters of red supergiants in dwarf irregular galaxies in the Local Group\thanks{Based on observations made with ESO Telescopes at the La Silla Paranal Observatory under program ID 095.D-0313 and observations made with the Gran Telescopio Canarias (GTC), installed in the Spanish Observatorio de El Roque de Los
Muchachos of the Instituto de Astrof\'{\i}sica de Canarias, in the island of La Palma; program ID: 93-MULTIPLE-2/14B.}}

\author{N. E. Britavskiy \inst{1,2,3} \and A. Z. Bonanos \inst{3} \and A. Herrero\inst{1,2} \and M. Cervi\~no \inst{4} \and D. Garc\'{\i}a-\'Alvarez\inst{1,2,5} \and M. L. Boyer \inst{6} \and \\T. Masseron\inst{1,2} \and  A. Mehner \inst{7} \and K. B. W. McQuinn \inst{8}}

\offprints{N. Britavskiy}

\institute{Instituto de Astrof\'{\i}sica de Canarias, Avenida V\'{\i}a L\'actea, E-38205 La Laguna, Tenerife, Spain\\
 \email{britavskiy@iac.es}
 \and
Universidad de La Laguna, Dpto. Astrof\'isica, E-38206 La Laguna, Tenerife, Spain
\and
IAASARS, National Observatory of Athens, GR-15236 Penteli, Greece
\and 
Centro de Astrobiolog\'ia (CSIC/INTA), 28850 Torrej\'on de Ardoz, Madrid, Spain\
\and
Grantecan S.\,A., Centro de Astrof\'{\i}sica de La Palma, Cuesta de San Jos\'e, E-38712 Bre\~na Baja, La Palma, Spain
\and
Space Telescope Science Institute, 3700 San Martin Dr., Baltimore, MD 21218, USA
\and
European Southern Observatory, Alonso de C\'{o}rdova 3107, Casilla 19001, Santiago de Chile, Chile
\and
Rutgers University, Department of Physics and Astronomy, 136 Frelinghuysen Road, Piscataway, NJ 08854, USA
}
\date{Received 6 February 2019 / Accepted 24 September 2019}

\authorrunning{Britavskiy et al.}
\titlerunning{Red supergiants in the Local Group.}

\abstract
{Increasing the statistics of evolved massive stars in the Local Group enables investigating their evolution at different metallicities. During the late stages of stellar evolution, the physics of some phenomena, such as episodic and systematic mass loss, are not well constrained. For example, the physical properties of red supergiants (RSGs) in different metallicity regimes remain poorly understood. Thus, we initiated a systematic study of RSGs in dwarf irregular galaxies (dIrrs) in the Local Group.}
{We aim to derive the fundamental physical parameters of RSGs and characterize the RSG population in nearby dIrrs.}
{The target selection is based on 3.6 $\mu$m and 4.5 $\mu$m photometry from archival {\it Spitzer} Space Telescope images of nearby galaxies. We selected 46 targets in the dIrrs IC 10, IC 1613, Sextans B, and the Wolf-Lundmark-Melotte (WLM) galaxy that we observed with the GTC--OSIRIS and VLT--FORS2 instruments. We used several photometric techniques together with a spectral energy distribution analysis to derive the luminosities and effective temperatures of known and newly discovered RSGs.}
{We identified and spectroscopically confirmed 4 new RSGs, 5 previously known RSGs, and 5 massive asymptotic giant branch (AGB) stars. We added known objects from previous observations. In total, we present spectral classification and fundamental physical parameters of 25 late-type massive stars in the following dIrrs: Sextans A, Sextans B, IC 10, IC 1613, Pegasus, Phoenix, and WLM. This includes 17 RSGs and 8 AGB stars that have been identified here and previously.}
{Based on our observational results and PARSEC evolutionary models, we draw the following conclusions: (i) a trend to higher minimum effective temperatures at lower metallicities and (ii) the maximum luminosity of RSGs appears to be constant at $\log(L/L_{\sun}) \approx$\,5.5, independent of the metallicity of the host environment (up to $\mathrm{[Fe/H]}$\,$\approx$\,$-1$ dex).}


\maketitle

\section{Introduction}

Red supergiants (RSGs) belong to a critical but short-lived ($\lesssim 35$~Myr) stage of massive star evolution. It is considered that all stars with initial masses of about $8-40~M_{\odot}$ are passing through this stage, which is identified as the core helium burning phase. Because only few objects are studied in detail so far, the predictions of stellar evolutionary models for their physical parameters, such as temperature and luminosity, still differ. An important open question is the effect of metallicity on the evolution of RSGs, that is, whether the Hayashi \citep{Hayashi} and Humphreys-Davidson \citep{Humphreys1979} limits depend on metallicity. Observationally, the average effective temperature of RSGs varies with metallicity \citep{Elias1985,Levesque2006,LM2012}. Some theoretical works support the dependency of the mixing length parameter on metallicity \citep[e.g.,][]{chun_2017}.  
Another factor that significantly affects the evolution of massive stars is mass loss \citep{Smith_2014, Meynet_2015, Groenewegen_2018}, which is difficult to measure in the RSG phase. Whereas the mass loss is driven by the iron content in hot stars, it is driven by the dust content in cool stars, and hence depends on different chemical species \citep{van_Loon_2005, Goldman_2017}. Available evolutionary models cannot reliably predict the RSG evolution.

In order to answer how various stellar physical parameters depend on metallicity, it is important to survey RSGs in host environments with different metallicities. However, only a few RSGs beyond the Milky Way, especially in more metal-poor host galaxies, are spectroscopically confirmed. 
The exceptions are the massive M31 and M33 galaxies and the Magellanic Clouds (MCs). In the past ten years, nearly 200 RSGs were discovered and spectroscopically confirmed in the Small Magellanic Cloud (SMC), and 250 RSGs in the Large Magellanic Cloud (LMC; \citealt{Massey_2002_MC,new_smclmc}). The situation is different in more distant dwarf irregular galaxies (dIrrs): only 53 RSGs are known in 6 dIrrs in the Local Group. There are 11 spectroscopically confirmed RSGs in the Wolf-Lundmark-Melotte (WLM) galaxy \citep{Bresolin_2006,LM2012}, 26 in NGC 6822 \citep{LM2012,Patrick15}, 7 in Sextans A, 6 in IC 1613 \citep{Tautvai2007,britavskiy14,britavskiy15}, 2 in NGC 3109 \citep{Evans_2007}, and 1 in Sagittarius \citep{2018_Sag}. Several RSG candidates lie in three distant spiral galaxies \citep{chun_rsgs2017}. Each additional RSG beyond the Milky Way is statistically significant as an observational reference point to constrain stellar evolution theories at the late stages of massive star evolution. The sample of dIrr galaxies and the MCs provides an ideal laboratory for investigating the physical properties of RSGs over a wide range of host galaxy metallicities from $\mathrm{[Fe/H]}$\,$\approx$\,$-0.4$ dex (LMC) to $\mathrm{[Fe/H]}$\,$\approx$\,$-1$ dex (Sextans A). 

\citet[][hereafter Paper I and Paper II, respectively]{britavskiy14,britavskiy15} probed mid-infrared (mid-IR) selection techniques for RSGs in star-forming dIrr galaxies in the Local Group (Sextans A, IC 1613, the WLM, Pegasus, and Phoenix). We here complete our census of RSGs by adding four dIrrs: Sextans B, IC 10, IC 1613, and the WLM. We derive the physical parameters for all discovered RSGs. In IC 10 and Sextans B we observed several RSG candidates for the first time. IC 1613 and the WLM were included in our previous surveys (Paper I and Paper II), but the selection process and the observations were repeated for consistency.
For the RSGs and massive asymptotic giant branch (AGB) stars for which we have a calibrated spectral energy distribution (SED), a systematic physical parameters analysis was performed using different spectroscopic and photometric techniques. Some targets that we previously classified as RSGs in Paper II appear to be massive AGB stars after the repeated analysis. We conclude that it is necessary to use the luminosity-type classification. 

The paper is organized in the following way. In Section 2 we describe the target selection criteria, observations, and basic spectral classification analysis. In Section 3 we present the physical parameter analysis, using SED fitting and three photometric approaches. Section 4 presents an interpretation and discussion of the obtained results, and in Section 5 we close with the conclusions. The appendix contains the information about the observed targets and the SED fitting curves.

\begin{table*}
{\small
\caption{Properties of the program galaxies.}
\label{tab:galax}
\begin{threeparttable}
\begin{tabular}{llcccccc}
\hline\hline
Name   & DDO &    Distance        &     Distance modulus  & Radial velocity &  $\mathrm{[Fe/H]}$ & SFR \textdagger \\
      & identifier   &          (kpc)            &  (mag)   &  ($\mathrm{km} \, \mathrm{s}^{-1}$) & (dex) & $\mathrm{M}_\sun \,\mathrm{yr}^{-1}$ \\
\hline
IC 10  &                        &     794$\pm$44        &        24.27$\pm$0.18  & $-348$$\pm$1                                 & $-0.50$\tnote{a}  & 0.05\tnote{b} \\
IC 1613 &DDO 8           &    755$\pm$42         &    24.39$\pm$0.12      & $-233$$\pm$1                                 & $-0.67$\tnote{h}  & $0.0029$\tnote{c} \\      
Pegasus & DDO 216    &   920$\pm$30            & 24.82$\pm$0.07       &$-183$$\pm$5                                 & $-0.80$\tnote{a}  & 0.00035\tnote{d}\\
Phoenix  &                    &   415$\pm$19           & 23.09$\pm$0.10     & $-13$$\pm$9, $-$52$\pm$6\tnote{j} & -- & --  \\
Sextans A& DDO 75    &  1432$\pm$53          & 25.60$\pm$0.03         &$+324$$\pm$2                                 &  $-1.0$\tnote{i} &   0.002\\
Sextans B &DDO 70    &    1426$\pm$20        &     25.60$\pm$0.03     & $+304$$\pm$1                                 &  -- & 0.002\tnote{e} , (0.0008)\\
WLM & DDO 221         &    933$\pm$34           & 24.95$\pm$0.03 & $-130$$\pm$1                                         &  $-0.87$\tnote{f} &  0.00047\tnote{g} , (0.003)*\\
\hline                                                                                                                   
\end{tabular}
\tablefoot{
* -- The distance, distance moduli, systemic radial velocities, and metallicities are taken from \cite{all_galax}. \textdagger -- star formation rates are taken from \cite{Mateo1998}. References: $^a$ \cite{Bergh_2000}, $^b$ \cite{IC10_sfr}, $^c$ \cite{IC1613_sfr}, $^d$ \cite{Pegasus_sfr}, $^e$ \cite{SextansB_sfr}, $^f$ \cite{WLM_Z}, $^g$ \cite{WLM_sfr1}, $^h$ \cite{Tautvai2007}, $^i$ \cite{Kaufer_2004}, $^j$ \cite{gallart2012}. }
\end{threeparttable}
}
\end{table*}

\section{Target selection and observations}

\subsection{Target selection}

We selected RSGs candidates in four nearby dIrrs with relatively high star formation rates (SFRs; $\ga 0.003~\mathrm{M}_\sun \,\mathrm{yr}^{-1}$), based on color ($[3.6]-[4.5] < 0$) and brightness (M$_{[3.6]} < -9$ mag) criteria; see Paper I and Paper II for details. These selection criteria are empirical and are based on the spectroscopic survey of massive stars in the LMC and SMC \citep{BMS09,BLK10}. We used the mid-IR colors because RSGs are very bright in the infrared due to their dusty envelopes. We used published {\it Spitzer/IRAC} photometry \citep[DUST in Nearby Galaxies with {\it Spitzer}, DUSTiNGS survey][]{dustings} of the four nearby dIrrs IC 10, IC 1613, Sextans B, and the WLM. In total, we observed 46 targets for follow-up observations. For the selected targets in the WLM, we also included six previously known RSGs from Paper II and \citet{LM2012}. The observations were carried out in October of 2014 and in August of 2015, when the final DUSTiNGS survey was not yet published. We therefore used an unpublished version of the survey that differs slightly from the final version for the selection process. This slightly affects the colors of the selected targets (see Section 2.3).

The basic properties, that is, the galaxy name, distance, radial velocity, metallicity, and SFR, of our program galaxies, together with the galaxies in which we have previously found RSGs, are listed in Table \ref{tab:galax}. The literature estimates of the metallicities are mainly based on the metallicities of blue supergiants (BSGs). We set the RSG metallicities equal to the metallicity estimates obtained using BSGs. Both $\mathrm{[Fe/H]}$ and Z (mostly oxygen-based) abundances are relevant for the stellar evolution and mass-loss properties. We note that the metallicities refer to averages and that metallicity variations exist inside these galaxies \citep[e.g.,][]{Berger_2018}. The average values allow us to map (within uncertainties of 0.2 dex) the metallicity dependence of the RSG population in different dIrrs.

\subsection{Observations and data reduction}
The targets in IC 10, IC 1613, and Sextans B were observed with the 10.4 m Gran Telescopio Canarias (GTC) using the Optical System for Imaging and low Resolution Integrated Spectroscopy (OSIRIS) in multi-object spectroscopy (MOS) mode in September and December 2014. Twenty-three targets were observed with the OSIRIS R1000R grism with the 1.2$\arcsec$ slit. The wavelength range was 5100 \AA\ to 10000 \AA\ with a resolving power of $R \approx1100$. The field of view of the MOS OSIRIS masks (7.5'x6') was suitable to cover each dIrr galaxy with one field. The spectra were reduced by standard IRAF\footnote{IRAF is distributed by the National Optical Astronomy Observatory, which is operated by the Association of Universities for Research in Astronomy (AURA) under cooperative agreement with the National Science Foundation.} routines: bias subtraction, division by the flat fields, wavelength calibration, flux calibration, and spectrum extraction. The accuracy of the wavelength solution is approximately 1 \AA. Spectra were flux calibrated using a spectroscopic standard star (usually BA spectral type) that was observed during the same observing run. The standard was taken in long-slit mode and a wider slit width (2.5$\arcsec$). At this slit width, the typical flux due to instrumental uncertainties is about 10\%. The difference in seeing between the science and calibration observational blocks did not exceed 0.25$\arcsec$. The spectrum of a standard star was used to determine the response curve of the spectrograph, which we used to obtain the relative flux calibration of the science spectra. We list the seeing value at the beginning of observations for each observational block in Table \ref{tab:gtc_mos}.

The targets in the WLM were observed with the FORS2 spectrograph at ESO's Very Large Telescope (VLT) in August 2015. In Paper II we have discussed 31 targets in this galaxy. We found 4 RSGs that were previously identified by \citet{LM2012}. We here selected 23 targets in this galaxy. To avoid slit overlaps in the compact WLM field, we created three masks for the same field (6.8'x6.8').  Only 23 were observed because the (service mode) program was not completed.

The data reduction was performed with the FORS2 ESO pipeline version 4.9.23 with the {\it Reflex} workflow version 2.6 \citep{reflex}. The reduction process includes standard procedures such as bias subtraction, flat field division, background subtraction, and wavelength and flux calibration. For each target, four spectra are combined using the IRAF routine {\it scombine}. The observation of the flux standard target (NGC 7293) was performed in the same night and all science spectra are absolute-flux calibrated using standard FORS2 pipeline routines. The spectra have a wavelength range from 4300~\AA\ to 9000~\AA\ and an average signal-to-noise ratio (S/N) of $\approx30$. We did not achieve this wavelength range for all targets because space on the CCD in MOS mode was limited. Thus, some of targets have a shorter wavelength coverage, which makes the spectroscopic analysis difficult. The resolving power varies from $R \approx400$ at 5000~\AA\ to $R \approx680$ at 8600~\AA. The journal of observations is provided in Table \ref{tab:gtc_mos}. In Paper II all targets in Pegasus, Phoenix, Sextans A, and the WLM  have been observed with the FORS2 instrument and have been processed in the same way.

\begin{table*}
{\small
\caption{Journal of observations with GTC-OSIRIS and VLT-FORS2.}
\label{tab:gtc_mos}
\begin{tabular}{lccccc}
\hline\hline
ID    & MJD                 & Seeing &    Exposure time       &  Observed &  Spectrograph\\
      & (days)  &   &     (s)      & targets &     \\
\hline
IC 10    &   56928.88963     &1.0  & 2000  &   12  & GTC--OSIRIS \\
IC 1613  &   57007.91051    & 1.2--1.5 & 1300  & 6  & GTC--OSIRIS\\
Sextans B &   57014.17851   & 1.5  & 2000  &   5 & GTC--OSIRIS \\
\hline
WLM  & 57260.15575 & 1.3 & 4x1200   & 23 & VLT--FORS2 \\
\hline                                                                                                                   
\end{tabular}
}
\end{table*}

\subsection{Spectral classification}

We used the same algorithm for the analysis as described in Paper I and Paper II to determine the spectral type, luminosity class, and measure the radial velocity.  
For the low-resolution spectroscopic data, we used the ESO UVES Paranal Observatory Project (POP) spectral library, which we degraded from $R$\,$\approx$\,70\,000 to $R = 1000$. To determine the spectral type, we used mainly the TiO bands, which dominate the optical wavelength region in spectra of RSGs. The luminosity class and radial velocities were determined based on the Ca~II line profiles ($\lambda \lambda$ 8380 -- 8800~\AA) by comparing the strengths and position of these gravity-sensitive features with giant and supergiant template spectra. 
In Tables \ref{tab:ic10} -- \ref{tab:wlm_p95} we present the target ID number, ID from the DUSTiNGS catalog, coordinates, radial velocities, absolute $[3.6]$~mag ($M_{[3.6]}$, computed using the distance in Table \ref{tab:galax}), $[3.6]-[4.5]$ colors, and the spectral classification. 

\begin{figure}
\begin{center}
\resizebox{\hsize}{!}{\includegraphics{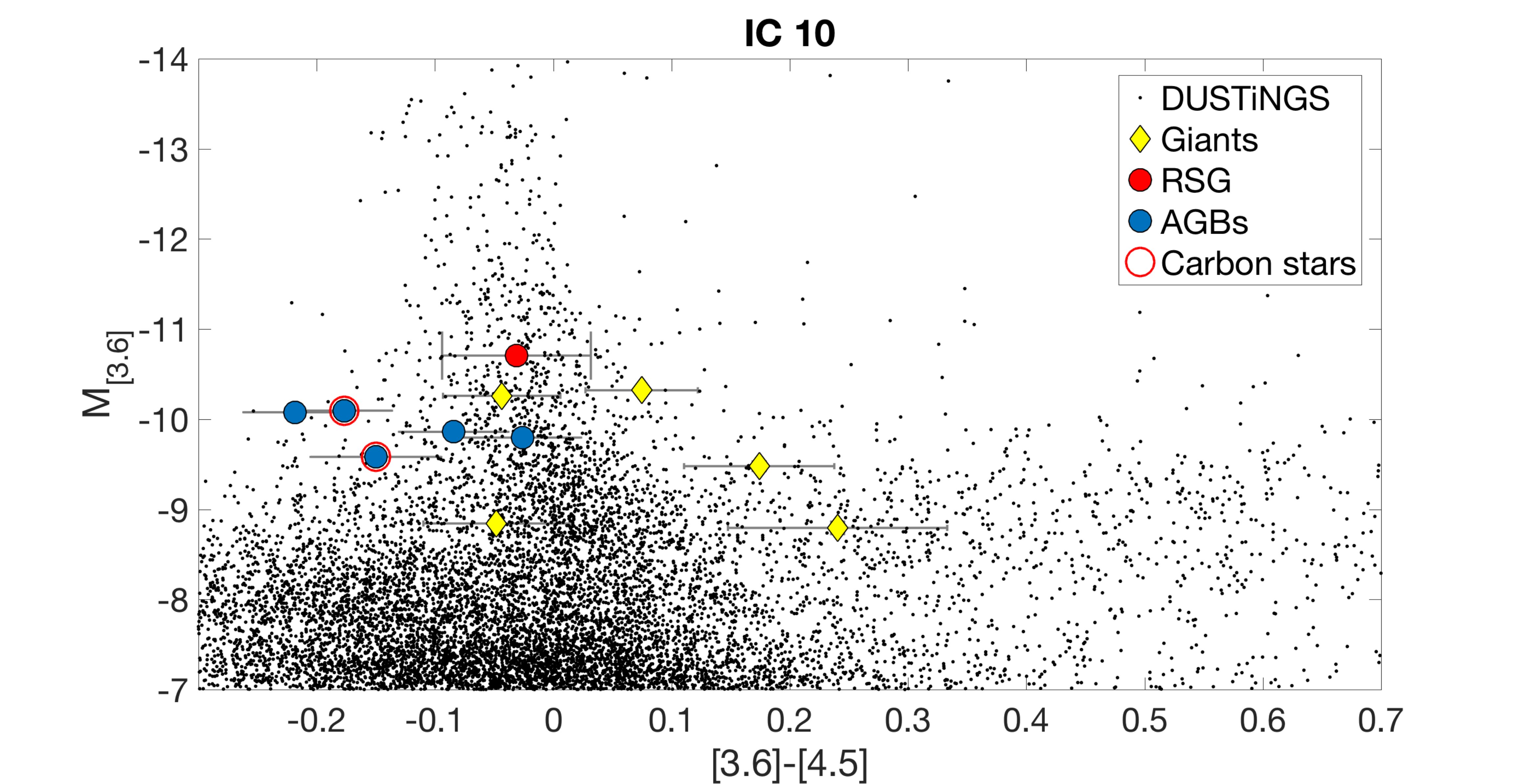}}
\includegraphics[width=0.65\linewidth]{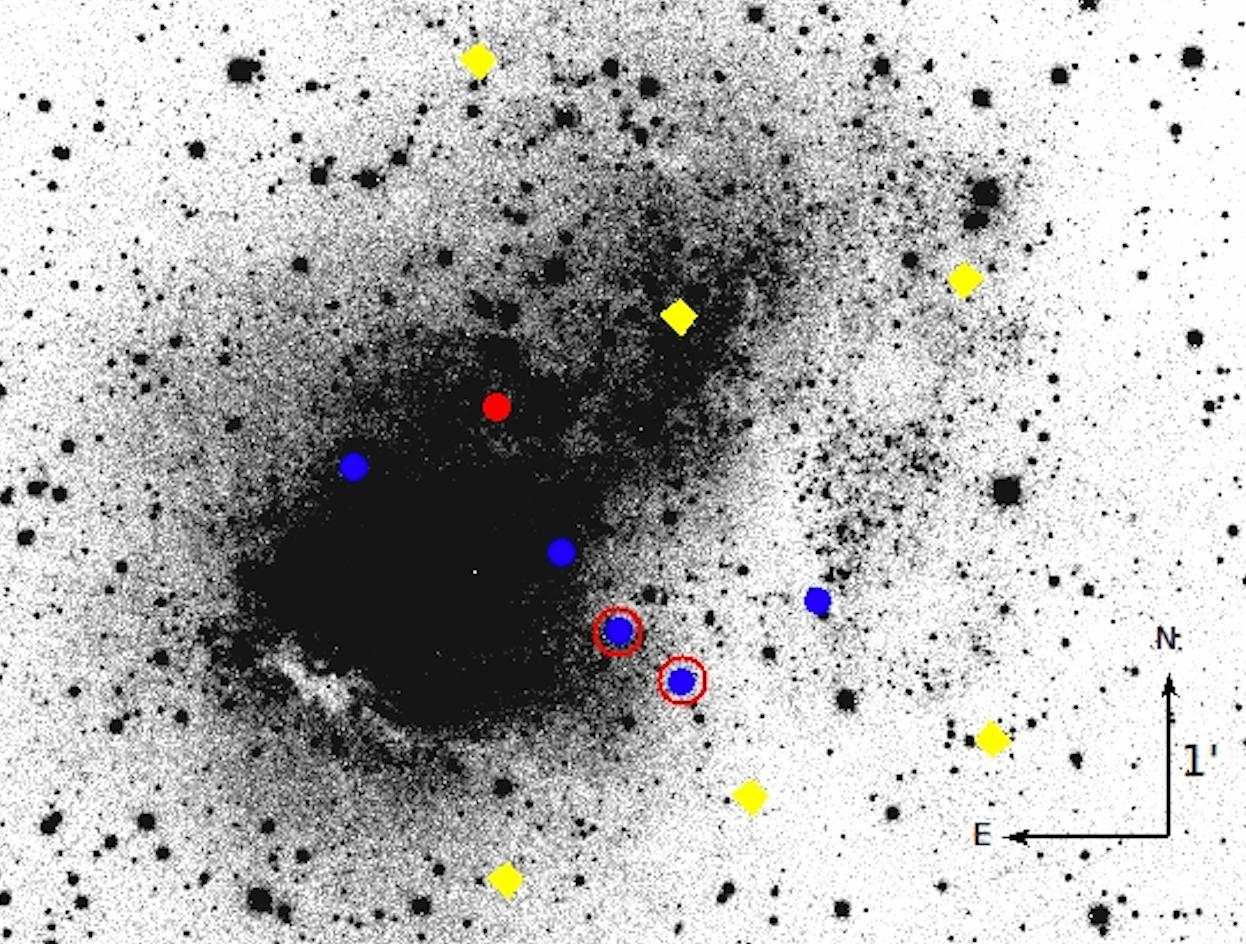}
\end{center}
\caption[]{Top panel: M$_{[3.6]}$ vs. $[3.6]-[4.5]$ CMD for the dIrr galaxy IC 10.
Observed stars are labeled with different symbols according to their classification, see Table \ref{tab:ic10}. The foreground late-type giant stars are labeled "giant". The error bars for colors and magnitudes are shown with gray lines. Bottom panel: Spatial distribution of the observed targets, superposed on $V$-band images of the galaxy IC 10 \citep{Massey_Images}.}
\label{Fig5_ic10}
\end{figure}

\begin{figure}
\begin{center}
\resizebox{\hsize}{!}{\includegraphics{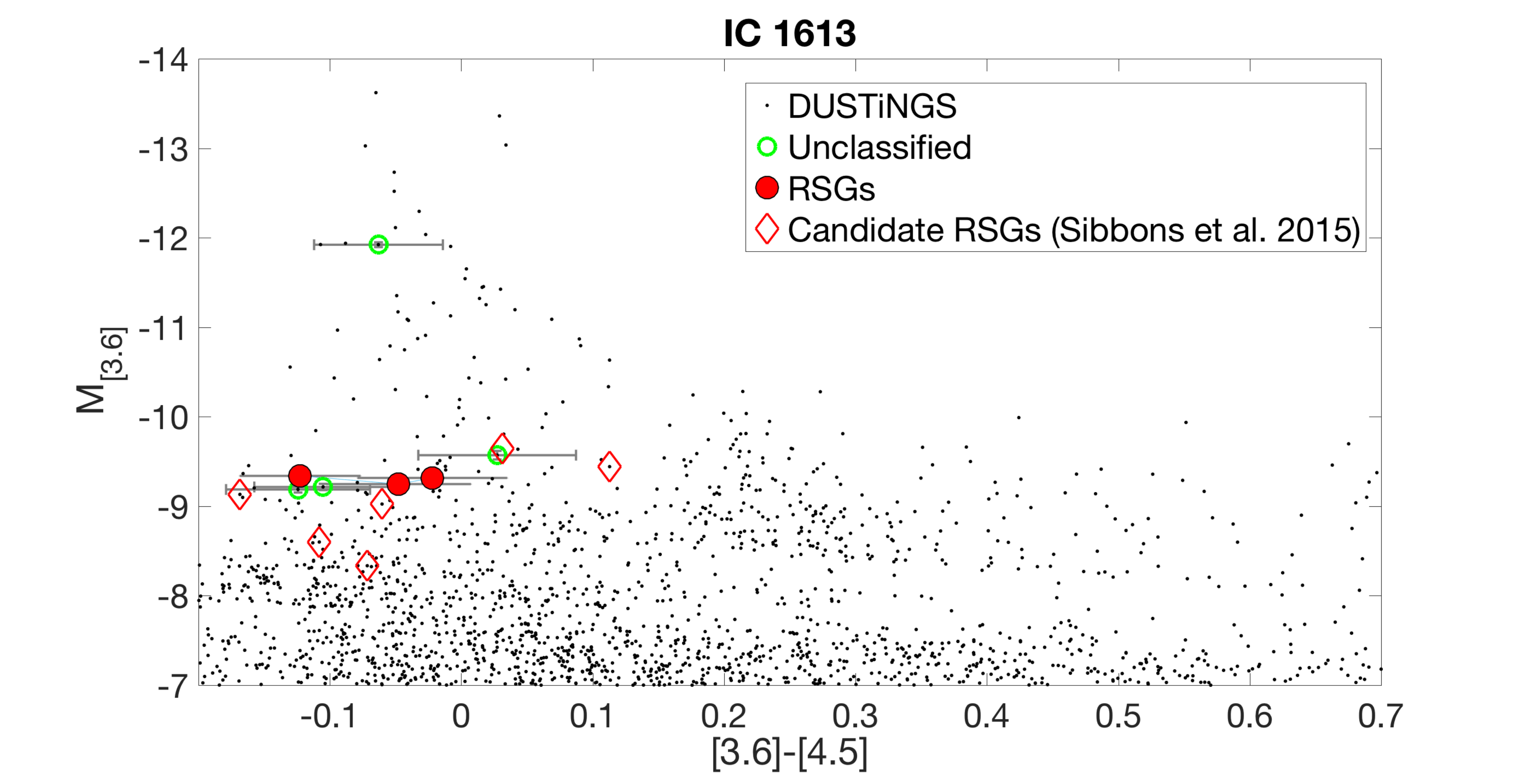}}
\includegraphics[width=0.65\linewidth]{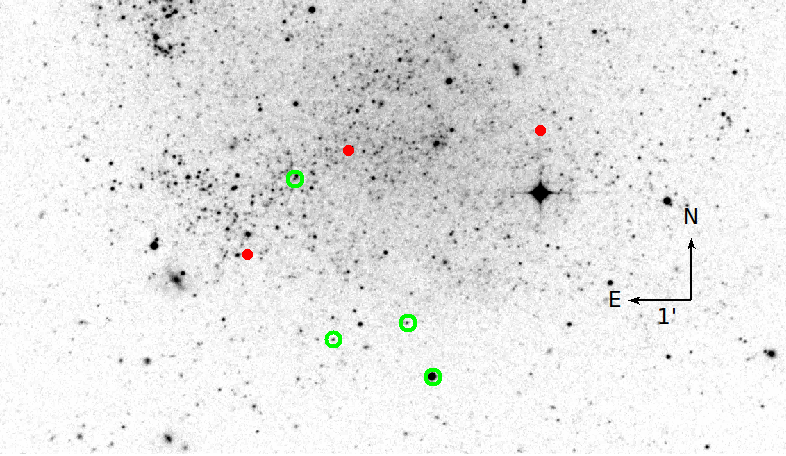}
\end{center}
\caption[]{Same as Figure \ref{Fig5_ic10}, but for the dIrr galaxy IC 1613. Details are provided in Table \ref{tab:ic1613}. In addition, we plot RSG candidates based on the JHK selection technique \citep{Sibbons_2015}.}
\label{Fig6_ic1613}
\end{figure}

\begin{figure}
\begin{center}
\resizebox{\hsize}{!}{\includegraphics{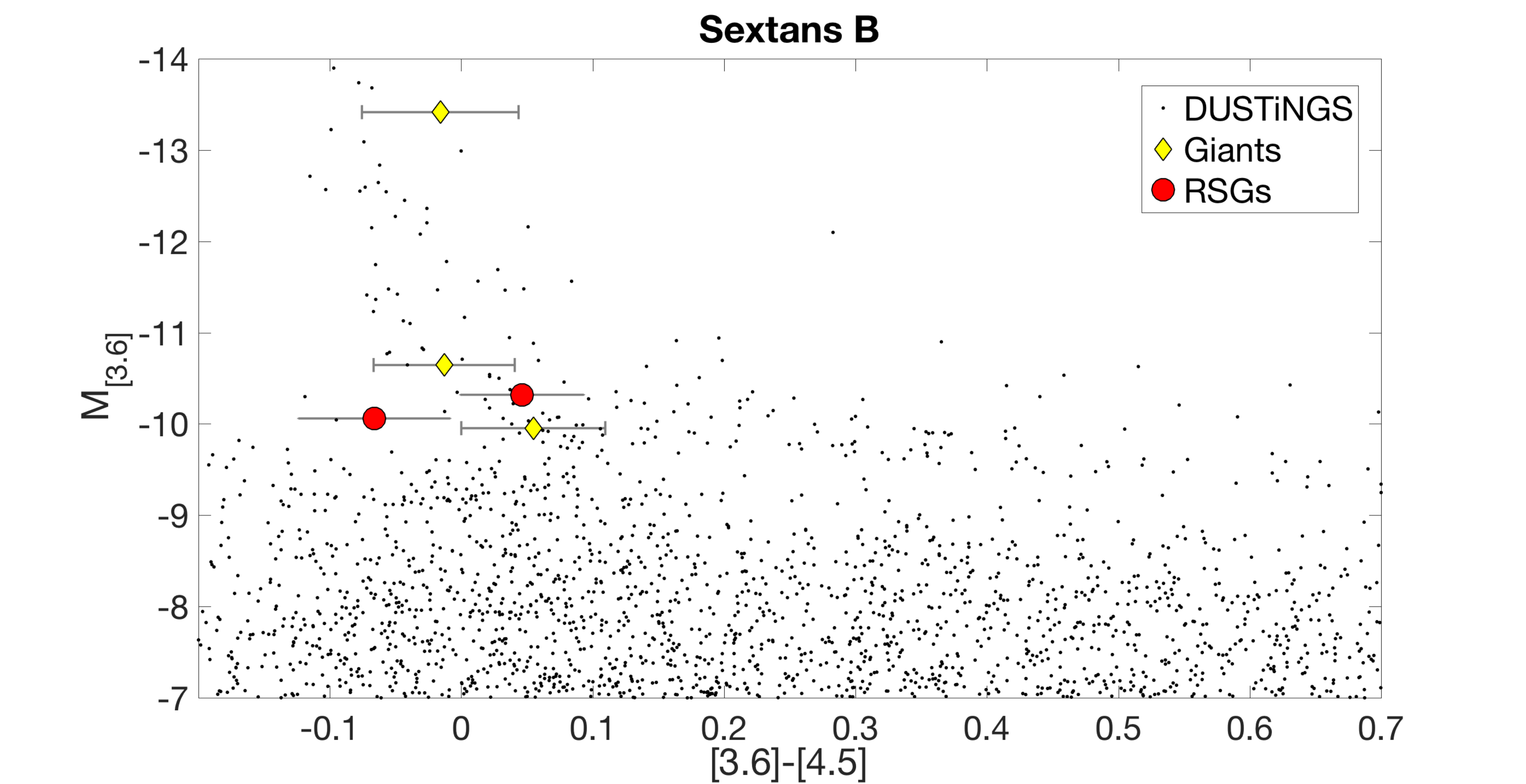}}
\includegraphics[width=0.65\linewidth]{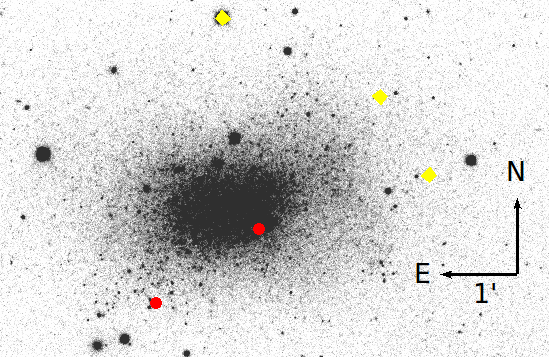}
\end{center}
\caption[]{Same as Figure \ref{Fig5_ic10}, but for the dIrr galaxy Sextans B. Details are provided in Table \ref{tab:sexb}.}
\label{Fig7_sexb}
\end{figure}

\begin{figure}
\begin{center}
\resizebox{\hsize}{!}{\includegraphics{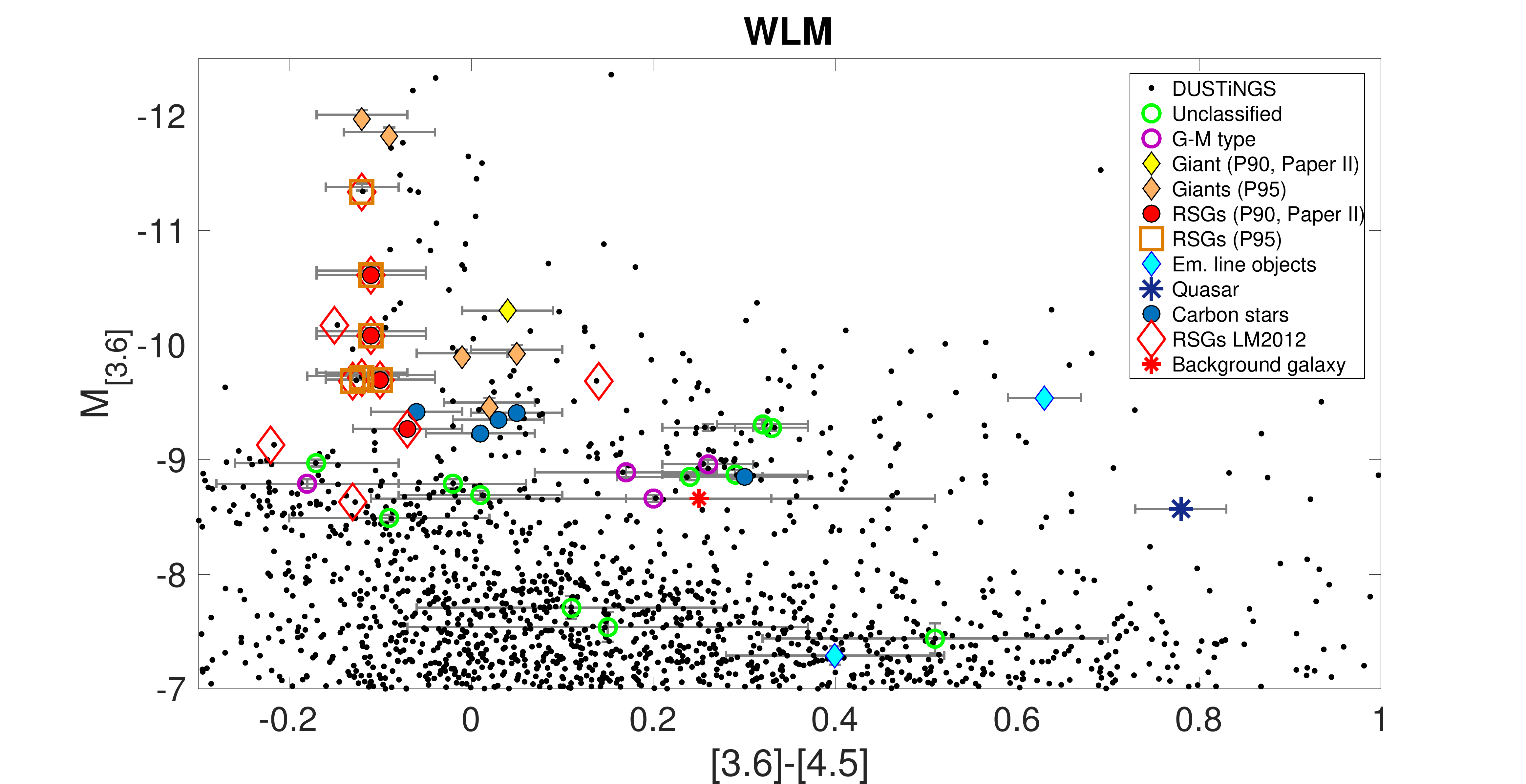}}
\includegraphics[width=0.65\linewidth]{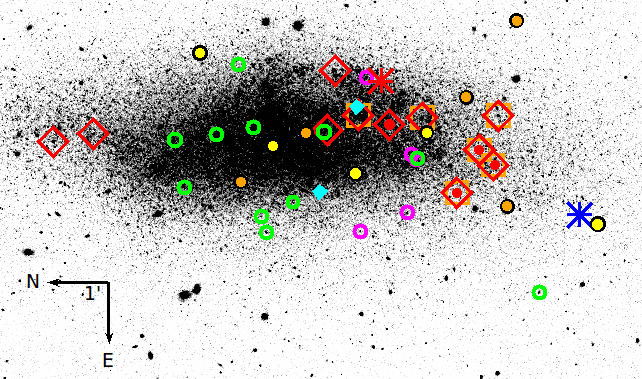}
\end{center}
\caption[]{Same as Figure \ref{Fig5_ic10}, but for the dIrr galaxy WLM. For this galaxy we combined the results of Paper II and the present work. Details are provided in Table \ref{tab:wlm_p95} and Table A.4 in Paper II.}
\label{Fig7_wlm}
\end{figure}

Figures \ref{Fig5_ic10} -- \ref{Fig7_wlm} present the color-magnitude diagrams (CMDs) and spatial distribution of our targets in the four program galaxies, IC 10, IC 1613, Sextans B, and the WLM. In the $M_{[3.6]}$ versus $[3.6]-[4.5] $ CMDs we used the same formalism of object spectral classification as in Tables \ref{tab:ic10} -- \ref{tab:wlm_p95}. The values of magnitudes and colors from the DUSTiNGS survey \citep{dustings} are different from those that we used for the initial target selection. This explains why the RSGs in some cases have [3.6]--[4.5]\,>\,0, which is not in agreement with our selection criteria. However, it shows that our color cut is not efficient in distinguishing RSGs from the late-type foreground giants because some of the field giants have the same [3.6]--[4.5] colors as the RSGs.
The analysis of the spatial distribution of observed targets shows that the majority of foreground and background objects are located outside the main body of dIrrs \citep[see also the spatial distribution of RSGs in Sextans A in][]{britavskiy15}. Moreover, we note that individual RSGs are located mainly at the edge of the galaxies. This is due to an observational bias: the crowded regions in the central part of the galaxies do not allow properly observing the targets in the MOS mode at optical wavelength.


\section{Determination of the fundamental parameters of RSGs}

The list of all identified RSGs and AGB stars in seven dIrr galaxies is presented in Table \ref{tab:hr_diag}. The table contains information on the 25 RSGs candidates that we identified in the present work and in Paper I and Paper II.

We used different approaches to obtain the luminosities and effective temperatures ($T_{\rm eff}$) of the newly discovered RSGs, that is, the SED fitting and several photometric techniques. We present each of these techniques in the following subsections. Deriving physical parameters of extragalactic RSGs is challenging, and no routine procedures have been developed so far. We therefore describe the techniques we have applied to our sample in detail. 

\subsection{SED fitting technique}

Before proceeding to the SED analysis, we fit the relative flux-calibrated spectra to the Johnson $BVI$ bands. 
Absolute flux calibration in MOS mode is quite challenging because of light loss from the slit or parallactic angle differences between standard and scientific targets, for instance. The best method for reliable absolute flux calibration therefore is to shift the flux to the known values from photometric bandpasses at the given wavelength.

We fit the flux-calibrated spectra of the RSGs (SED) at optical wavelength between 4500 -- 6500~\AA\ with a grid of synthetic spectra with the same wavelength binning as the observed spectra. To build this grid we proceeded in a similar way to \citet{Davies15}.

We used the local thermal equilibrium (LTE) MARCS code \citep{Gustafsson} to construct the stellar atmosphere models in spherical geometry over 56 depth points. For consistency, we follow the standard MARCS physical parameters recommended for red giants for all the models, that is, we used 1.5 for the mixing length parameter, 0.076 for the temperature distribution within the convective elements, and 8 for the  energy  dissipation  by turbulent viscosity. From the model atmosphere grid, high-resolution synthetic spectra were calculated using the turbospectrum 1D LTE radiative transfer code \citep{Turbospectrum}, including atomic and molecular line lists (VO, CaH, FeH, CrH, SiO, MgH, CH, C2, and CN) and most importantly, TiO, which commonly serves as an important indicator for spectral types in the optical region. The solar-scaled abundance ratios were taken from
\citet{Grevesse2007}. We degraded the resolution of the synthetic MARCS spectra from $R$\,=\,500\,000 to $800$, which is the average resolution of our flux-calibrated RSG spectra. In addition, we added artificial noise to the synthetic spectra, which corresponds to an S/N\,$\approx$\,30 (the average S/N of the observed spectra). The effective temperatures of the grid (from 3200~K to 5000~K in steps of 50~K) have a wider range than the Davies grid. For the mass ($M$), microturbulence velocity ($v_{mic}$), and surface gravity ($\log~g$) parameters, \citet{Davies15} argued that changes in mass within the range of typical RSGs (8-25 M$_\odot$) do not significantly affect the atmosphere structure. They also demonstrated that microturbulent velocities in RSGs are nearly constant and that surface gravity and metallicity are strongly degenerate when the parameters are derived by spectral fitting. Therefore, we fixed these parameters such that $M$ = $15~M_\odot$ \citep[typical of RSGs;][]{Davies15}, $\log~g = 1$, and $v_{mic} = 4~\mathrm{km} \, \mathrm{s}^{-1}$.  Finally, we assumed $\mathrm{[Fe/H]}$ = $-1$ dex as the most appropriate for RSGs in dIrr galaxies. We only varied the effective temperature and extinction (A$_{V}$) in order to obtain the best match of the modeled and observed RSG SEDs.


With a precomputed grid of synthetic spectra, we fitted the observed SED using the $\chi^{2}$ minimization that returns the final values of $T_\mathrm{eff}^\mathrm{SED}$ and A$_{V}$. We varied the $T_{\rm eff}$ from 3200~K to 5000~K with a step of 50~K and the extinction in a range from 0.1 to 3 with a step 0.1 mag, but in some cases (e.g., for targets in IC 10), we modified the range of extinction up to 4 mag. To calculate the extinction across all wavelength ranges, we used the extinction law from \cite{2014_new_extlaw}. We used a constant value of the total to selective absorption extinction $R_{V}$ = 3.1, which is a best-fit value for RSGs, as was shown in \citet{Levesque05}. The resulting uncertainties in derived parameters indicate the goodness of fit based on chi-squared statistic. 

The goodness of the $\chi^{2}$ minimization and the resulting best-fit SEDs for the RSGs of each galaxy are presented in Figures \ref{Fig_peg1} -- \ref{Fig_ic101}. In each figure we present the final fit of observed and modeled MARCS SEDs together with archival optical $BVI$ band photometry. The names of the RSGs in each plot are those given in Table \ref{tab:hr_diag}. For the majority of targets in IC 10, we were unable to fit the observed SED properly. The derived values of the effective temperature, luminosity, and extinction therefore cannot be considered reliable. The reason is that the wavelength regions in the observed SEDs are limited, and we were unable to find a reliable fitting solution (see Figure \ref{Fig_ic101}). We indicate these targets by a question mark in Table \ref{tab:hr_diag}. The superposed photometric bands for each of the SED fitting solutions show the goodness of the flux calibration of the observed RSG candidates in terms of the target color. The photometric and spectroscopic observations do not differ significantly.

Based on the best-fit MARCS model for each individual RGS SED, we calculated the luminosity ($L^\mathrm{SED}$) for these targets through the integrated flux of the synthetic SED,
\begin{equation}
\log \frac{L^{\mathrm{SED}}}{L{_\mathrm{\sun}}} = \log{(4 \pi  \, d^2 \, flux / L{_\mathrm{\sun}})}
,\end{equation}
where $d$ is the average distance to the host galaxy, and $flux$ is the integrated flux for the best-fit MARCS model.
We used Monte Carlo simulations that vary the distances, $T_\mathrm{eff}^\mathrm{SED}$ and A$_{V}$, within the uncertainties in order to calculate the uncertainties in derived luminosity for each target.

With the determined effective temperatures and luminosities, we derived the final values of radii using the Stefan-Boltzmann equation:
\begin{equation}
R^{\mathrm{SED}}/R_{\odot}=(L^\mathrm{SED}/L_{\odot})^{0.5}(T_\mathrm{eff}^\mathrm{SED}/5770)^{-2}.
\end{equation}

The preliminary values of $T_\mathrm{eff}^\mathrm{SED}$,  $L^\mathrm{SED}$, $A_{V}^\mathrm{SED}$, $R_{V}$ , and $R^{\mathrm{SED}}$ for 25 targets are presented in Table \ref{tab:hr_diag_L}.


\subsection{Alternative photometric techniques for determining RSG luminosities}

To determine the reliability of the SED fitting approach for deriving RSG luminosities and temperatures, we applied several empirical photometric techniques. They gave us estimates of the bolometric corrections (BCs) and the effective temperatures. All of them have been tested in the literature for various samples of RSGs in the MCs. Here, we briefly describe these techniques. More detailed information is presented for a VLT-FLAMES sample of RSGs in the 30 Doradus region \citep{Britavskiy_2019}.

\begin{enumerate}
\item{The single-band technique (I band).}
One photometric approach is the empirical near-IR band-calibration technique presented by \citet[][equation 2]{Davies13}. Based on the assumption that in the MCs the bolometric correction for RSGs is constant for each given band, the authors presented a BC calibration for several optical and near-IR bands. We chose the I band for this analysis because the effect of extinction is relatively weak and this band is available in all photometric surveys of our targets. Moreover, the maximum of the RSGs SED is located near the I band, which is important for the accuracy of the photometry for our faint-target sample. When the apparent I-band magnitudes and distance modulus to the host galaxy are known, the luminosity estimation is straightforward.
The luminosities $L^{(I-band)}$ are presented in Table \ref{tab:hr_diag_L}.

\item{The J-K technique.}
This method uses the $(J-K_{s})_{0}$ color and is based on the bolometric correction for the spectroscopically late-type long periodic variables \citep{Bessell_1984}. This method is relatively insensitive to extinction. Moreover, using the RSG sample in the LMC and SMC from \citet{Tabernero_2018}, we can obtain an effective temperature calibration based on the $(J-K_{s})_{0}$ color:  ${\rm T}_{\rm eff}^{(J-K)} = -1432 \times (J-K_{\rm s})_{0} + 5449),$ which is based on the RSG sample in the SMC \citep[see Section 3.1.2 in][]{Britavskiy_2019}. This relation is very similar to the relation presented in \citet{Neugent_2012} for the LMC: ${\rm T}_{\rm eff} = -1746.2 \times (J-K_{\rm s})_{0} + 5638)$. These similarities occur because of the sampling of the targets. The RSG sample of \citeauthor{Tabernero_2018} consisted only of red targets: with a range of colors 0.9 < $(J-K_{\rm s})_{0}$ < 1.4 for the LMC, and 0.7 < $(J-K_{\rm s})_{0}$ < 1.1 for the SMC.  \citeauthor{Neugent_2012} used a larger sample of RSGs in the LMC, however, that included some yellow RSGs: 0.7 <$ (J-K_{\rm s})_{0} $< 1.4.
The resulting values $L^{(J-K)}$ and $T_\mathrm{eff}^{(J-K)}$ are presented in Table \ref{tab:hr_diag_L}.

\item{The V-K technique.}
This is the classic method based on the optical spectroscopy of RSGs in the MCs \citep{Levesque05,Levesque2006}. With this technique it is possible to estimate the effective temperatures and BCs of RSGs using the $V-K$ color. We used the $V-K$ relation, which was adapted for the SMC metallicity, as the most metal-poor calibration published. The main disadvantage of this method is that it is highly sensitive to extinction, which is usually unknown. We used the $A_{V}^\mathrm{SED}$, which we determined from the SED fitting. 
The resulting values $L^{(V-K)}$ and $T_\mathrm{eff}^{(V-K)}$ are presented in Table \ref{tab:hr_diag_L}.
\end{enumerate}

Taking into account that these photometric techniques are based on a limited number of RSGs and were calibrated only to the LMC and SMC metallicities, the accuracies presented in the reference studies for each of the methods most likely underestimate the true accuracy. In order to estimate the errors of the derived luminosities for each of these techniques, we therefore took the dispersion of the methods together with errors of the target photometry and extinction values into account using Monte Carlo simulations (I-band, J-K, and V-K techniques). 

\begin{figure*}
\includegraphics[width=0.5\linewidth]{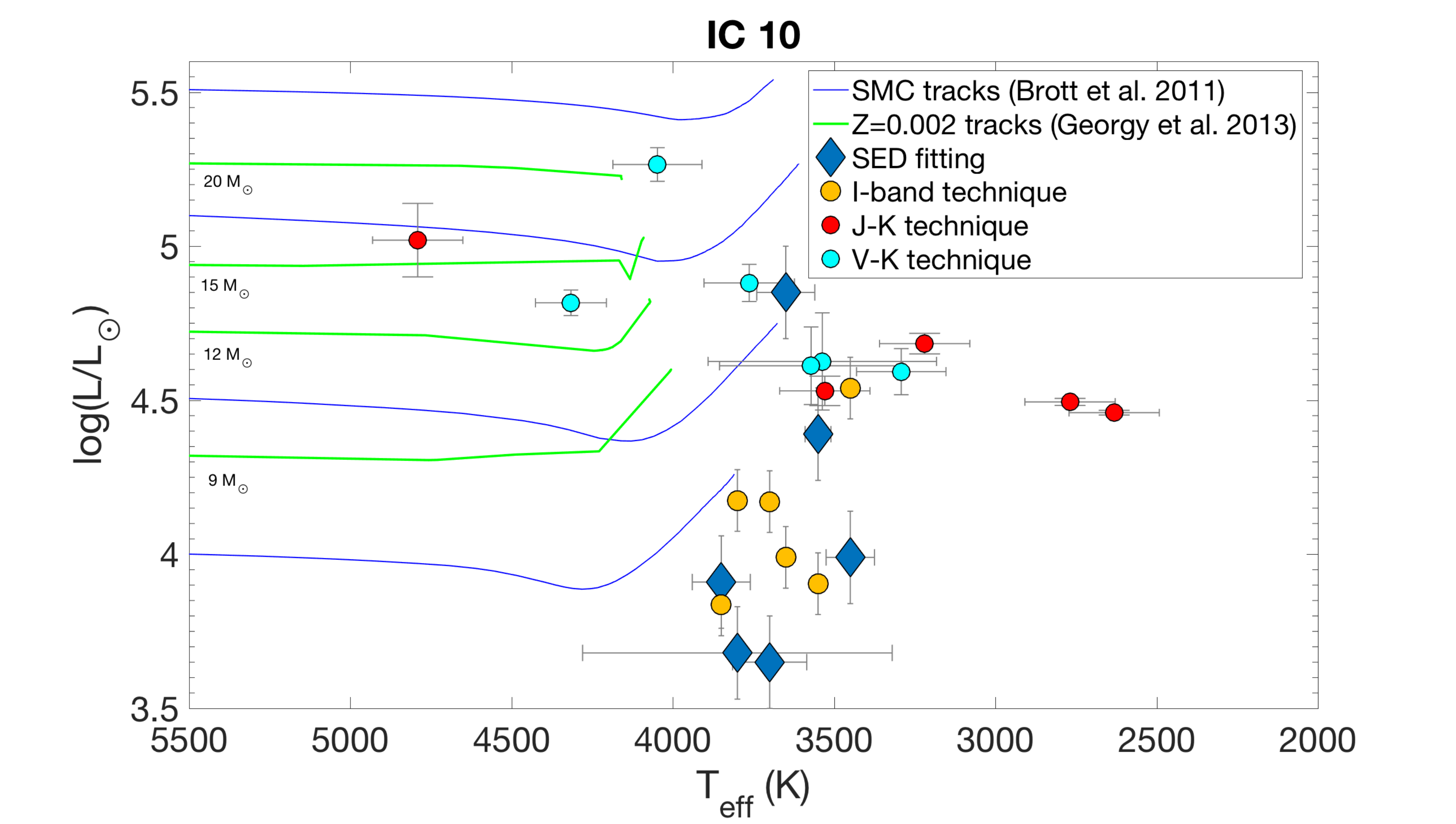}
\includegraphics[width=0.5\linewidth]{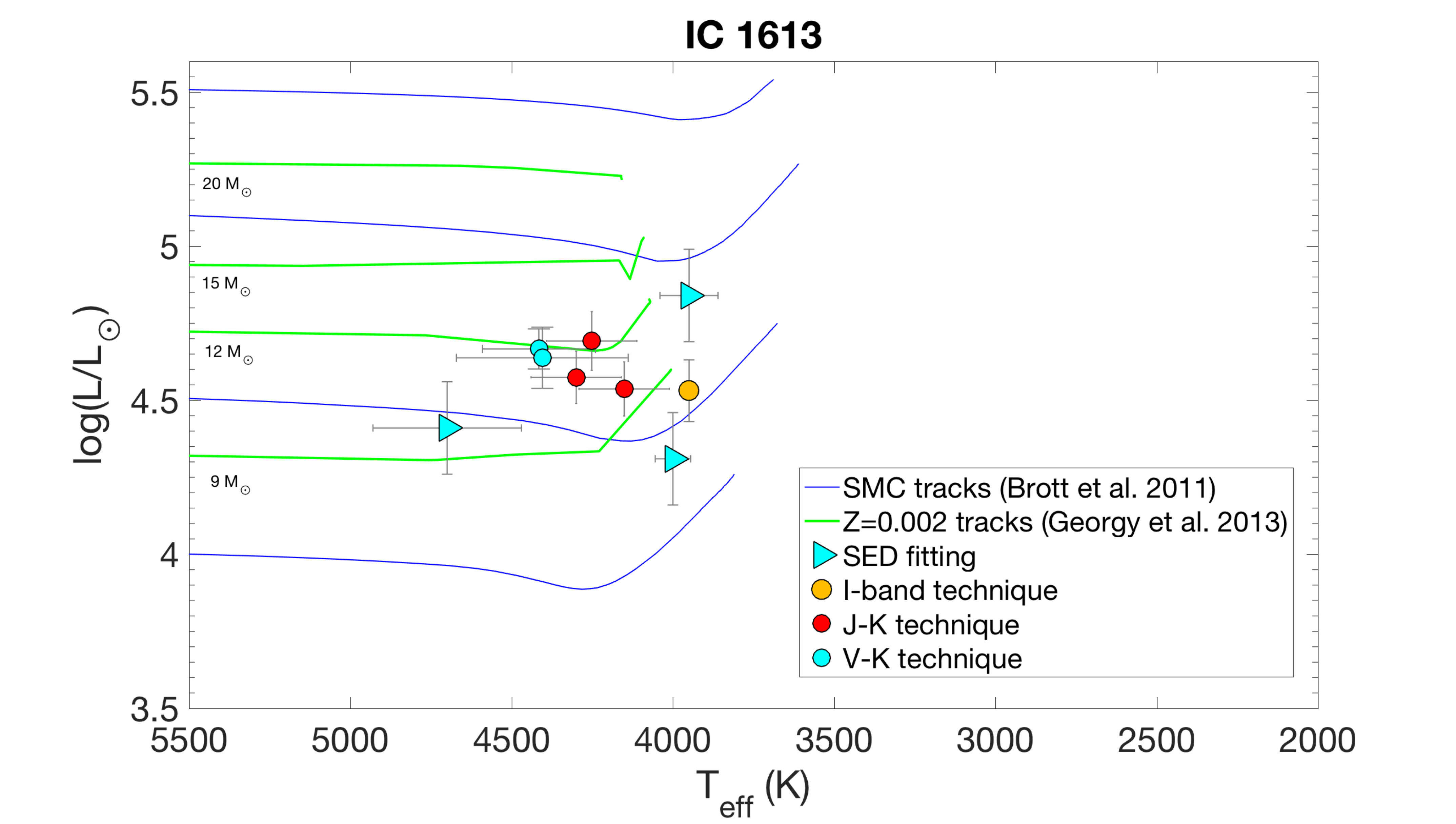}
\includegraphics[width=0.5\linewidth]{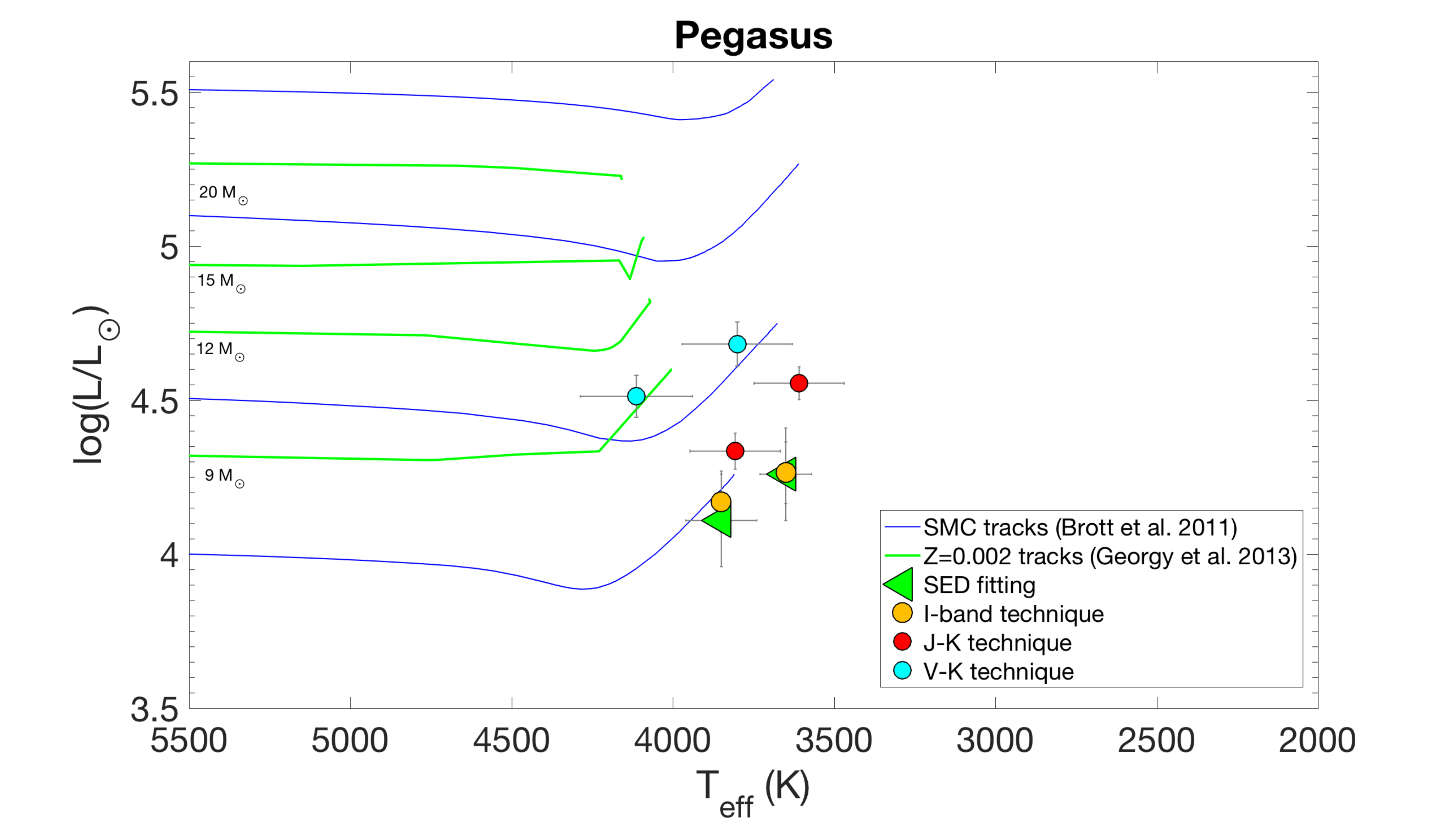}
\includegraphics[width=0.5\linewidth]{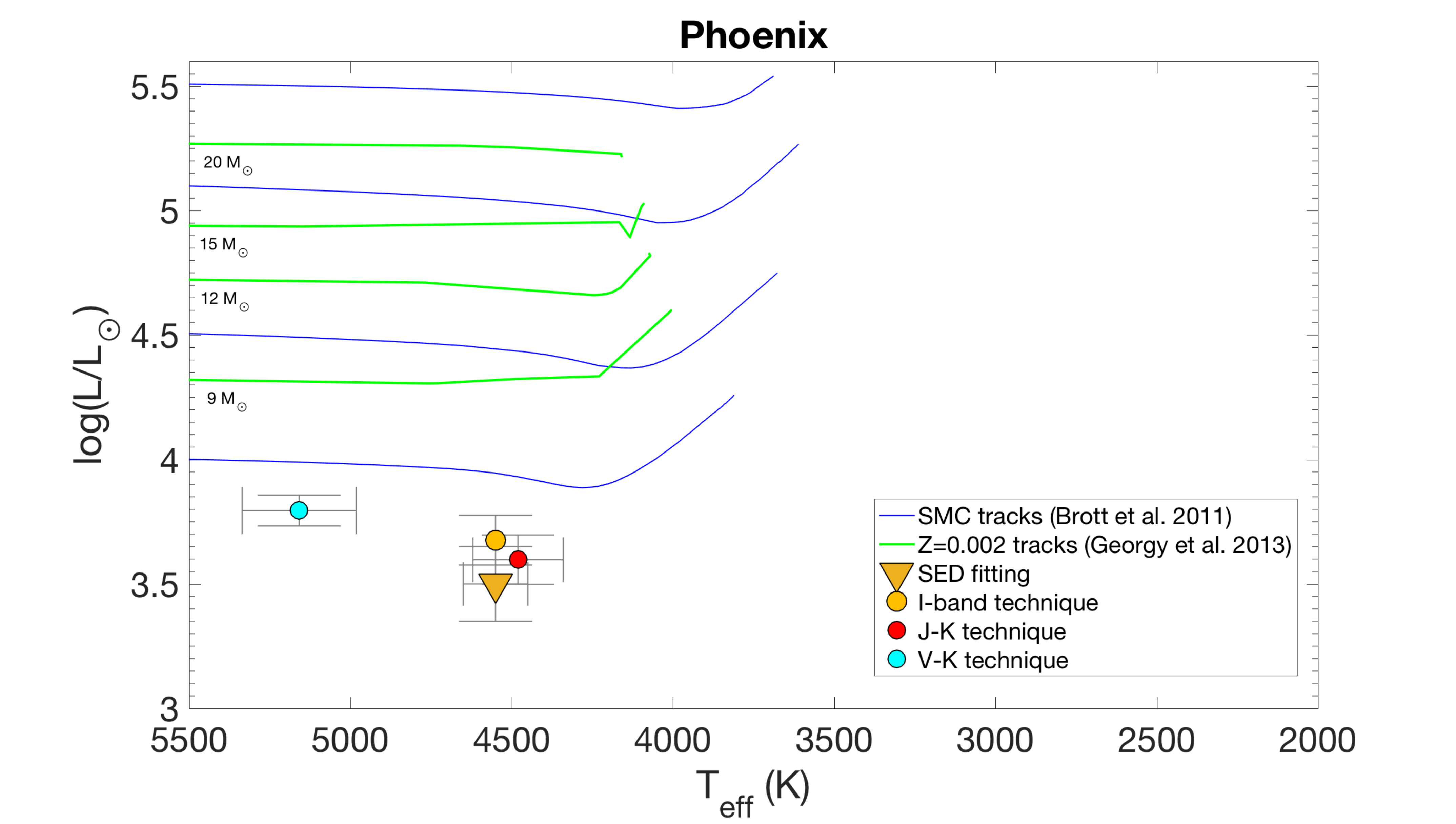}
\includegraphics[width=0.5\linewidth]{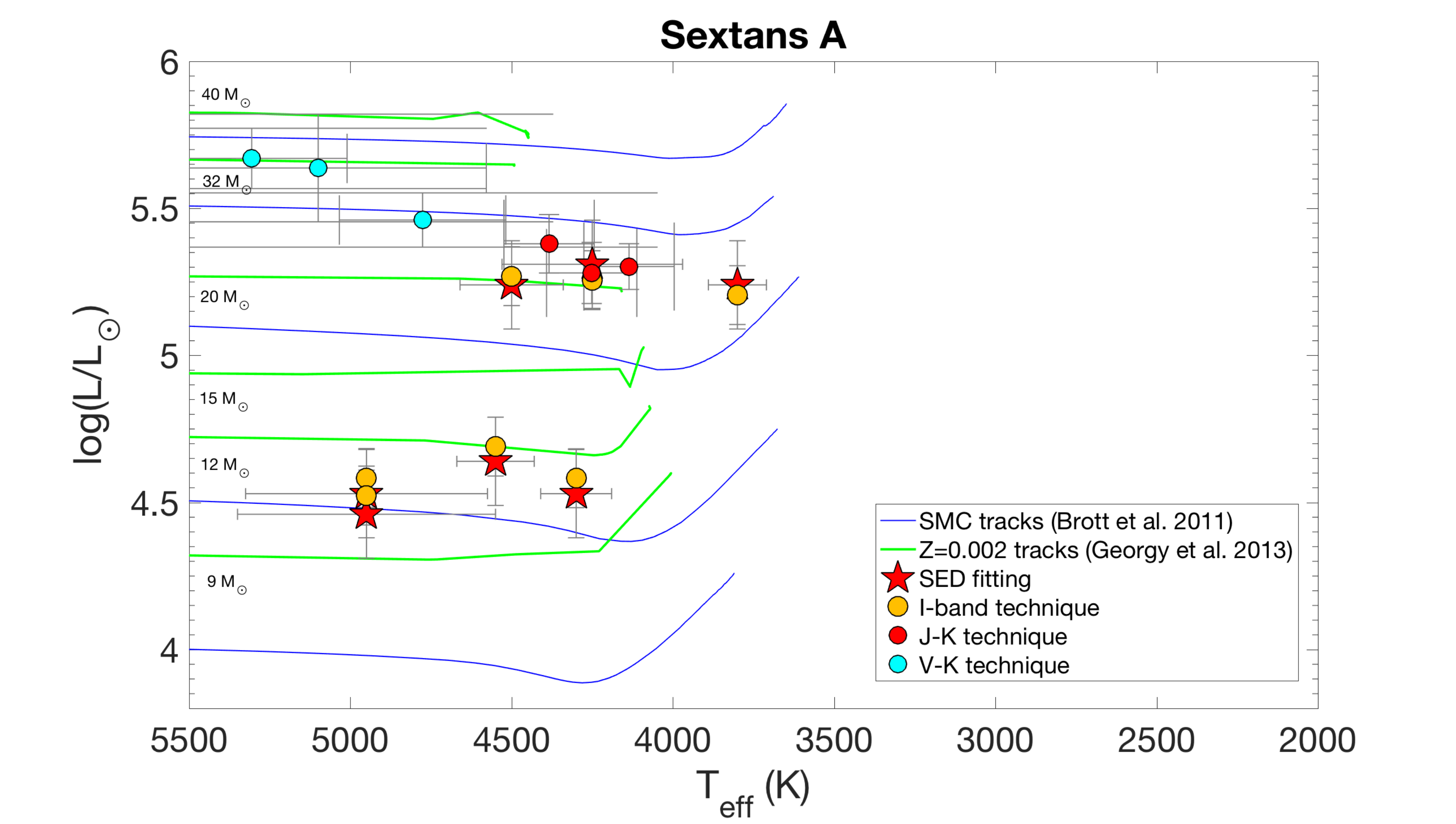}
\includegraphics[width=0.5\linewidth]{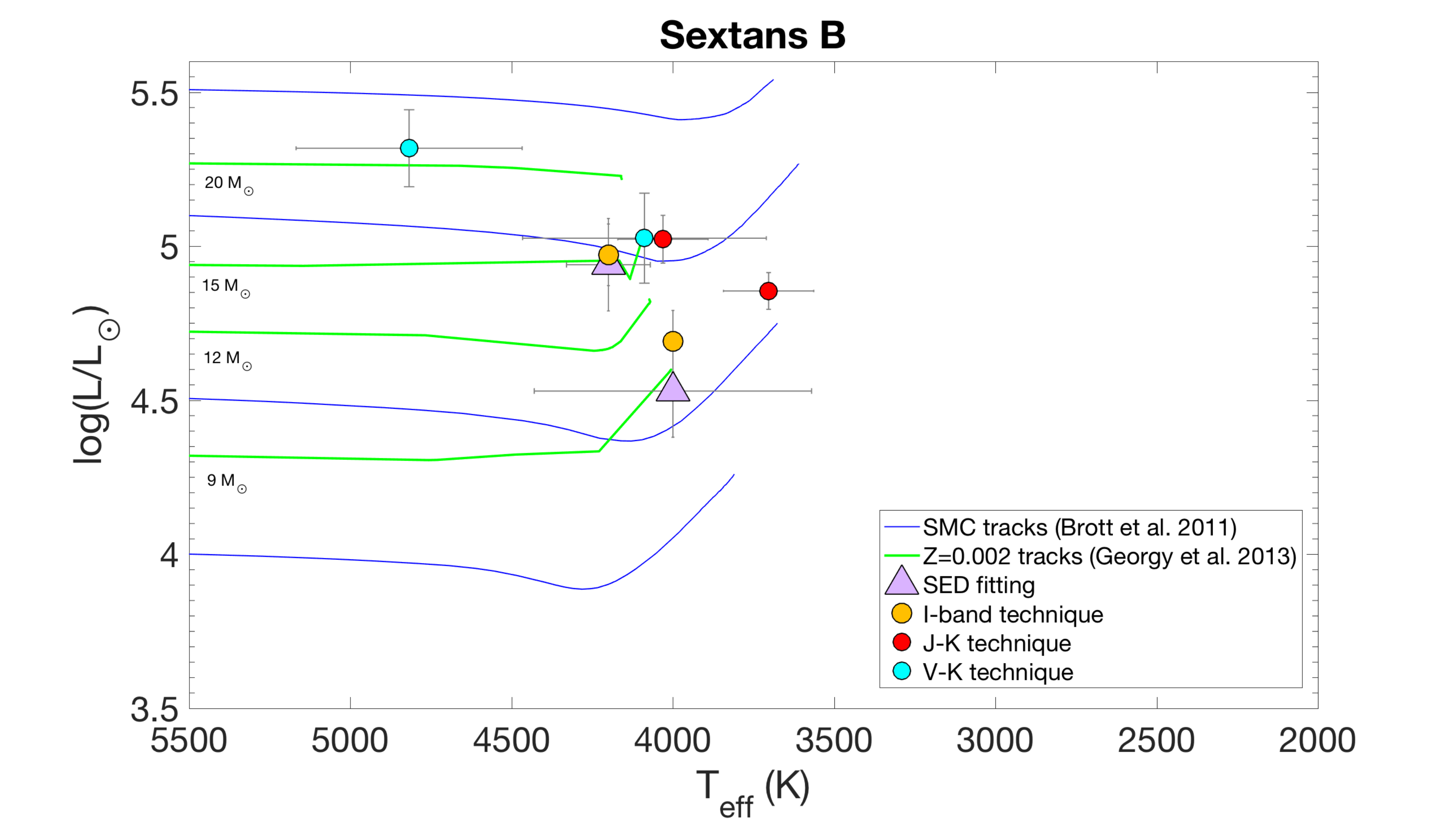}
\includegraphics[width=0.5\linewidth]{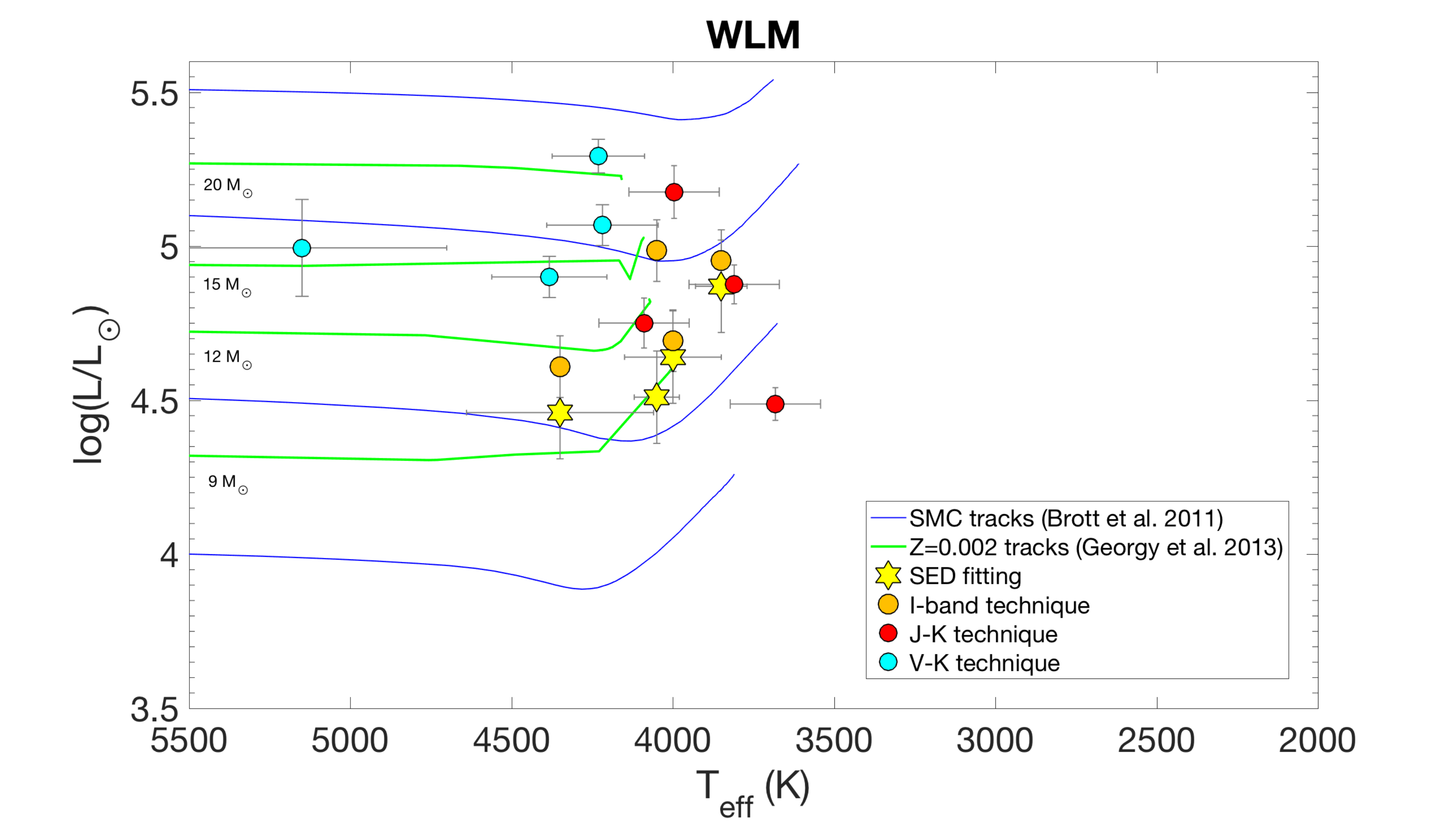}
\caption{Red supergiant region of the Hertzsprung--Russell diagram for all identified RSG candidates in each program galaxy compared with the evolutionary tracks from \cite{Brott2011} for the SMC metallicity and the evolutionary tracks from \citet{Georgy_2013} for Z=0.002. For each target we present the results obtained from different techniques:  the SED technique ($T_\mathrm{eff}^\mathrm{SED}$,  $L^\mathrm{SED}$), I- band technique ($T_\mathrm{eff}^\mathrm{SED}$, $L^{(I-band)}$), J-K technique ($T_\mathrm{eff}^{(J-K)}$, $L^{(J-K)}$), and the V-K technique ($T_\mathrm{eff}^{(V-K)}$, $L^{(V-K)}$ ).}
\label{Fig_HR_All}
\end{figure*}

The obtained luminosities and effective temperatures for a program RSG, derived with each method, are presented in Table \ref{tab:hr_diag_L}. In Figure \ref{Fig_HR_All} we present the Hertzsprung--Russell (H--R) diagram for each galaxy with all RSGs and AGB stars, together with the SMC evolutionary tracks from \cite{Brott2011} and the evolutionary tracks for Z\,=\,0.002 from \citet{Georgy_2013}. Both samples of the evolutionary tracks are for a set of rotating stellar atmosphere models. We plot the luminosity and effective temperature estimates that we derived from the different methods. Using $L^{\mathrm{SED}}$ as a reference, we find interesting objects in the H--R diagram in Figure \ref{Fig_HR_All}. For instance, the three very luminous RSGs in Sextans A appear to be very massive, with radii of $\approx$\,900\,$R_{\odot}$. These two different RSG populations in Sextans A illustrate the different star formation regions of the galaxy, or the red straggler phenomena \citep[as discussed in][]{Beasor_2019,Britavskiy_2019}. In addition, we can clearly see a high discrepancy in the physical parameters of IC 10 targets and also in the accuracy of the SED fitting for these targets (see Fig. \ref{Fig_ic101}). The reason of this unreliable fitting is a high interstellar extinction toward IC 10 because it lies near the Galactic plane.

\begin{table*}                                                                                          
{\small                                                                                         
\caption{Basic information of the identified RSG candidates in dIrr galaxies.}                                                                                              
\label{tab:hr_diag}                                                                                             
\begin{tabular}{lcccccccc}                                                                                              
\hline\hline                                                                                            
  RSG Name   &   DUSTiNGS &      RA  &          DEC  &     V    &          I      &          J    &           K  &          Sp. Class \\
                       &             ID        &         (deg)    &      (deg)          &        (mag)           &       (mag)            &       (mag)            &      (mag)         & Present work and Paper II \\
\hline
IC 10    1&       103677        &       5.0575594&      59.2875137&      22.043  &       19.038& 16.811& 14.871&     K3-5 -- Early M / AGB star (CN bands)    \\
IC 10    2&       117107        &       5.0198626&      59.2903671&      22.556  &       18.852& 15.751& 15.248&     Late M I / AGB candidate     \\
IC 10    3&       95408 &       5.0804367&      59.3092765&      20.229 &         17.136& 15.269& 13.753&     M1-3 I    \\
IC 10    4 &      99773 &       5.0684976&      59.2951965&      21.082 &         18.132& 16.175& 14.832&     M0-2 I  / AGB candidate    \\
IC 10    5 &      85592 &       5.1077690&      59.3035316&      21.415 &         18.122& 16.032& 14.483&     M1-3 I / AGB candidate \\
IC 10    6&       107961        &       5.0455651&      59.2827033&      21.911  &       18.621& 16.520& 14.671&     K3-5 / AGB star (CN bands)   \\
\hline                                  
IC 1613  1 &    161666& 16.158911&      2.112404&        19.046 &       17.36&  16.272& 15.343& IC1613-1 in Paper I, Late K I --  M0-2 I    \\
IC 1613  2 &    119457& 16.210361&      2.106991&        18.941 &       17.279& 16.213& 15.383&    IC1613-2 in Paper I,  M2-4 I    \\      
IC 1613  3 &      97761    &    16.237533&      2.078947&        18.623 &         17.107& 16.080& 15.218&    K1-3 I     \\
\hline                                  
Pegasus 1 &     116602& 352.14938&              14.73709 &      20.722& 18.428& 17.165& 16.007&  M0-2 I / AGB candidate  \\
Pegasus 2 &     136539& 352.12616&              14.74971 &      20.689& 18.684& 16.640& 15.350&   K4-5 I / AGB candidate  \\
\hline                                                          
Phoenix 3 &     119803& 27.79501&               $-$44.41927&    19.507& 18.29&  17.470& 16.760&   K1-2  / AGB / RGB star \\
\hline                                          
Sextans A  4  & 77330&  152.76654&              $-$4.70795 &    20.031& 18.35&    --       &      --      &         K1-3 I     \\
Sextans A  5  & 72683&  152.77316&              $-$4.69916 &     18.322 &         16.530& 15.661& 14.722&    K1-3 I     \\
Sextans A  6  & 70373&  152.77670&              $-$4.70510 &     19.588 &         18.059&    --    &              --       &         Late G -- Early K I \\
Sextans A  7  & 106505& 152.72426&              $-$4.68539 &     18.295 &         16.493& 15.584& 14.810&    K1-3 I     \\
Sextans A  8  & 102187& 152.73050&              $-$4.71217 &     19.935 &         18.347&   --     &              --       &         K1-3 I     \\
Sextans A  9  & 98470&  152.73587&              $-$4.70284 &     19.985 &         18.509&   --     &              --       &         K1-3 I     \\
Sextans A  10 & 98112&  152.73636&       $-$4.67753  &   18.596 &       16.667& 15.758& 14.896&    K3-5 I     \\
\hline          
Sextans B 1  &    100179         &      149.994064&     5.326573&        18.997  &       17.297& 16.298& 15.289&     K1-3 I   \\
Sextans B 2  &  82970&  150.017166&     5.309929&        19.979 &       18.055& 16.672& 15.445&     K1-3 I   \\                                                                                                  
\hline                                                  
WLM    11 &     101523& 0.48976&          $-$15.48786  &         19.266 &         17.628& 16.959& 15.718&    K1-3 I     \\
WLM    12 &     90263&  0.50340&          $-$15.52166  &         18.690 &         17.40&  15.901& 14.745&    K1-3 I     \\
WLM    13 &     94581&  0.49837&          $-$15.51678   &        18.980 &         16.61&  16.257& 15.287&     K1-3 I     \\
WLM    14 &     83414&          0.51268&          $-$15.50950   &        18.676  &       16.698& 15.294& 14.262&    K4-5 I     \\        
\hline  
\hline                                                                          
\end{tabular}
{\bf Notes.} In Paper I we identified two RSGs in IC 1613. However, there is an error in their coordinates: the IDs are IC1613-1 and IC1613-2. This table contains the correct coordinates for these objects. }                                                                                   
\end{table*}                                                                                            

\section{Discussion}

The reliability of the obtained effective temperatures and luminosities and the preferred methods are the first points to discuss. We cannot give a preference to one of the methods yet because we work in a narrow wavelength range and simply measure the temperature and luminosity at a given depth in the extended atmosphere of RSGs. The SED fitting and V-K techniques are based on MARCS stellar atmosphere models, and the strengths of the TiO band depths are not connected with a temperature based on the state of the atomic lines in the spectra \citep[i.e., CaT or J-band spectroscopic techniques,][]{Tabernero_2018,Patrick15}.

The results obtained using the V-K technique deserve particular attention. They systematically overestimate the values of $T_{\mathrm{eff}}$ and $L$ in comparison with the other methods. The main reason for these discrepancies are uncertain values of A$_{V}$, which were derived by SED fitting. The RSGs in our sample are optically faint, which results in large uncertainties in $V$-band values and significant uncertainties of the SED fitting of the spectra with low S/N with synthetic spectra. In addition, the V-K technique is only suitable for SMC metallicity; our targets are more metal poor than the SMC. These reasons explain the large differences in the resulting values. 

In order to minimize the uncertainties in the derived parameters, we first of all suggest that the optical bands should be avoided for deriving the temperature and luminosity of RSGs. They are significantly affected by extinction. A reliable extinction is difficult to derive without spectra that cover a wide wavelength range. Second, we suggest that photometry in the $H$ and $K$ bands is not included in such analyses because these bands show effects of mass-loss excess, in particular, from episodic mass-loss events. For our analysis we adopted the $L^{\mathrm{SED}}$ because it is more reliable to derive luminosities from observational spectra, and the resulting values are in agreement with the photometric methods that are relatively free of extinction, that is, the J-K and I-band techniques.

\begin{figure}
\begin{center}
\resizebox{\hsize}{!}{\includegraphics{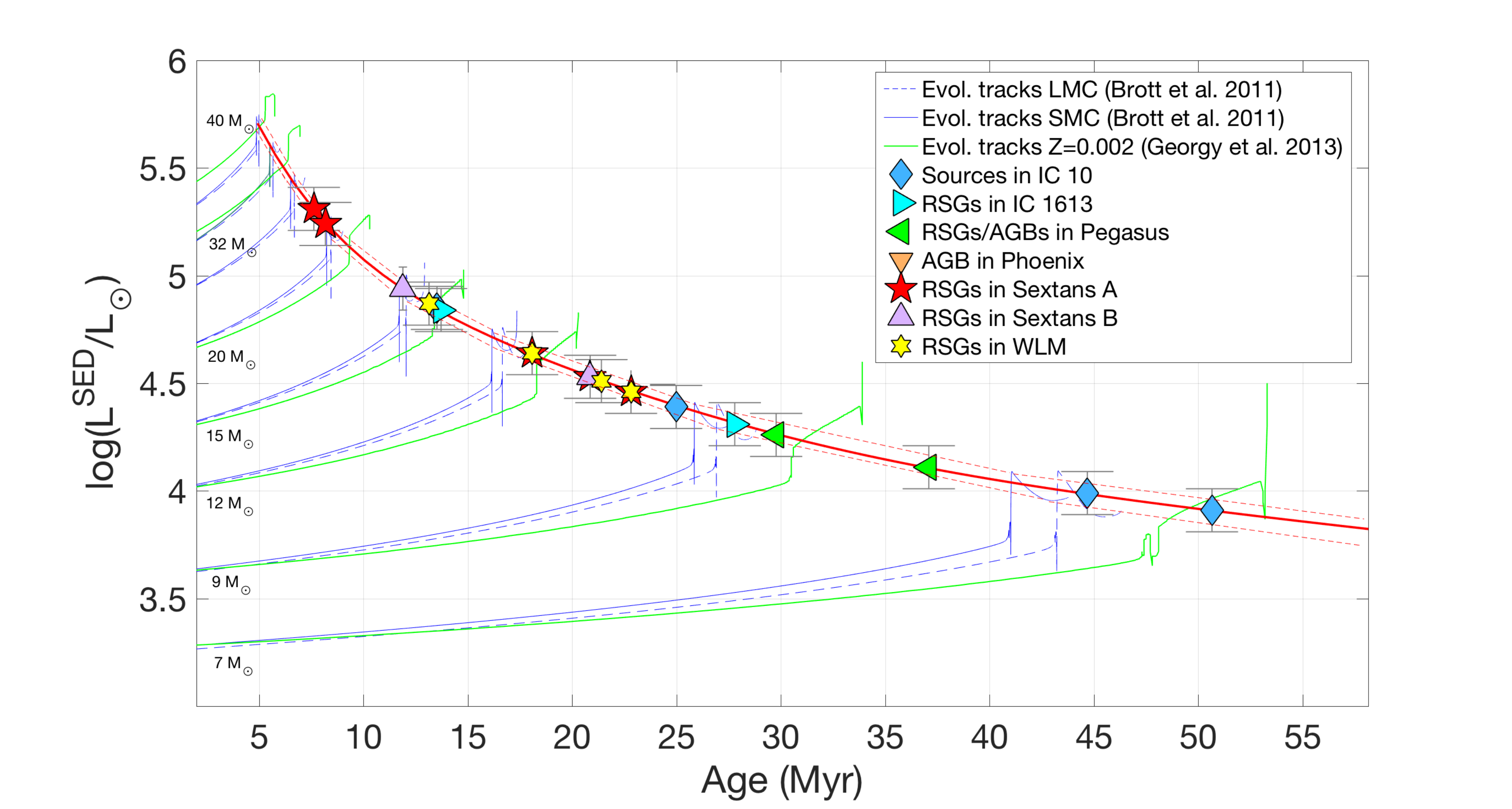}}
\end{center}
\caption[]{Luminosity-age diagram for the RSG region based on the LMC and SMC evolutionary tracks \citep{Brott2011}, together with the evolutionary tracks from \citet{Georgy_2013}. The evolutionary stage of the RSG region, i.e.,  the He-burning phase, is marked by the red dashed lines. The solid red line corresponds to the weighted polynomial fit of the He-burning phase based on the SMC evolutionary tracks.}
\label{Fig_Z_Age}
\end{figure}

Targets that are located below the $8~M_{\odot}$ evolutionary track (see Fig.\ref{Fig_HR_All}) are probably massive AGB stars and are labeled AGB star candidates in Table \ref{tab:hr_diag}. As an additional check for possible AGB stars in our sample, we placed our 25 RSG candidates on the luminosity-age (L-Age) diagram (Figure~\ref{Fig_Z_Age}). This diagram is based on the LMC and SMC evolutionary tracks by \citet{Brott2011} with an initial rotation rate of 150 ($\mathrm{km} \, \mathrm{s}^{-1}$). In addition, we placed the evolutionary tracks from \citet{Georgy_2013} for Z\,=\,0.002 with rotation (150 -- 300 $\mathrm{km} \, \mathrm{s}^{-1}$) there. The advantage of this method is that at a given luminosity, independently of effective temperature estimates, it is easy to fit the position of the sources to the narrow theoretical RSG region that corresponds to the final evolutionary phase (He-burning phase) of massive stars with initial masses from 5 to $40~M_{\sun}$. We fit the luminosities of each RSG and AGB star to the SMC evolutionary tracks as to the closest sample of tracks in terms of the average metallicity of all program galaxies. However, this analysis strongly depends on the model: Figure~\ref{Fig_Z_Age} shows that the position of the RSG region significantly varies depending on which model we used \citep[see also Fig. 7 in][]{Britavskiy_2019}. This clearly shows how sensitive the RSG evolution is to the different underlying physics in the stellar atmosphere models. This diagram shows that all bona fide RSGs are located above the log$(L/L_{\sun}) = 4.3$ limit. Other targets are likely massive AGB stars (e.g., two targets in Pegasus). Targets below log$(L/L_{\sun}) = 4.3$ correspond to an age $\approx$30~Myr or older and belong to the AGB or RGB stellar population at the SMC metallicity.

\begin{figure}
\begin{center}
\resizebox{\hsize}{!}{\includegraphics{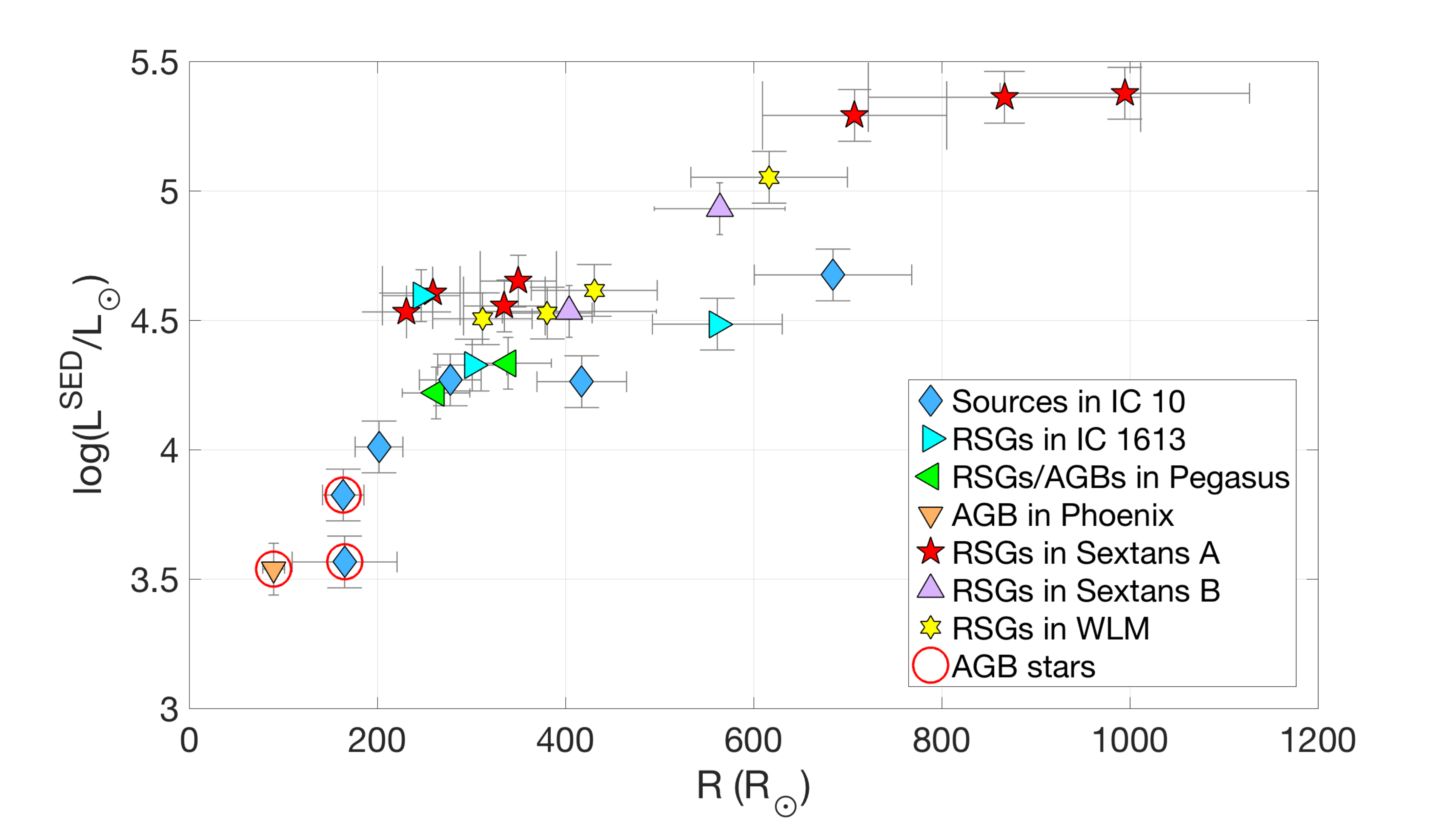}}
\end{center}
\caption[]{Distribution of the program RSG radii according to their luminosities. Different symbols are used to label targets from different host galaxies. All bona fide RSG are located above the log$(L/L_{\sun}) =4.3$ luminosity limit.}
\label{Fig_ZR}
\end{figure}

Figure \ref{Fig_ZR} shows the relation of stellar radii with luminosities ($L^{\mathrm{SED}}$) for all RSG candidates. The targets are divided into two groups: RSG and AGB stars, with a separation at luminosity log$(L/L_{\sun}) = 4.3$. For some extremely low-luminosity targets, the CN bands become visible in the spectra, which indicates the carbon-rich population of AGB stars. These targets were considered as carbon stars; we indicate them in Table \ref{tab:hr_diag} and Figure \ref{Fig_ZR}. 
Most of the RSGs, except for the most luminous ones, have a typical radius of $R^{\mathrm{SED}}$\,$\approx$\,300\,$R_{\sun}$, which is in agreement with studies of type II supernova progenitor radii \citep[e.g.,][]{Garnavich_2016}. The systematic accuracy of our radius measurements is not higher than 50~$R_{\sun}$, mainly because of the uncertainties in target distances, which we assumed to be constant for each dIrr galaxy.

\newgeometry{left=1cm}                                                                                                                                                  \begin{table*}                                                                                          
{\scriptsize                                                                                            
\caption{Fundamental physical parameters for all identified RSG candidates in dIrr galaxies derived with different techniques.}                                                                                                
\label{tab:hr_diag_L}                                                                                           
\begin{tabular}{lcccccccccccc}                                                                                          
\hline                                                                                          
\hline                                                                                          
 RSG Name       &        RA     &       DEC&       $T_\mathrm{eff}^\mathrm{SED}$ &           A$_{V}^\mathrm{SED}$  &          $R^{\mathrm{SED}}$   &                $T_\mathrm{eff}^{(V-K)}$   &   $T_\mathrm{eff}^{(J-K)}$        &         $L^\mathrm{SED}$ &       $L^{(I-band)}$  &            $L^{(J-K)}$                  &       $L^{(V-K)}$    \\
                  &     (deg)      &          (deg)          &     (K)    &           (mag)        &        (R$_{\sun}$)  &                          (K)          &        (K)       &        log($L/L_{\odot}$)  &       log($L/L_{\odot}$)        &        log($L/L_{\odot}$)                       &       log($L/L_{\odot}$)     \\                 
                         &         &                &      &                 &                           &            &   $\pm$ 140        &        $\pm$ 0.10 &  $\pm$ 0.10       &                               &              \\                                                                         
\hline                                                                  
IC 10    1&     5.0575594&      59.2875137&      3800$\pm$480 (?)&         2.8$\pm$1.1 (?)&      165$\pm$60&       3540 $\pm$   350   &  3200&            3.68  (?)&      3.84&      4.46 $\pm$     0.10   &        4.63 $\pm$    0.16 \\ 
IC 10    2&     5.0198626&      59.2903671&      3450$\pm$75 (?)&            4.9$\pm$0.2 (?)&    280$\pm$30 &      4320 $\pm$   110  &   4400&            3.99  (?)&      3.90&      4.83 $\pm$     0.12   &        4.82 $\pm$    0.05 \\ 
IC 10    3&     5.0804367&      59.3092765&      3650$\pm$90 &     3.4$\pm$0.3 &       685$\pm$90 &      4050 $\pm$   140  &    3550&          4.85 &  3.99&      4.99 $\pm$      0.04   &        5.27 $\pm$    0.05 \\ 
IC 10    4 &    5.0684976&      59.2951965&      3850$\pm$90 (?)&            1.8$\pm$1.0 (?)&    200$\pm$25 &      3570 $\pm$   280  &    3700&           3.91  (?)&      3.84&      4.53 $\pm$     0.05   &        4.61 $\pm$    0.13 \\ 
IC 10    5 &    5.1077690&      59.3035316&      3550$\pm$40&        3.3$\pm$0.4 &       420$\pm$50 &      3760 $\pm$   140  &   3530&           4.39  &         3.90&      4.68 $\pm$      0.05   &        4.88 $\pm$    0.06 \\ 
IC 10    6&     5.0455651&      59.2827033&      3700$\pm$115 (?)&          1.8$\pm$0.7 (?)&     160$\pm$25 &      3790 $\pm$   140    &   3280&          3.65 (?)&       4.17&      4.49 $\pm$     0.10   &        4.59 $\pm$    0.07 \\ 
\hline                                                  
IC 1613  1 &    16.158911&      2.112404&        4000$\pm$75   &          1.0$\pm$0.3 &  300$\pm$40 &      4420 $\pm$   170  &           4150 &    4.31 & 4.50&      4.54 $\pm$     0.10   &        4.67 $\pm$    0.07 \\ 
IC 1613  2 &    16.210361&      2.106991&        3950$\pm$90  &   0.8$\pm$0.4 &       560$\pm$70 &       4400 $\pm$   270   &         4300 &    4.84 &        4.53&      4.57 $\pm$      0.09   &        4.64 $\pm$    0.10 \\ 
IC 1613  3 &    16.237533&      2.078947&        --     &         --  & --    &       5950  $\pm$  350  &   4250          &  --  & 4.59&      4.69 $\pm$     0.10   &         5.24 $\pm$    0.12 \\ 
\hline
Pegasus 1 &     352.14938&              14.73709      &          3650$\pm$80   &      1.6$\pm$0.4 &  340$\pm$50 &       4110 $\pm$   170 &         3800&              4.26  &         4.27&      4.33 $\pm$     0.07   &        4.51 $\pm$    0.07 \\ 
Pegasus 2 &     352.12616&              14.74971      &          3850$\pm$110   &       1.6$\pm$0.4 &260$\pm$40 &        3800 $\pm$   170  &  3600 &           4.11  &         4.17&      4.56 $\pm$     0.06   &        4.68 $\pm$    0.07 \\ 
\hline                                                                  
Phoenix 3 &     27.79501&               $-$44.41927 &     4550$\pm$ 100   &        0.9$\pm$0.2 & 90$\pm$15 &        5160 $\pm$   170 &   4480 &  3.50&   3.68&      3.60 $\pm$      0.11  &         3.79 $\pm$    0.06 \\ 
\hline                                                                  
Sextans A  4  & 152.76654&              $-$4.70795   &   4300$\pm$110    &        1.6$\pm$0.2    &       335$\pm$40&              --                    &   --    &       4.53&   4.58&     --                                 &    --                       \\
Sextans A  5  & 152.77316&              $-$4.69916 &     4250$\pm$280   &         1.7$\pm$0.6    &       870$\pm$145 &   5100 $\pm$   520 &     4130  &  5.31&   5.26&      5.30 $\pm$      0.08   &        5.64 $\pm$    0.18 \\ 
Sextans A  6  & 152.77670&              $-$4.70510 &     4550$\pm$120    &         1.9$\pm$0.2    &      350$\pm$40 &      --                            &         --     &       4.64&   4.69&     --                                  &   --                        \\
Sextans A  7  & 152.72426&              $-$4.68539 &     4500$\pm$160     &        1.9$\pm$0.3    &      710$\pm$100 &   5300 $\pm$   290 &       4380 &  5.24&   5.27&      5.38 $\pm$     0.10   &        5.67 $\pm$    0.10 \\ 
Sextans A  8  & 152.73050&              $-$4.71217 &     4950$\pm$375   &          1.9$\pm$0.5    &      260$\pm$60 &      --                            &          --     &       4.53 &  4.58&    --                                     &         --                        \\
Sextans A  9  & 152.73587&              $-$4.70284 &     4950$\pm$400   &        1.8$\pm$0.8   & 230$\pm$50 &             --                     &        --      &       4.46&   4.52&     --                                    &        --                       \\
Sextans A  10 & 152.73636&       $-$4.67753 &    3800$\pm$100  &           1.5$\pm$0.3   &       995$\pm$130 &   4770 $\pm$   250  &      4250 &         5.24&   5.21&      5.28 $\pm$      0.10   &        5.46 $\pm$    0.09 \\ 
\hline                                                          
Sextans B 1  &  149.994064&     5.326573&        4200$\pm$130  &          1.5$\pm$0.5  & 565$\pm$70 &               4820 $\pm$   350  &  4030&   4.94&   4.97&      5.02 $\pm$      0.08   &        5.32 $\pm$    0.12 \\ 
Sextans B 2  &  150.017166&     5.309929&        4000$\pm$430  &          1.4$\pm$0.7 &  405$\pm$90&                4080 $\pm$   380  &   3700&  4.53&   4.69&      4.85$\pm$       0.06   &        5.03 $\pm$    0.15 \\ 
\hline                                                                                          
WLM    11 &     0.48976&          $-$15.48786  &         4350$\pm$290 &   1.7$\pm$0.5 &  310$\pm$50 &               5150 $\pm$   450  &  3680 &           4.46  &         4.61&      4.49 $\pm$     0.07   &        4.99 $\pm$    0.16 \\ 
WLM    12 &     0.50340&          $-$15.52166   &        4000$\pm$150  &           0.9$\pm$0.4 & 430$\pm$70 &               4220 $\pm$   170  &  3810 &           4.64&   4.69&      4.88 $\pm$     0.07   &        5.07 $\pm$    0.07 \\ 
WLM    13 &     0.49837&          $-$15.51678  &         4500$\pm$70 &    0.9$\pm$0.3  & 380$\pm$50 &               4380 $\pm$   180 &   4080 &           4.51&   4.99&      4.75 $\pm$     0.09   &        4.90 $\pm$    0.07 \\ 
WLM    14 &     0.51268&          $-$15.50950 &  3850$\pm$80  &   1.5$\pm$0.3 &       610$\pm$80 &               4230 $\pm$   140  &  3990 &          4.87  &      4.95&      5.18 $\pm$     0.08   &        5.29 $\pm$    0.06 \\ 
\hline
\hline                                                                                                                                                          
\end{tabular}                                                                                           
{\bf Notes.} $T_\mathrm{eff}^\mathrm{SED}$,  $L^\mathrm{SED}$ were determined used the optical SED fitting technique. $T_\mathrm{eff}^{(V-K)}$, $L^{(V-K)}$ are based on the V-K techniques.  $T_\mathrm{eff}^{(J-K)}$, $L^{(J-K)}$ are based on the J-K techniques. $L^{(I-band)}$ is based on the single I-band technique. See text for details.}                                                                                            
\end{table*}                                                                                            
\restoregeometry

\subsection{Spectral types of RSGs}

The effect of different metallicities on the observed properties of RSGs is significant. This was first reported in \citet{Elias1985} and \citet{MasseyOlsen03} for a sample of RSGs in the SMC and the LMC, and later for a sample of more metal-poor galaxies in the Local Group \citep{LM2012}. The average spectral types of RSGs move toward earlier types at lower host galaxies metallicities. The average spectral type of RSGs in the Milky Way is M2, RSGs in the SMC  have an average spectral type of K5--7, and RSGs in the WLM, as the most metal-poor galaxy in this sample, have an average spectral type K1--3 \citep{LM2012}.
The explanation for this effect can be found in the behavior of the TiO bands, which are used as the primary indicator of spectral type classification. At lower metallicities these molecular bands become weaker, and as a result, the comparison of observed RSG spectra with Atlas 9 \citep{Kurucz1993} or MARCS \citep{Gustafsson} stellar atmosphere models suggests early spectral types.

Our newly identified RSGs in galaxies more metal poor than the WLM (e.g., Sextans A) follow this trend of RSG spectral types. The majority of RSGs that we identified in our program galaxies have early-K spectral types (see Table \ref{tab:hr_diag}), with the exception of targets in IC 10, for which we identified five AGB stars with spectral types later than the typical RSG type at this metallicity. 

\begin{figure*}
\begin{center}
\includegraphics[width=0.45\linewidth]{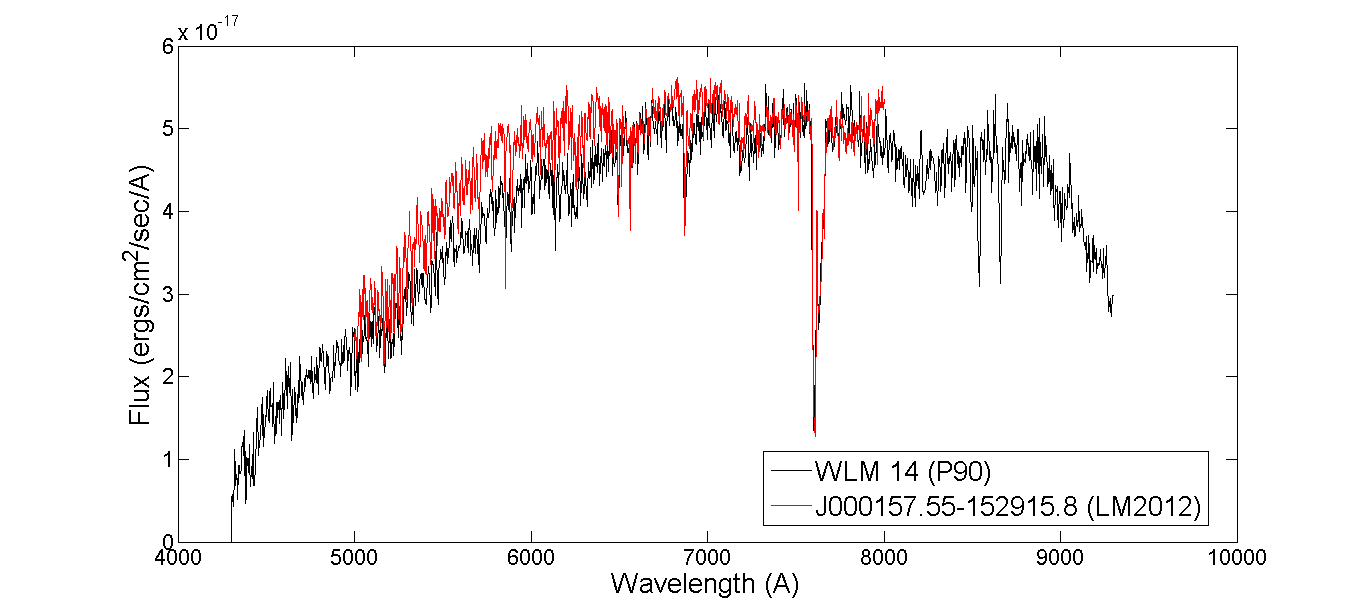}
\includegraphics[width=0.45\linewidth]{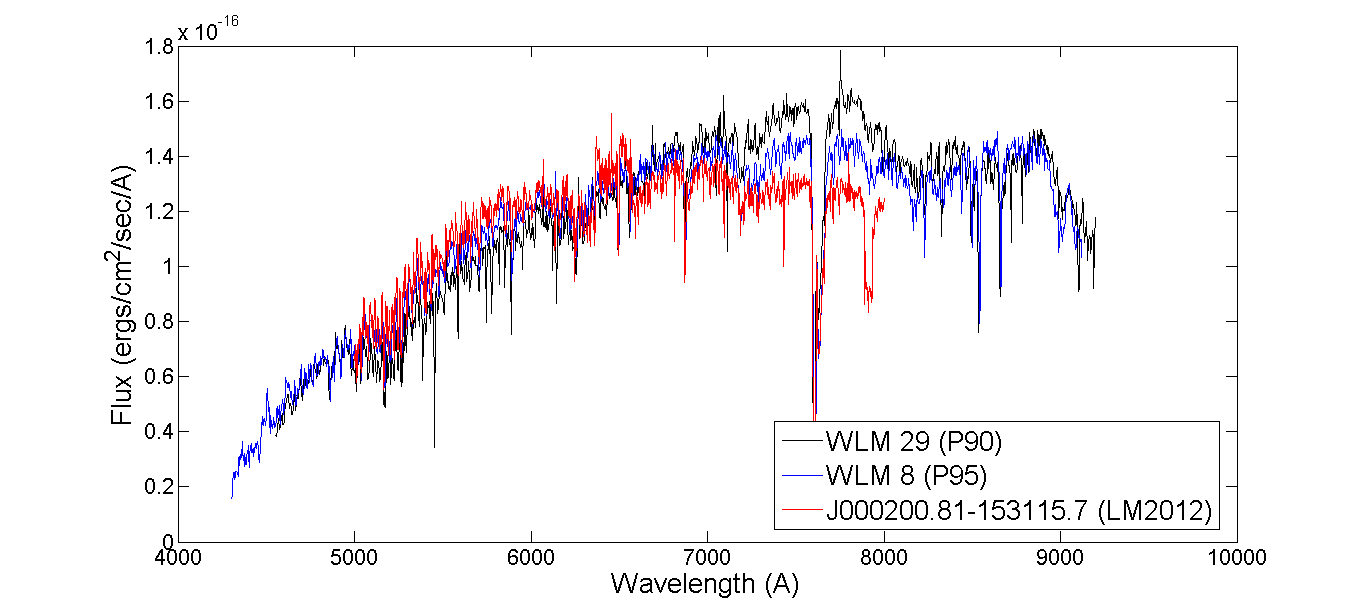}
\includegraphics[width=0.45\linewidth]{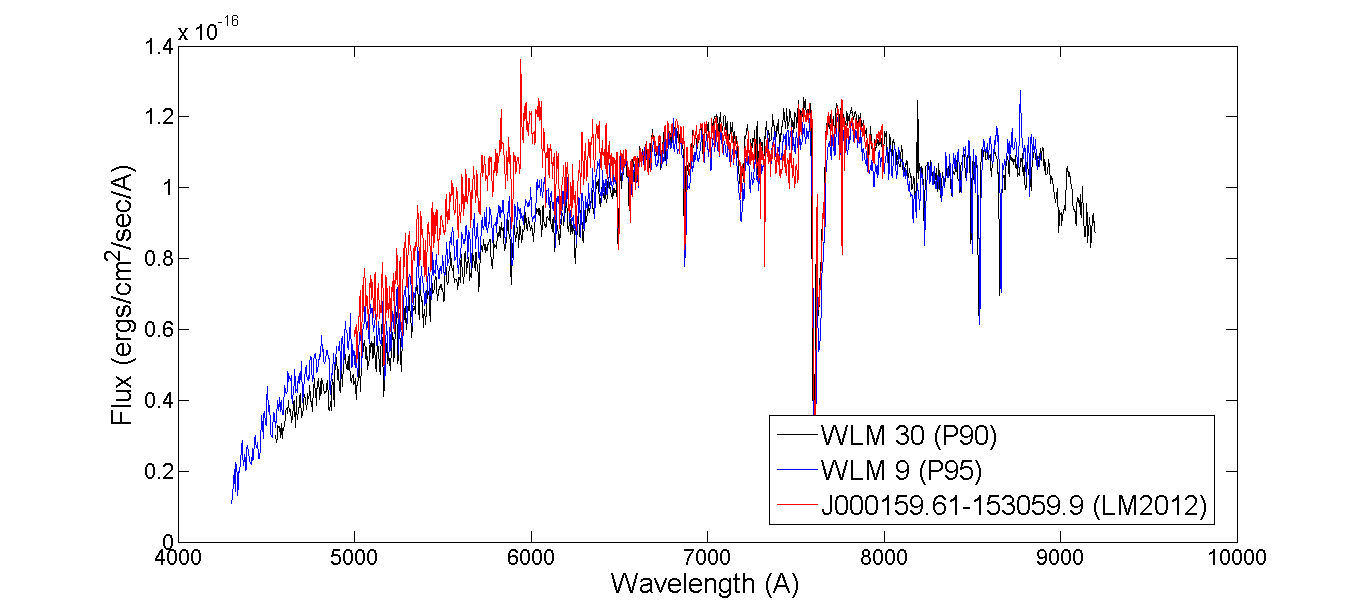}
\includegraphics[width=0.45\linewidth]{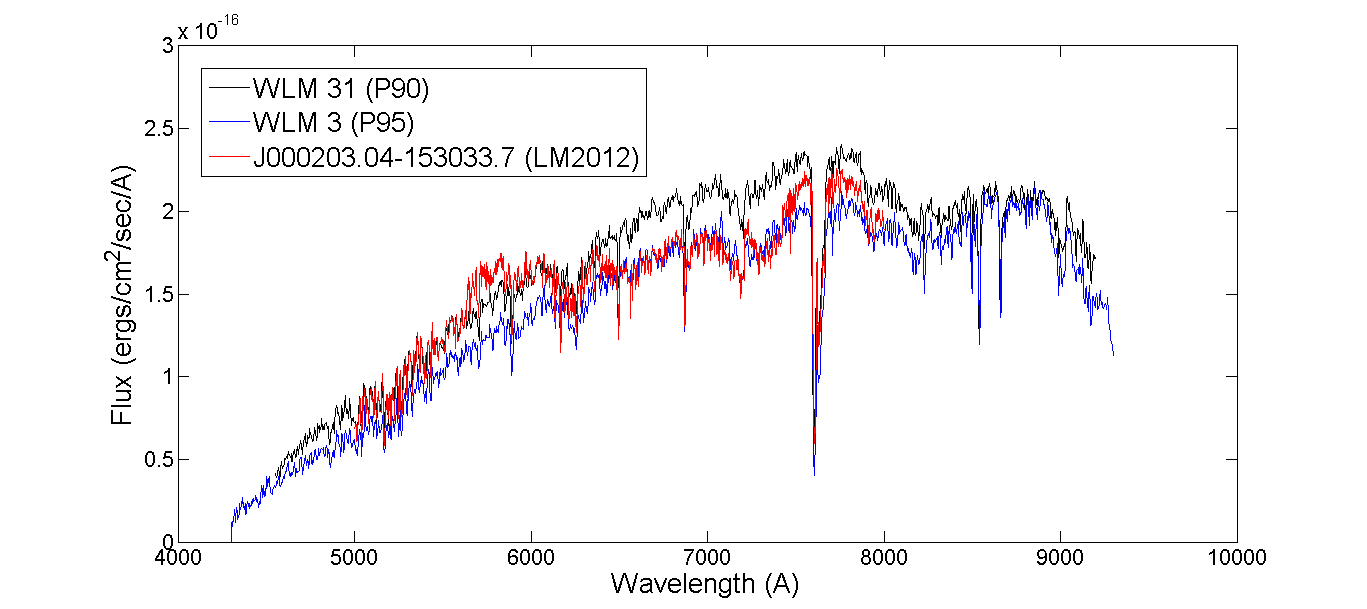}
\includegraphics[width=0.45\linewidth]{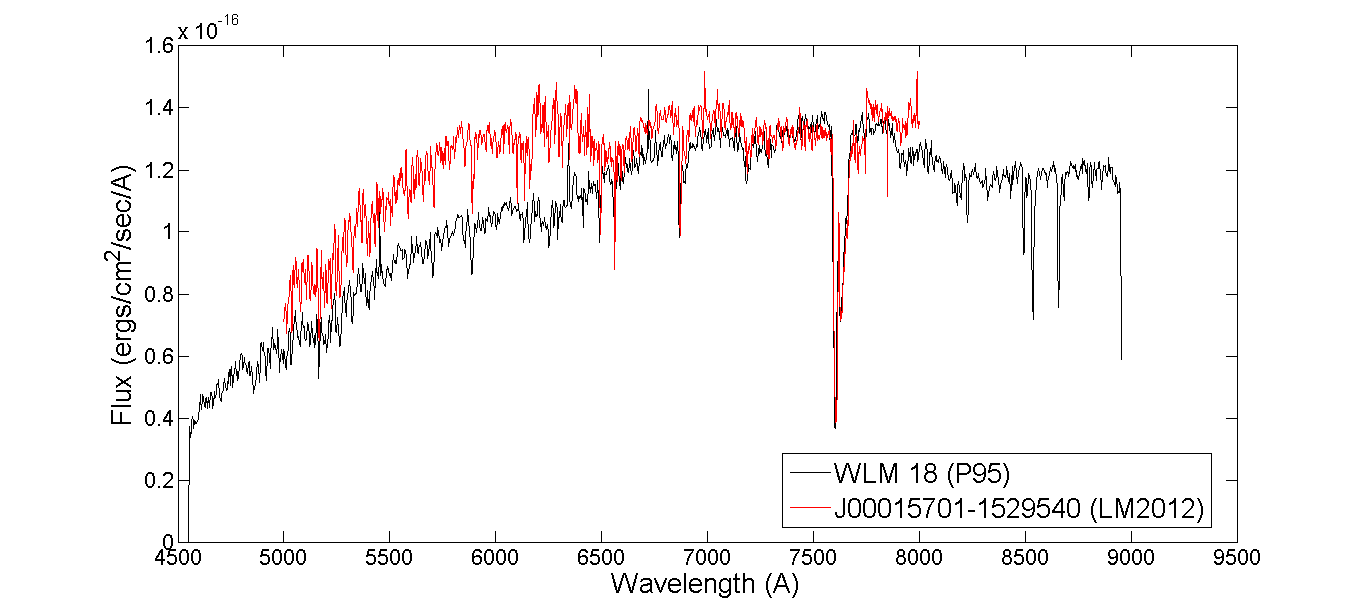}
\includegraphics[width=0.45\linewidth]{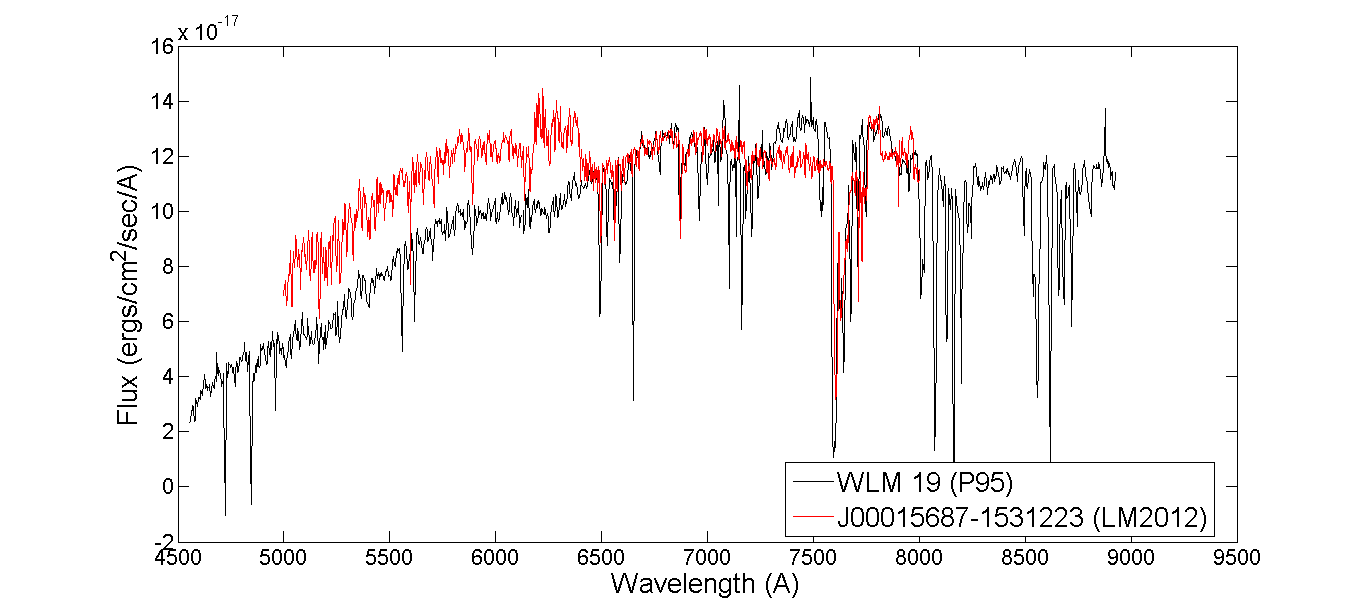}
\includegraphics[width=0.45\linewidth]{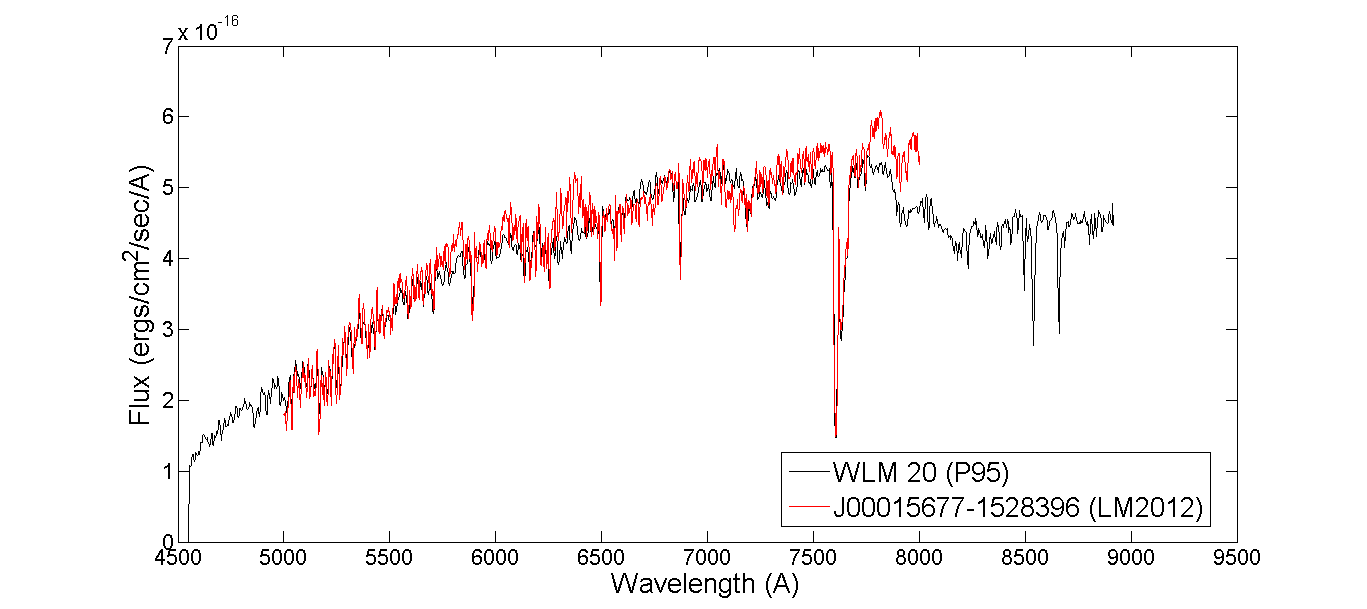}
\end{center}
\caption[h]{Comparison of FORS2 spectra of 7 previously known RSGs in WLM (black: P90; blue: P95) with the spectra of \cite{LM2012} (red). The continuum variations in the reference spectra are the instrumental artifacts. }
\label{Fig9a}
\end{figure*}

Applying our mid-IR selection criteria, we selected seven RSGs in WLM independently of \cite{LM2012}. Spectroscopic observations of these seven objects, carried out four years apart, enable us to compare their spectral types and identify spectral variability \citep{Massey_var,Levesque10_phys}. Our spectroscopic analysis (Figure \ref{Fig9a}) of flux-calibrated spectra does not show any significant difference in spectral type.

\subsection{Nature of RSGs in dIrr galaxies}

This paper, together with Paper I and Paper II, increases the sample of spectroscopically confirmed RSGs in dIrr galaxies in the Local Group by 13 (30\%) by employing mid-IR criteria. As we mentioned, prior to these works, 43 RSGs were spectroscopically confirmed in dIrrs of the Local Group (in NGC 3109, NGC 6822, IC 1613, the WLM, and the Sagittarius dIrr).

\begin{figure*}
\begin{center}
\resizebox{\hsize}{!}{\includegraphics{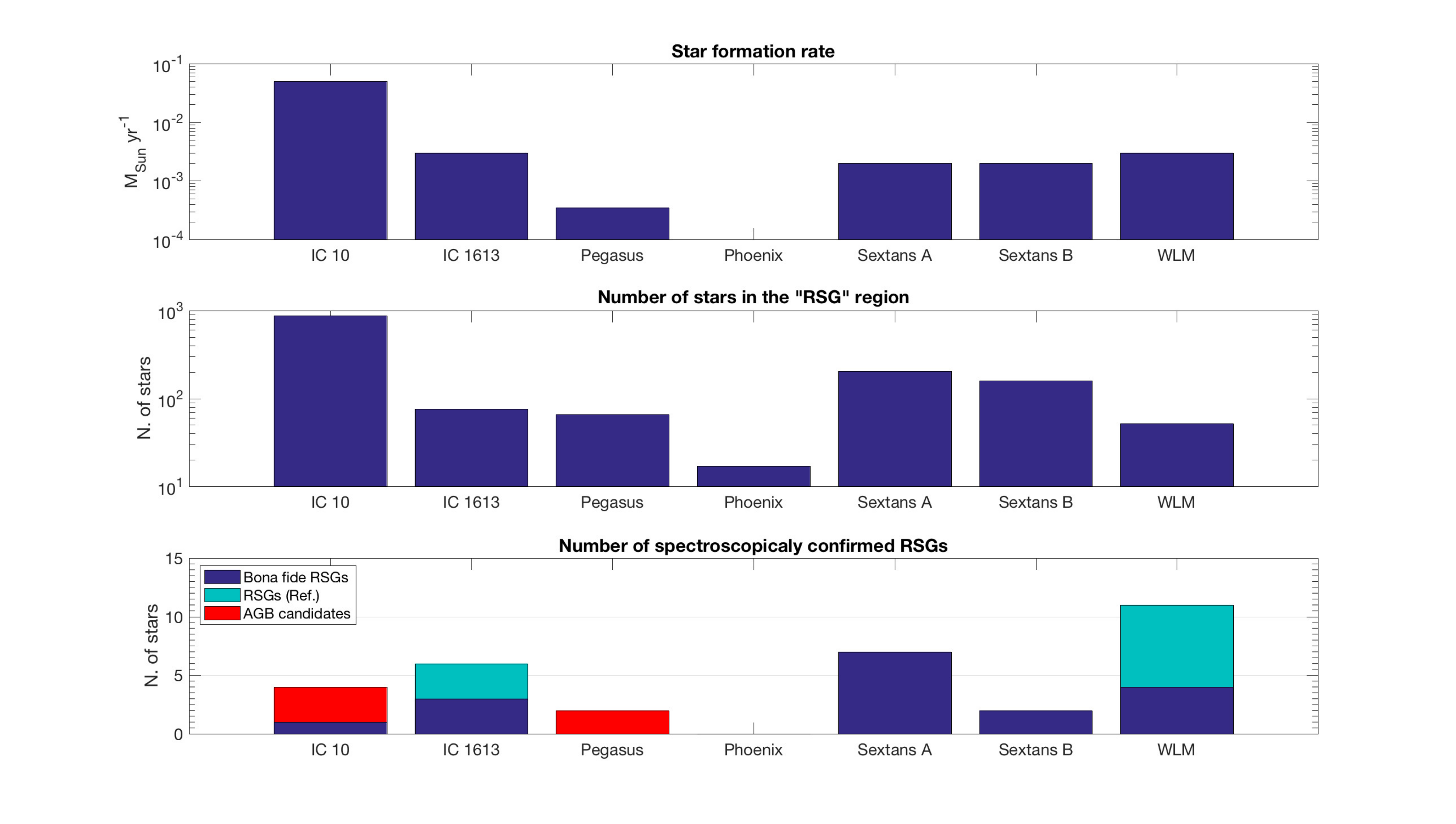}}
\end{center}
\caption[]{Total number of discovered RSGs in each dIrr galaxy (bottom panel), total number of sources in the RSG region, i.e., $[3.6]-[4.5] < 0$ and $M_{[3.6]} < -9$ (middle panel), and SFR estimate in each dIrr galaxy (top panel). The number of AGB candidates and the number of RSGs that have been identified by other authors \citep{Tautvai2007,LM2012} are labeled.}
\label{Fig_dirr}
\end{figure*}

An important question is the completeness of the RSG sample in each of the dIrr galaxies discussed here. How many more RSGs do we expect? As a first-order estimate, we counted the total number of sources in the so-called RSG region, that is, $[3.6]-[4.5] < 0$ and $M_{[3.6]} < -9$ mag according to \citet{BMS09}, in the CMD and compared it with the SFR of each galaxy, which is tabulated in Table \ref{tab:galax} (see Figure \ref{Fig_dirr}). The estimates of the SFRs are based on the HII regions and the most luminous stars \citep{Mateo1998} and indicate the most recent SFR ($\approx$\,10 Myr). We assumed that past star formation rates are proportional to the most recent one. This assumption is not always true, especially in case of dIrr galaxies \citep{Weisz_2014}, but as a zeroth-order assumption, it can be used for our purposes. A strong correlation is observed between the SFR and the total number of sources. On average, we identified 3-5 RSGs in each galaxy. They are among the brightest and most massive stars, but taking into account that the WLM hosts 11 known RSGs \citep{LM2012}, it indicates that at least twice as many RSGs lie in each galaxy if the RSG population is independent of metallicity. However, as we show below, the RSG properties depend on metallicity. When the lifetimes of RSGs decrease with metallicity, the observed RSGs would be almost complete for the lowest metallicity galaxies. An additional problem is the low number statistics, which is related to the low SFR. It is therefore difficult to establish robust conclusions on the completeness and RSG lifetimes. IC 10 deserves particular attention because, as we mentioned before, this galaxy is located near the Galactic plane. It is therefore significantly foreground contaminated, and based on its high SFR, further discoveries are expected and studies of its RSG population are encouraged.

\begin{table}[!t]
\caption{Expected number of RSGs for the program galaxies depending on the assumed SFRs.}              
\label{table_3}      
\centering                 
\begin{tabular}{l c c c c c}          
\hline\hline 
\multicolumn{5}{c}{Number of RSGs}& \\
\cline{2-5}
\hline  
 Galaxy       &  Observed    & SFR$_{H\alpha}$ &  SFR$_{FUV}$   & SFR$_{CMD}$ \\
 \hline
IC 10         &   4 (?)     &  1.93       &   0.00           &  3.35 \\
IC 1613       &  6      &   3.04      &   6.21           & 2.11  \\
Pegasus  &   2 (?)      &  4.94E$-4$  &   3.34E$-3$    & 6.58E$-3$ \\
Phoenix     &   0     &   0.00        &  7.66E$-4$    & 0.00 \\
Sextans A  &   7     &   1.27        &  3.27           & 0.65 \\
Sextans B &    2     &   7.73E$-3$ &  2.50E$-2$   & 1.31E$-2$ \\
WLM          &  11    &   1.18         & 3.32          &  1.73 \\
\hline  
\hline  
\end{tabular}
\tablefoot{The evolutionary tracks underestimate the number of RSGs, the theoretical numbers must be considered as a lower limits. The question marks indicate possible AGB contamination in the listed number of observationally confirmed RSGs.}
\end{table}


In Table~\ref{table_3} we show the observed number of RSG and the maximum number of RSGs estimated using the PARSEC evolutionary tracks \citep{Bressan_2012,Chen_2015} assuming a constant SFR over the last 50 Myr. For this analysis we used three different SFR estimates: the estimates based on the CMD analysis (listed in Table~\ref{tab:galax}), and SFRs obtained from H$\alpha$ and far-ultraviolet (FUV) analysis obtained by \citet{Karachentsev_2013}. We note that the estimates are always lower than the observed number of RSG. To explain these results, we recall that evolutionary computations have a two problems regarding RSGs. The first problem is the position of RSGs in the H--R diagram, which we studied here, and the second problem is the lifetime of the RSG phase, which defines the possible number of RSG in a given population. This requires a correct consideration of the previous step to reproduce the position of RSGs in the H--R diagram and additional considerations about their evolution. This is beyond the scope of this paper.

For our RSG analysis, we adopted the mean metallicities of the BSGs population (if the data were available) in these galaxies as a reference metallicity of our RSG sample. According to Table \ref{tab:galax}, the BSGs metallicity measurements are available only for IC 1613, Sextans A, and the WLM. For IC 1613 we used direct measurements of RSG metallicities from \citet{Tautvai2007}. In addition, we assumed that the mean metallicity of Sextans B is approximately the same as for Sextans A. For Pegasus and IC 10 we used the [O/H] abundances from \citet{Bergh_2000}. Very importantly, the metallicities of dIrr galaxies are not homogenous. There are some effects, such as different star formation regions and different chemical evolution histories, which cause the metallicity spread thoruhgout the galaxy \citep[e.g.,][]{WLM_Z,Patrick_2017}. Moreover, in case of IC 1613, the metallicity of the B- and A-type supergiants is bimodal \citep{Berger_2018}. To conclude, appropriate reliable measurements of RSG metallicities in the program galaxies are difficult to obtain, but our assumptions are enough to separate the sample of RSGs by mean metallicities of the host galaxies within 0.2 dex error bars.

\begin{figure*}
\begin{center}
\includegraphics[width=1\textwidth]{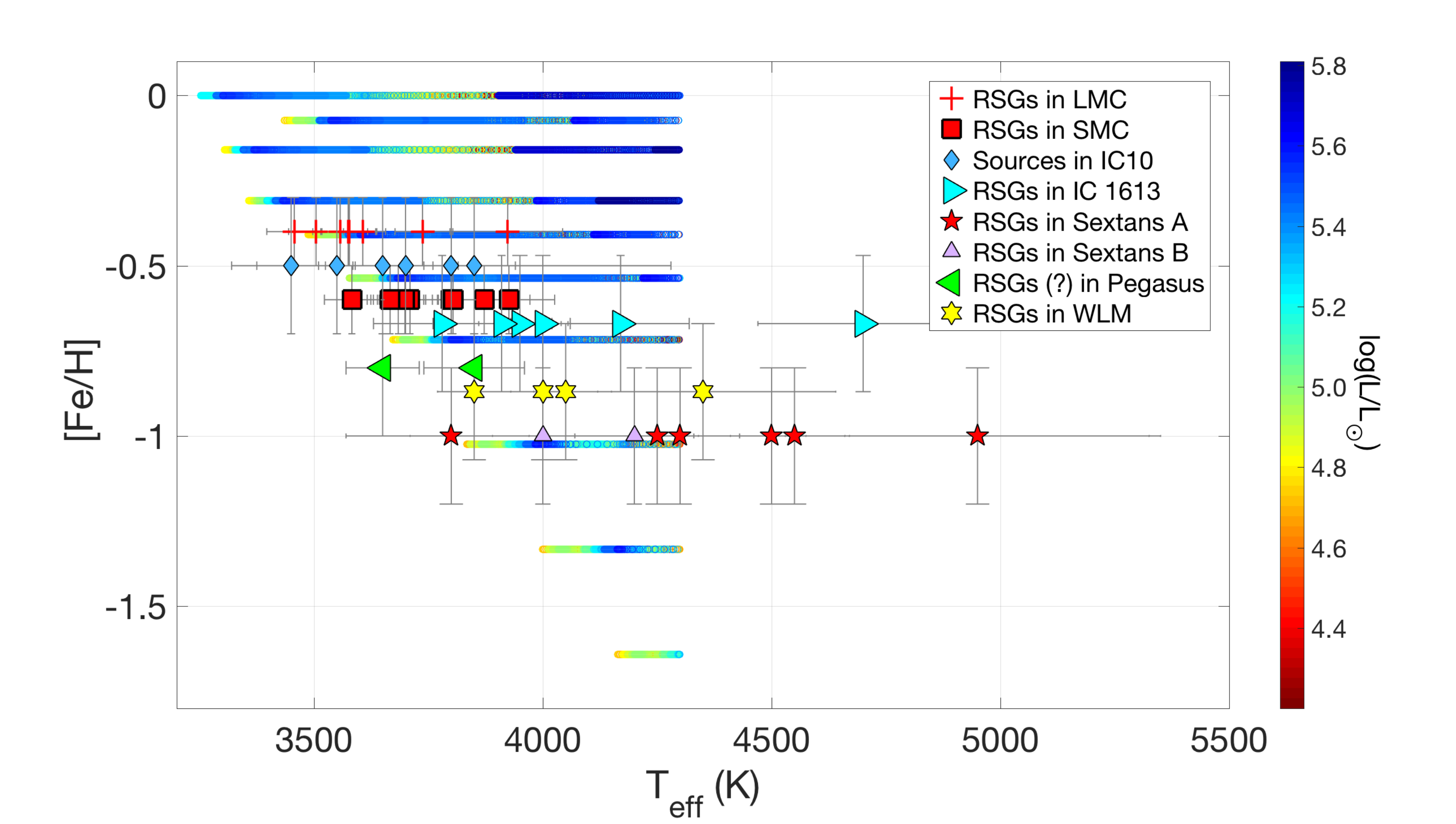}
\end{center}
\caption[]{Distribution of the RSG effective temperatures ($T_\mathrm{eff}^\mathrm{SED}$) depending on host galaxy metallicity ($\mathrm{[Fe/H]}$). We add sample RSGs from the LMC and SMC \citep{Davies13}, and three well-studied RSGs in IC 1613 from \citet{Tautvai2007}. The horizontal lines correspond to possible effective temperatures for RSGs with initial masses of $8-40~M_{\odot}$ for each available metallicity in the PARSEC evolutionary tracks \citep{Bressan_2012}. The range of possible theoretical luminosities for a given range of temperatures is presented in the color bar.} 
\label{Fig_ZT}
\end{figure*}

In Figure \ref{Fig_ZT} we plot the relation between effective temperature ($T_\mathrm{eff}^\mathrm{SED}$) and metallicities of the host galaxies ($\mathrm{[Fe/H]}$), adding the $T_\mathrm{eff}^{(J-K)}$ values for our targets. We added three known RSGs in IC 1613 with adopted physical parameters derived by \citet{Tautvai2007} to our sample, as well as RSGs from the LMC and SMC \citep{Davies13}, for which the temperatures were calculated by TiO-band SED fitting. We adopted $\mathrm{[Fe/H]}_{LMC}$\,=\,$-0.4$ and $\mathrm{[Fe/H]}_{SMC}$\,=\,$-0.6$ dex with an uncertainty 0.2 dex, which is in agreement with the mettallicity estimates of the RSG population in these galaxies according to \citet{Davies15}. A trend of increasing RSG effective temperatures toward lower $\mathrm{[Fe/H]}$ is clearly visible, which implies decreasing depths of the TiO bands at lower metallicities. The $T_\mathrm{eff}^{(J-K)}$ values show the same trend, but weaker. This fact indicates that the Hayashi limit depends on the metallicity of the host environment. This result mainly shows the behavior of the TiO bands, but this trend is also observed if we were to use the photospheric atomic lines in spectra of RSGs \cite[as was shown on a sample of RSGs in the MCs][]{Tabernero_2018,Davies_2018}.

In order to test this observational trend with theory, we investigated the theoretical predictions of RSG physical parameters, such as effective temperatures for given $\mathrm{[Fe/H]}$ and luminosity, at low metallicities. We chose the PARSEC evolutionary tracks without rotation because they range to low metallicities (up to $\mathrm{[Fe/H]}$\,=\,$-1.65$ dex assuming a $\mathrm{[Fe/H]}$ to Z relation of $[Fe/H] = 1.024 \log(Z ) + 1.739$ and solar isotopic content). We considered as RSGs stars with initial masses of 8 to $40~M_{\sun}$ and temperatures from 3200~K to 4300~K. We selected all luminosities and all evolutionary points at ${\rm [Fe/H]} < 0$. In Figure \ref{Fig_ZT} we present the selected theoretical points, which indicate an RSG phase in the PARSEC evolutionary tracks. 
The theoretical trend of decreasing $T_{\rm eff}$ with decreasing ${\rm [Fe/H]}$ is clear, which is in agreement with our observations. The theoretical maximum of $T_{\rm eff}$ is 4300~K, in order to avoid contamination by other stellar types (e.g., AGB) that satisfy our selection criteria. Thus, we could not investigate the maximum temperature of RSG temperature at low ${\rm [Fe/H]}$.

\begin{figure*}
\begin{center}
\includegraphics[width=1\textwidth]{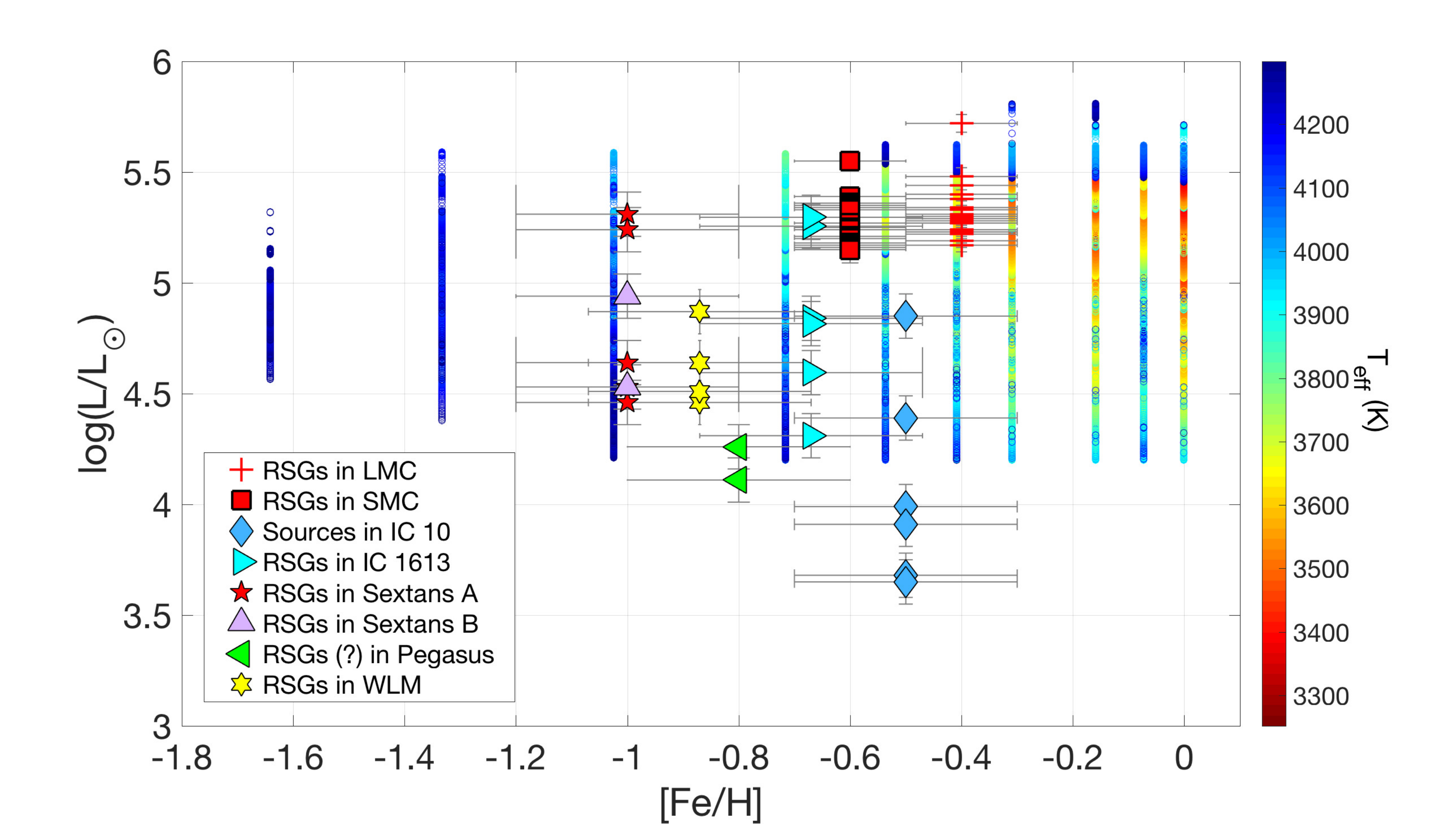}
\end{center}
\caption[]{Distribution of RSG luminosities depending on the metallicity of host galaxies ($\mathrm{[Fe/H]}$). The sample of the most luminous RSGs in the SMC and the LMC is from \citet{Davies_2018}, and three well studied RSGs in IC 1613 are from \citet{Tautvai2007}. The vertical lines correspond to possible luminosities for RSGs with initial masses $8-40~M_{\odot}$ for each available metallicity in the PARSEC evolutionary tracks \citep{Bressan_2012}. The range of possible theoretical effective temperatures for a given range of luminosities is presented in the colorbar.}
\label{Fig_ZL}
\end{figure*}

Our unique sample of RSGs in metal-poor dIrr galaxies gives us the possibility to investigate how the Humphreys-Davidson limit in the RSGs region depends on metallicity. According to theory \citep[e.g.,][]{Meynet_2015}, the luminosities of RSGs are expected to be higher at lower metallicity. The mass-loss rate at lower metallicity is predicted to be lower, therefore the lifetimes of RSGs are expected to be longer, and their luminosities should be higher. However, our analysis shows no such trend toward high luminosities of RSGs in a metal-poor dIrr galaxies. In Figure \ref{Fig_ZL} we show the relation between luminosities of RSGs with metallicities of their host galaxies ($\mathrm{[Fe/H]}$). We added the sample of the most luminous RSGs in the MCs \citep{Davies_2018}. Concerning theoretical predictions of RSG luminosities as a function of metallicity, we used the same evolutionary tracks as in the PARSEC models. In Figure \ref{Fig_ZL} we add vertical lines that correspond to the possible luminosities of RSGs at each given metallicity. The theoretical maximum luminosity at $\mathrm{[Fe/H]}$\,$<$\,$-0.4$ is about log$(L/L_{\sun})$\,$\approx$\,$5.5$ and remains constant down to $\mathrm{[Fe/H]}$\,$\approx$\,$-1.4$. This prediction is in agreement with RSGs in the LMC and SMC \citep{Davies_2018}. 
Our limited observational sample of RSGs in each of the dIrr galaxies prevents us from claiming a definite trend in luminosities. However, we can state that we have analyzed the most massive RSGs in Sextans A and that their luminosities do not display a significant trend in terms of their maximum luminosity (i.e., the Humphreys-Davidson limit), in agreement with the PARSEC evolutionary tracks.

Summarizing, we conclude that we did not observe the most luminous RSGs in IC 1613, Sextans B, and the WLM galaxies. The observed RSGs in these galaxies have absolute luminosities near the minimum selection cutoff criteria ($M_{[3.6]}$\,$\approx$\,$-9$, see the CMD in Figures \ref{Fig6_ic1613} and \ref{Fig7_wlm}). In Sextans A we investigated a more complete sample of the RSG population. This observational bias explains why we reach the maximum theoretical luminosity only for RSGs in Sextans A and not in the others galaxies. This should encourage future studies to find and analyze RSGs in poorly studied galaxies such as IC 10, IC 1613, and the WLM.

\section{Conclusions}
Together with Paper I and Paper II, we here expand the census of RSGs in seven star-forming dIrr galaxies. We spectroscopically confirmed 13 new RSGs and confirmed 4 RSGs that have been reported previously. We also identified 8 massive AGB star candidates. For all targets a comprehensive analysis of the physical parameters was performed.
We applied SED fitting and several photometrical techniques to obtain the physical parameters of RSGs. The derived parameters are in good agreement for the different methods when interstellar extinction is accurately determined. However, we suggest to use the results based on the SED fitting technique, which does not depend on the empirical photometric calibrations. These calibrations have only been tested for the RSGs in the MCs. Moreover, for the RSGs in the metal-poor galaxies we cannot confirm that the bolometric correction is uniform throughout all ranges of effective temperatures, as was assumed in several photometric calibrations.

The number of discovered RSGs in dIrr galaxies together with the well-studied RSG sample in the MCs is statistically significant for an investigation of the nature of RSG in metal-poor galaxies.
Comparison of the observational properties of RSGs with PARSEC evolutional tracks shows that (i) the minimum effective temperature of RSGs increases with decreasing metallicity, and (ii) the maximum luminosity of RGSs is constant (log$(L/L_{\sun})$\,$\approx$\,5.5 dex) with decreasing metallicity within a metallicity range $\mathrm{[Fe/H]}$$\sim$[0...$-$1]~dex. 

These statements were confirmed by an analysis of the physical parameters of RSGs in the most metal-poor dIrr galaxy Sextans A ($\mathrm{[Fe/H]}$$\approx$\,$-1$~dex). In this galaxy we found three of the most massive RSGs with the $R^{\mathrm{SED}}$\,$\approx$\,900\,$R_{\sun}$. The derived physical parameters of RSGs will be useful for supernovae progenitor studies in the studied dIrr galaxies. 
The absence of more luminous RSGs supports the idea that more luminous massive stars in the metal-poor environment do not evolve to the RSG phase and remain in the blue part of the H--R diagram as blue supergiants. The reasons can either be observational (very fast evolving to the RSG phase) or evolutionary (the upper limit of the initial mass of the RSGs phase decreases at low metallicities). The expected initial mass range of the RSG phase becomes smaller, that is, $\sim$[8...40]~$M_{\sun}$ at $\mathrm{[Fe/H]}$\,=\,0 versus $\sim$[8...15]~$M_{\sun}$ at $\mathrm{[Fe/H]}$\,=\,--1.74~dex \citep[according to the recent studies of][]{Limongi_2018,Groh_2019}.
In this context, a similar work to establish the census and physical properties of blue supergiants in these galaxies is required. This would help place observational constraints on the ratio of blue to red supergiants, which is important for understanding the nature of the supernova progenitors \citep{Langer_1995,Eggenberger_2002}.

With this work, we would like to stimulate further studies of RSGs in particular and of post main-sequence massive stars in general in metal-poor galaxies. Future optical and near-IR instruments will be available on the next generation of telescopes to conduct studies like this.

\begin{acknowledgements}
We thank the anonymous referee for constructive remarks that have significantly improved the manuscript.
The authors would like to thank Ben Davies for helpful comments on the investigation of RSG upper luminosities.
NB, AH and MC acknowledge support under MINECO projects AYA2015-68012-C2-1-P and SEV2015-0548, also acknowledge support  from the grant of Gobierno de Canarias (ProID2017010115). MC acknowledge support from the Spanish State Research Agency grant
MDM-2017-0737 (Unidad de Excelencia Maria de Maeztu, CAB). TM acknowledges support from the State Research Agency (AEI) of the
Spanish Ministry of Science, Innovation and Universities (MCIU) and the European Regional Development Fund (FEDER) under grant AYA2017-88254-P. NB and AZB acknowledge funding by the European Union (European Social Fund) and National Resources under the ``ARISTEIA'' action of the Operational Program ``Education and Lifelong Learning'' in Greece. This research has made use of NASA's Astrophysics Data System Bibliographic Services and the VizieR catalogue access tool, CDS, Strasbourg, France. Funding for SDSS-III has been provided by the Alfred P. Sloan Foundation, the Participating Institutions, the National Science Foundation, and the U.S. Department of Energy Office of Science. NB has devoted this work to Toma Balaeva for support in preparing the manuscript.
\end{acknowledgements}

\bibliographystyle{aa}
\bibliography{ref}

\onecolumn
\begin{appendix}
\section{Information for observed targets.}

\begin{table*}
{\small
\caption{Characteristics and spectral classification of observed targets in IC 10. GTC-OSIRIS observing run.}
\label{tab:ic10}
\begin{tabular}{lcccrrrr}
\hline\hline
Name/Obs. Name &  DUSTiNGS & R.A.(J2000) & Decl.(J2000)  & Radial velocity                     &$M_{[3.6]}$& $[3.6]-[4.5]$   & Notes     \\
     & ID        &   (deg)     &  (deg)        &    (km~s$^{-1}$)       & (mag)        &  (mag)       &          \\
\hline
1 / Ap3 CCD1 &128619 &4.9865770 &59.2769393&   44$\pm$12    & $-$10.26$\pm$0.03 & $-$0.04$\pm$0.05  & For.    \\
2 / Ap3 CCD2 &107761 &5.0460181 &59.3180999&   19$\pm$7     & $-$8.80$\pm$0.07 &  0.24$\pm$0.09 & For. F-G?  \\
3 / Ap2 CCD1 &112657 &5.0324745 &59.2714500&   54$\pm$25    & $-$8.84$\pm$0.04 & $-$0.05$\pm$0.06  &  For., K 1-3 III \\
4 / Ap5 CCD1 &103677 &5.0575594 &59.2875137&  --317$\pm$10    & $-$9.58$\pm$0.04 & $-$0.15$\pm$0.05  & IC 10 1, K 3-5 (AGB)     \\
5 / Ap6 CCD1  &117107 &5.0198626 &59.2903671&  --305$\pm$13   & $-$9.86$\pm$0.04 & $-$0.08$\pm$0.05  & IC 10 2, M0-2  (AGB)  \\
6 / Ap1 CCD1  &96020  &5.0788016 &59.2634048&   87$\pm$16     &$-$10.32$\pm$0.03 &  0.07$\pm$0.05  & For.    \\
7 / Ap4 CCD2 &126841 &4.9916567 &59.3216438&   57$\pm$10      & $-$9.48$\pm$0.05 &  0.17$\pm$0.06  &  For.  \\
8 / Ap5 CCD2 &94159  &5.0840334 &59.3429145&    --4$\pm$6      &$-$9.19$\pm$0.08 &  --  &  For., Late G - Early K \\
9 / Ap2 CCD2 &95408  &5.0804367 &59.3092765&  --250$\pm$6    &$-$10.71$\pm$0.04 & $-$0.03$\pm$0.06 &  IC 10 3, M 1-3 I     \\
10 / Ap7 CCD1 &99773  &5.0684976 &59.2951965&  --252$\pm$6    &$-$9.80$\pm$0.03 & $-$0.03$\pm$0.05 &   IC 10 4, M 0-2    (AGB)   \\             
11 / Ap1 CCD2 &85592  &5.1077690 &59.3035316&  --261$\pm$8    &$-$10.08$\pm$0.03 & $-$0.22$\pm$0.04 &  IC 10 5, M 1-3    (AGB)   \\
12 / Ap4 CCD1 &107961 &5.0455651 &59.2827033&  --321$\pm$17   &$-$10.09$\pm$0.03 & $-$0.18$\pm$0.04 &  IC 10 6, K 3-5   (AGB)   \\
\hline
\end{tabular}
\tablefoot{
For. means foreground giants. }
}
\end{table*}

\begin{table*}
{\small
\caption{Characteristics and spectral classification of observed targets in IC 1613. GTC-OSIRIS observing run.}
\label{tab:ic1613}
\begin{tabular}{lcccrrrr}
\hline\hline
Name/Obs. Name &  DUSTiNGS & R.A.(J2000) & Decl.(J2000)  & Radial velocity                     &$M_{[3.6]}$& $[3.6]-[4.5]$   & Notes   \\
     & ID        &   (deg)     &  (deg)        &    (km~s$^{-1}$)       & (mag)        &  (mag)       &         \\
\hline
1 / Ap3 CCD1  & 97761  &16.237533&2.078947& --& $-$9.34$\pm$0.03 & $-$0.12$\pm$0.05 & RSG IC1613-3, K1-3   \\
2 / Ap4 CCD1  & 107793 &16.224805&2.099375& --& $-$9.19$\pm$0.03 & $-$0.12$\pm$0.05 & --     \\
3 / Ap1 CCD1  & 115974 &16.214523&2.056277& --& $-$9.57$\pm$0.05 &  0.03$\pm$0.06   & M0-1   \\
4 / Ap2 CCD1  & 132449 &16.194620&2.060652& --& $-$9.22$\pm$0.03 & $-$0.11$\pm$0.05 & --    \\
5 / Ap1 CCD2  & 161666 &16.158910&2.112404& --& $-$9.25$\pm$0.04 & $-$0.05$\pm$0.05 &  RSG IC1613-1, BBM2014  \\
6 / Ap2 CCD2  & 138020 &16.187889&2.046187& $-$102$\pm$19 & $-$11.92$\pm$0.03 & $-$0.06$\pm$0.05 &  K1-3   \\
\hline
\end{tabular}
\tablefoot{
BBM2014 refers to targets previously observed by \cite{britavskiy14}}
}
\end{table*}

\begin{table*}
{\small
\caption{Characteristics and spectral classification of observed targets in Sextans B. GTC-OSIRIS observing run.}
\label{tab:sexb}
\begin{tabular}{lcccrrrr}
\hline\hline
Name/Obs. Name &  DUSTiNGS & R.A.(J2000) & Decl.(J2000)  & Radial velocity                     &$M_{[3.6]}$& $[3.6]-[4.5]$   & Notes   \\
     & ID        &   (deg)     &  (deg)        &    (km~s$^{-1}$)       & (mag)        &  (mag)       &         \\
\hline
1 / Ap2 CCD1 &  82970  &150.017166&5.309929& 390$\pm$20 & $-$10.06$\pm$0.04 &$-$0.07$\pm$0.06  & Sex B 2, K1-3 I           \\
2 / Ap1 CCD1 &  100179 &149.994064&5.326573& 337$\pm$10  & $-$10.32$\pm$0.03 & 0.05$\pm$0.05   & Sex B 1, K1-3 I       \\
3 / Ap2 CCD2 &  120428 &149.966735&5.356087& 25$\pm$11     & $-$10.65$\pm$0.04 &$-$0.01$\pm$0.05 & For., M1-3   \\
4 / Ap3  CCD2 &  94161  &150.002227&5.373697& 128$\pm$6   & $-$13.41$\pm$0.05 &$-$0.016$\pm$0.06 & For. \\
5 / Ap1 CCD2 &  128477 &149.955780&5.338477&  --  &  $-$9.95$\pm$0.03 &   0.05$\pm$0.05 & For., M1-3  \\
\hline
\end{tabular}
\tablefoot{For. means foreground giants.}
}
\end{table*}

\begin{landscape}
\begin{table*}
{\small
\caption{Characteristics and spectral classification of observed targets in the WLM. P95 observing run.}
\label{tab:wlm_p95}
\begin{tabular}{lcccccccr}
\hline\hline
Name  & DUSTiNGS & R.A.(J2000) & Decl.(J2000)  & Radial velocity &$M_{[3.6]}$& $[3.6]-[4.5]$      & Spectral & Notes    \\
   & ID       &   (deg)     &  (deg)       & (km~s$^{-1}$) & (mag)        &  (mag)    &  class &          \\
\hline
1-P95 &  78491 &0.519299& -15.454340& -- &$-9.36\pm$0.04&       $0.45\pm$0.05 &         &\\
2-P95 &  79964 &0.517339& -15.525577& $-124\pm$3 &$-9.46\pm$0.04&       $0.02\pm$0.05 & &\\
3-P95  &83414 &0.512683& -15.509500&$-94\pm$6 &$-10.61\pm$0.04&  $-0.11\pm$0.06 & K4-5I &K5I LM2012, BBM2015\\
4-P95  &83875 &0.512081& -15.465138& -- &$-9.31\pm$0.03&        $0.32\pm$0.05 &        & \\
5-P95  &86526 &0.508648& -15.450670& -- &$-9.77\pm$0.03&        $0.46\pm$0.05 &         &\\
6-P95  &86598 &0.508562& -15.439708&$-88\pm$13 &$-9.92\pm$0.04& $0.05\pm$0.05 & M5 &  For.\\
7-P95  &88720 &0.505814& -15.487849& -- &$-9.59\pm$0.05&        $0.53\pm$0.06 &         &\\
8-P95   &90598 &0.503403& -15.521162&$-108\pm$6 &$-10.08\pm$0.03&  $-0.11\pm$0.06 & K3I  & K0I LM2012, BBM2015  \\
9-P95 &94581 &0.498370& -15.516783&$-168\pm$4 &$-9.70\pm$0.03&  $-0.10\pm$0.06 &       K3I & K2-3I LM2012, BBM2015  \\
10-P95   &96709 &0.495804& -15.505663& -- &$-9.44\pm$0.04&      $0.34\pm$0.05 &         & \\
11-P95   &97051 &0.495405& -15.501128& -- &$-7.97\pm$0.10&        --          &   & \\
12-P95    &96974 &0.495471& -15.541072& -- &$-10.13\pm$0.03&    $0.41\pm$0.05 &       & \\    
13-P95   &97035 &0.495379& -15.470988& -- &$-9.15\pm$0.03&      $0.61\pm$0.05 &       & \\
14-P95  &98086 &0.494067& -15.435750& -- &$-9.00\pm$0.03&       $0.27\pm$0.04 &        & \\
15-P95 &99502 &0.492291& -15.461046& $16\pm$2 & $-11.97\pm$0.04&  $-0.12\pm$0.05 & & $H\alpha$ in emission \\
16-P95  &99581 &0.492176& -15.467833& -- &$-9.02\pm$0.04&       $0.42\pm$0.05 &        &   \\
17-P95  &101111 &0.490272& -15.518739& -- &$-10.02\pm$0.04&     $0.56\pm$0.05 &         &  \\
18-P95  &103310 &0.487502& -15.498477&$-165\pm$3  &$-9.72\pm$0.04&  $-0.12\pm$0.05 & K1I & K0I  LM2012   \\
19-P95   &103756 &0.486942& -15.522984& -- &$-9.69\pm$0.03&  $-0.13\pm$0.05 &       & K0I LM2012 \\
20-P95  &104054 &0.486522& -15.477794&$-94\pm$7 &$-11.34\pm$0.03&  $-0.12\pm$0.04 & & K2I  LM2012   \\
21-P95 &108689 &0.480663& -15.512545& -- &$-11.82\pm$0.04&  $-0.09\pm$0.05 &        &For. \\
22-P95   &114857 &0.472440& -15.474608& -- &$-9.73\pm$0.04&     $0.57\pm$0.05 &        &  \\
23-P95  &127157 &0.455256& -15.529089& -- &$-9.89\pm$0.03&  $-0.01\pm$0.05 & M3-5 & For.  \\
\hline
\end{tabular}
\tablefoot{
BBM2015  refers to targets previously observed by \cite{britavskiy15}, "LM2012" -- \cite{LM2012}. For. -- Foreground giants.}
}
\end{table*}
\end{landscape}

\section{Figures of the SED fitting for each RSG candidate from Table \ref{tab:hr_diag_L}.}

\begin{figure*}
\includegraphics[width=0.5\linewidth]{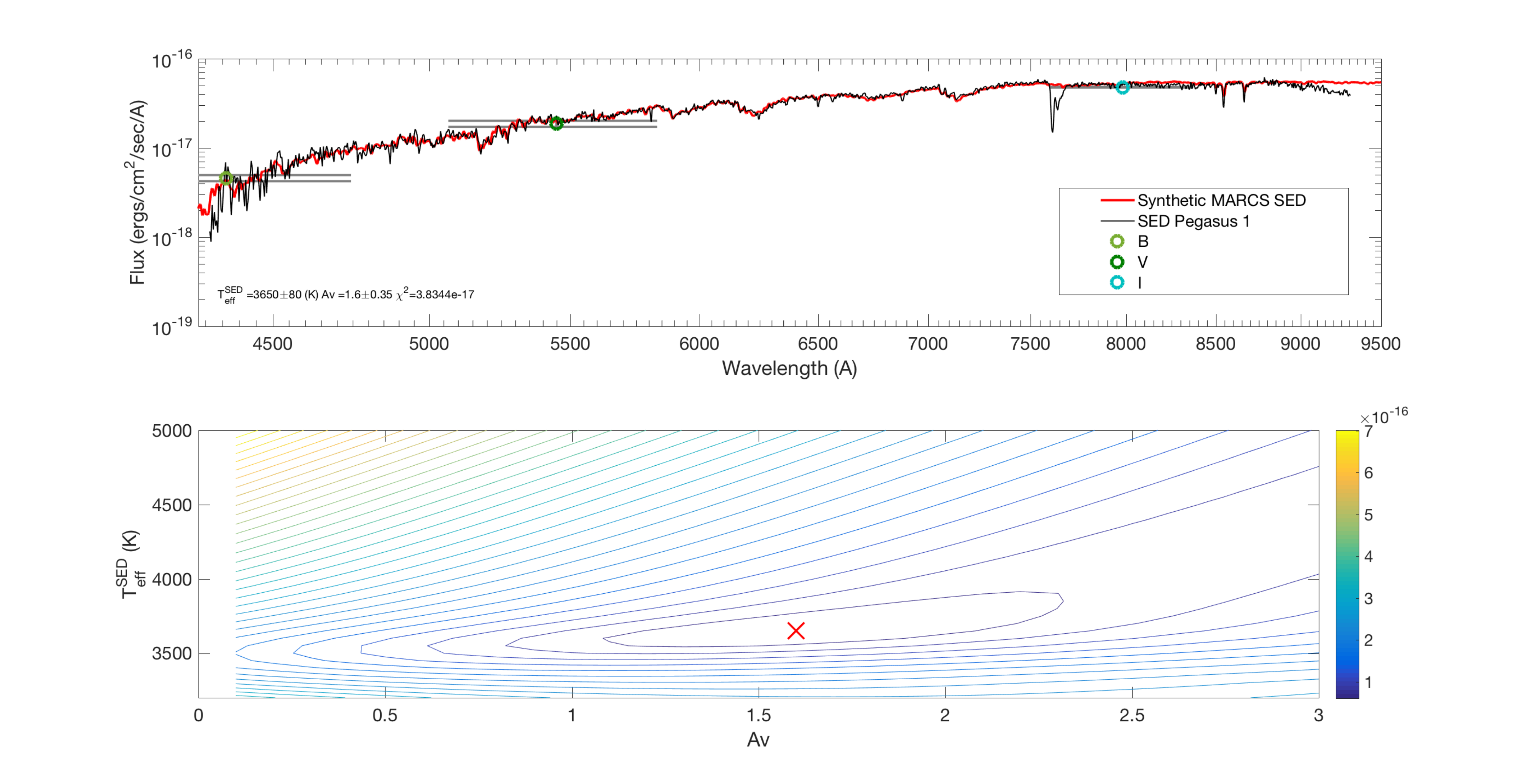}
\includegraphics[width=0.5\linewidth]{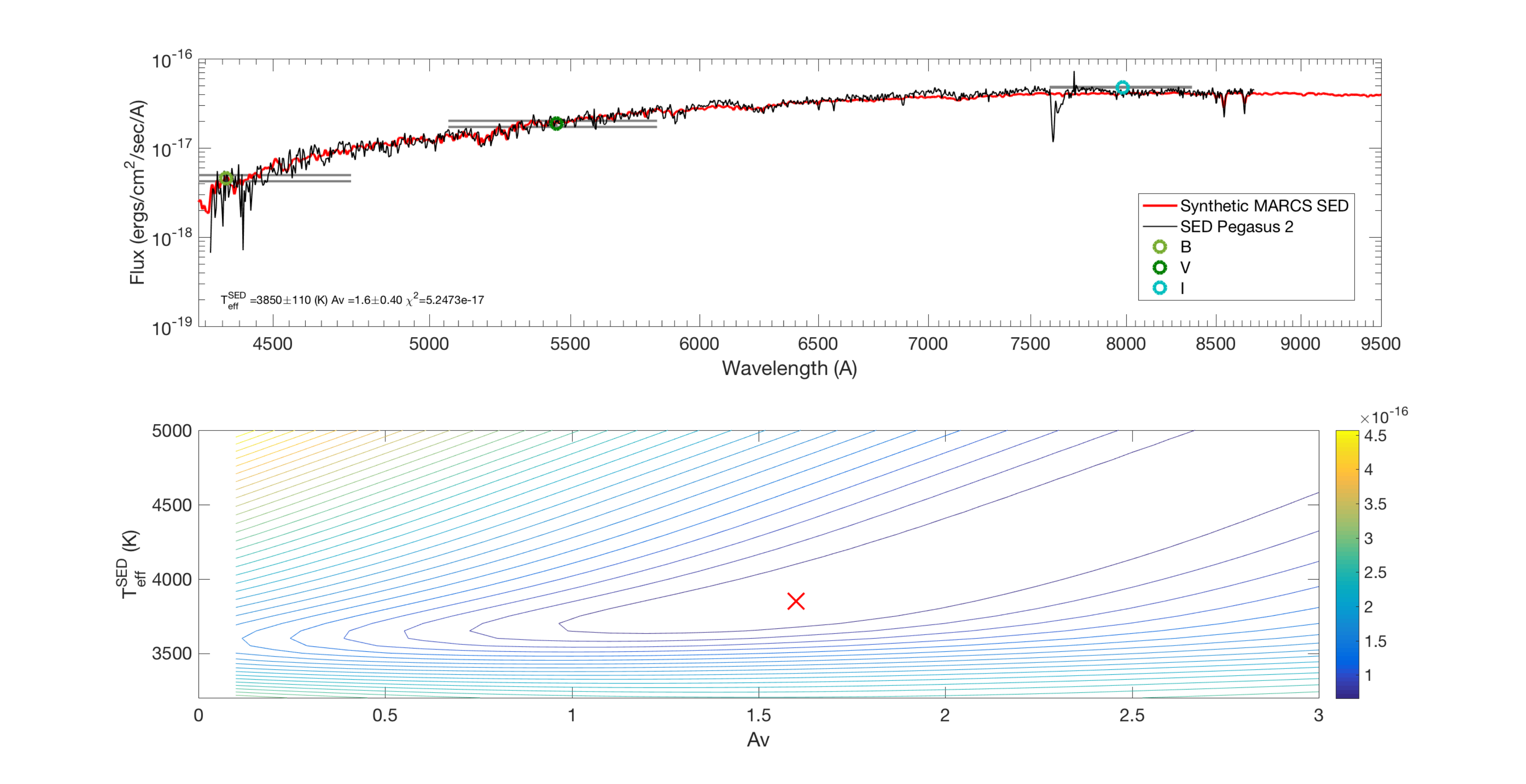}
\caption[h]{Best-fit MARCS model SED for the spectra of RSGs in the Pegasus galaxy and available optical $BVI$ photometry for each target.  Bottom panel: Result of the $\chi^{2}$ minimization of the MARCS model SED fitting by varying the values of $T^{SED}_{eff}$ and $Av$.}
\label{Fig_peg1}
\end{figure*}

\begin{figure*}
\includegraphics[width=0.5\linewidth]{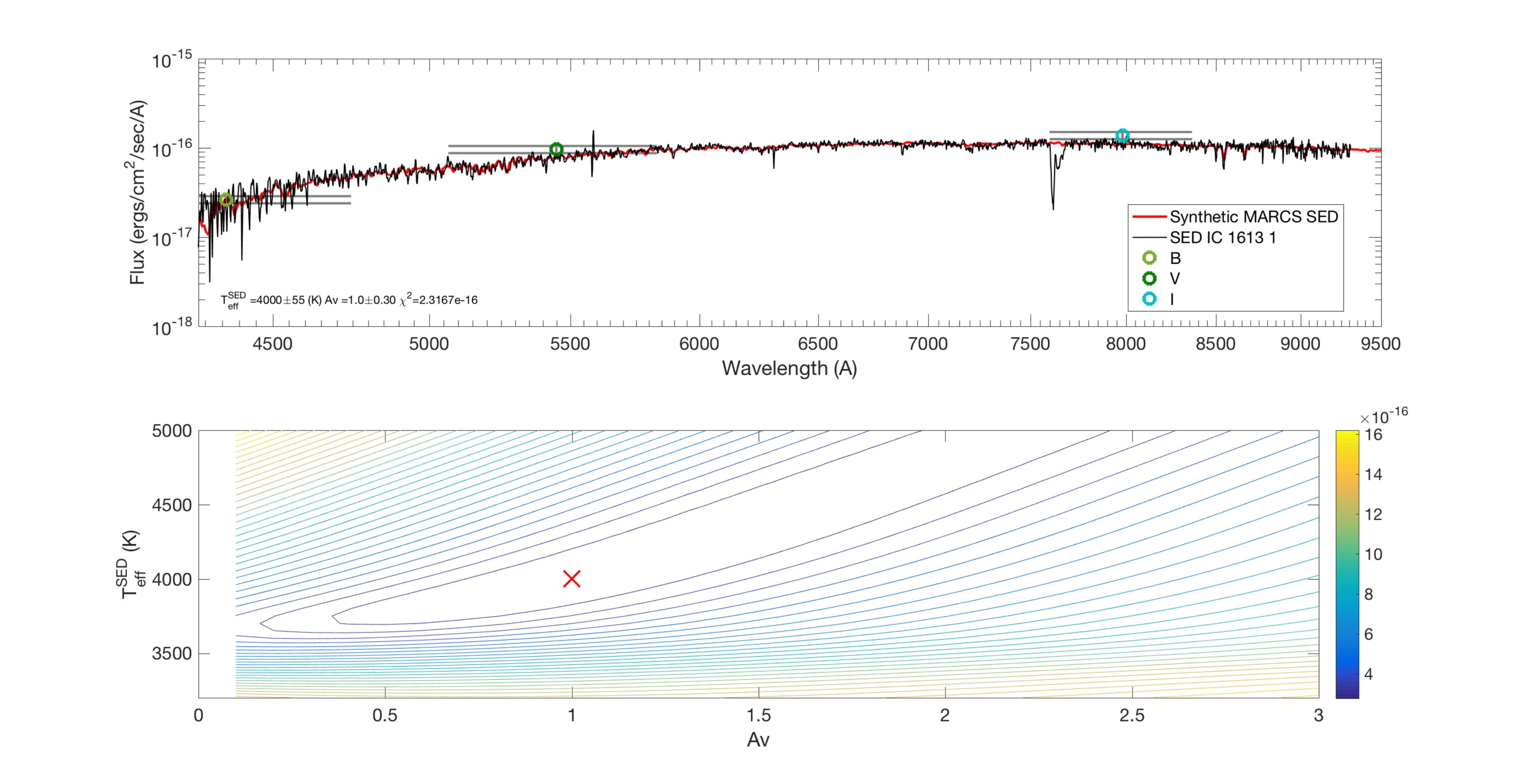}
\includegraphics[width=0.5\linewidth]{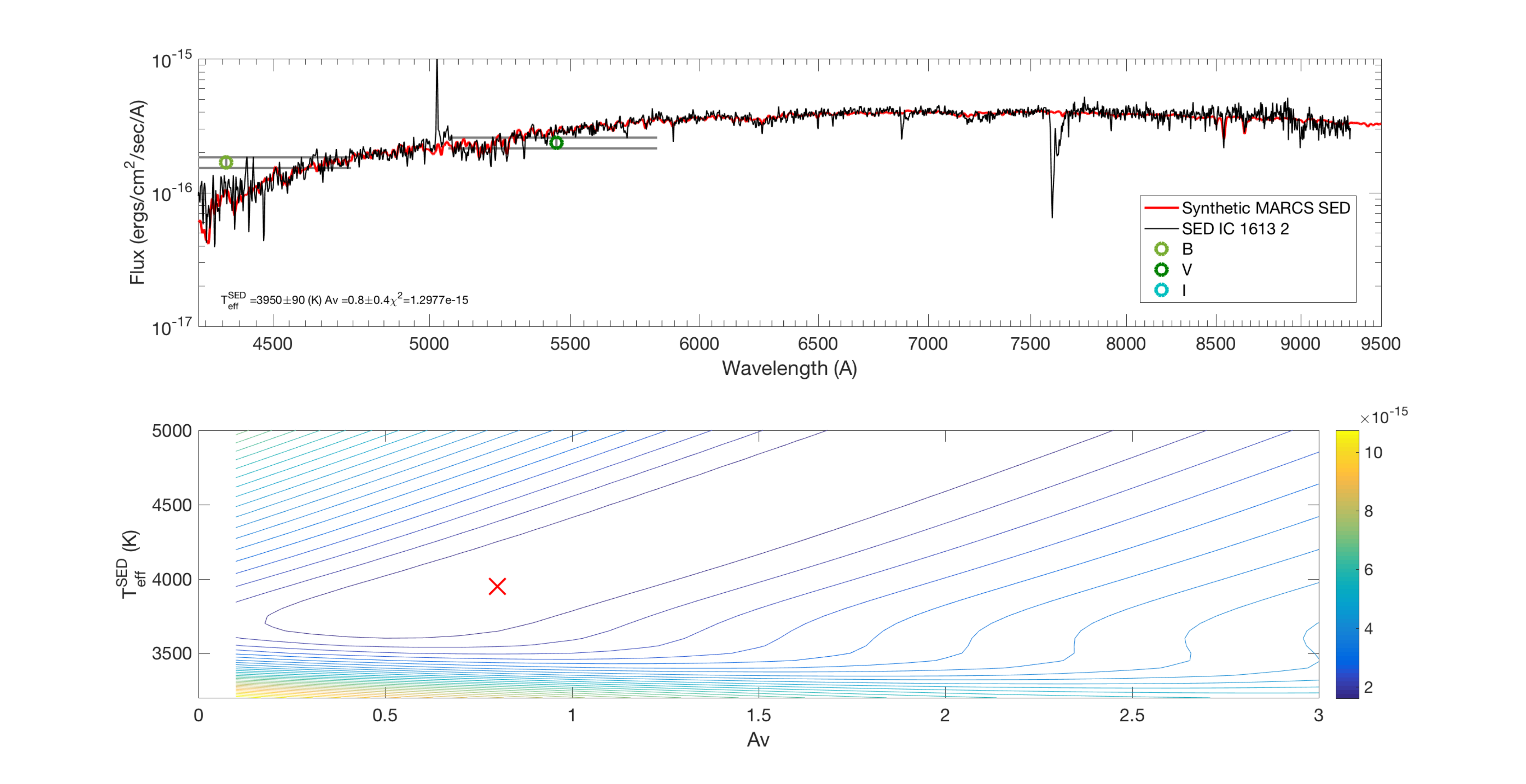}
\caption[h]{Best-fit  MARCS model SED for the spectra of RSGs in the IC 1613 galaxy. The bottom panel, the legend and the labels are the same as in Figure \ref{Fig_peg1}.}
\label{Fig_peg2}
\end{figure*}

\begin{figure*}
\includegraphics[width=0.5\linewidth]{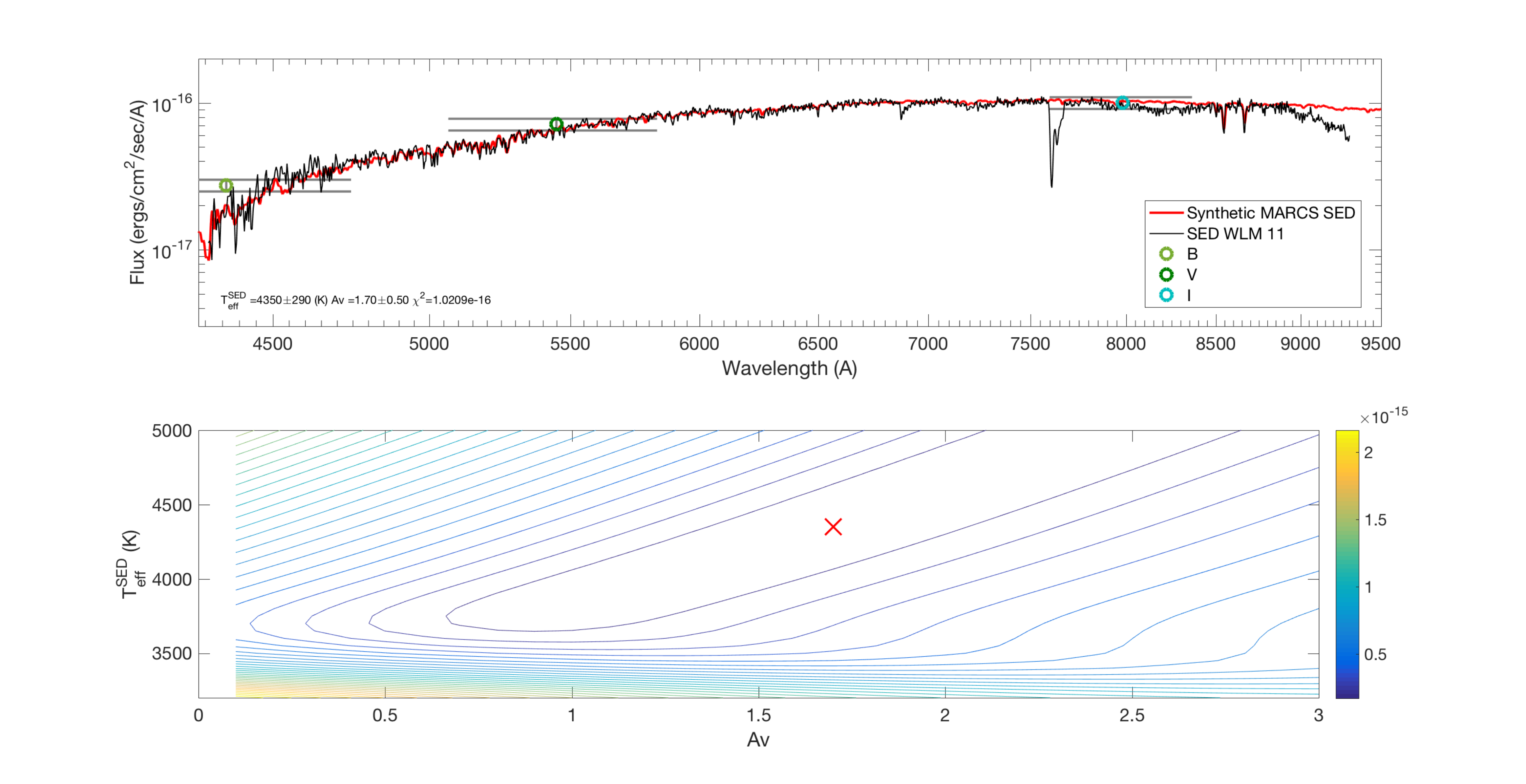}
\includegraphics[width=0.5\linewidth]{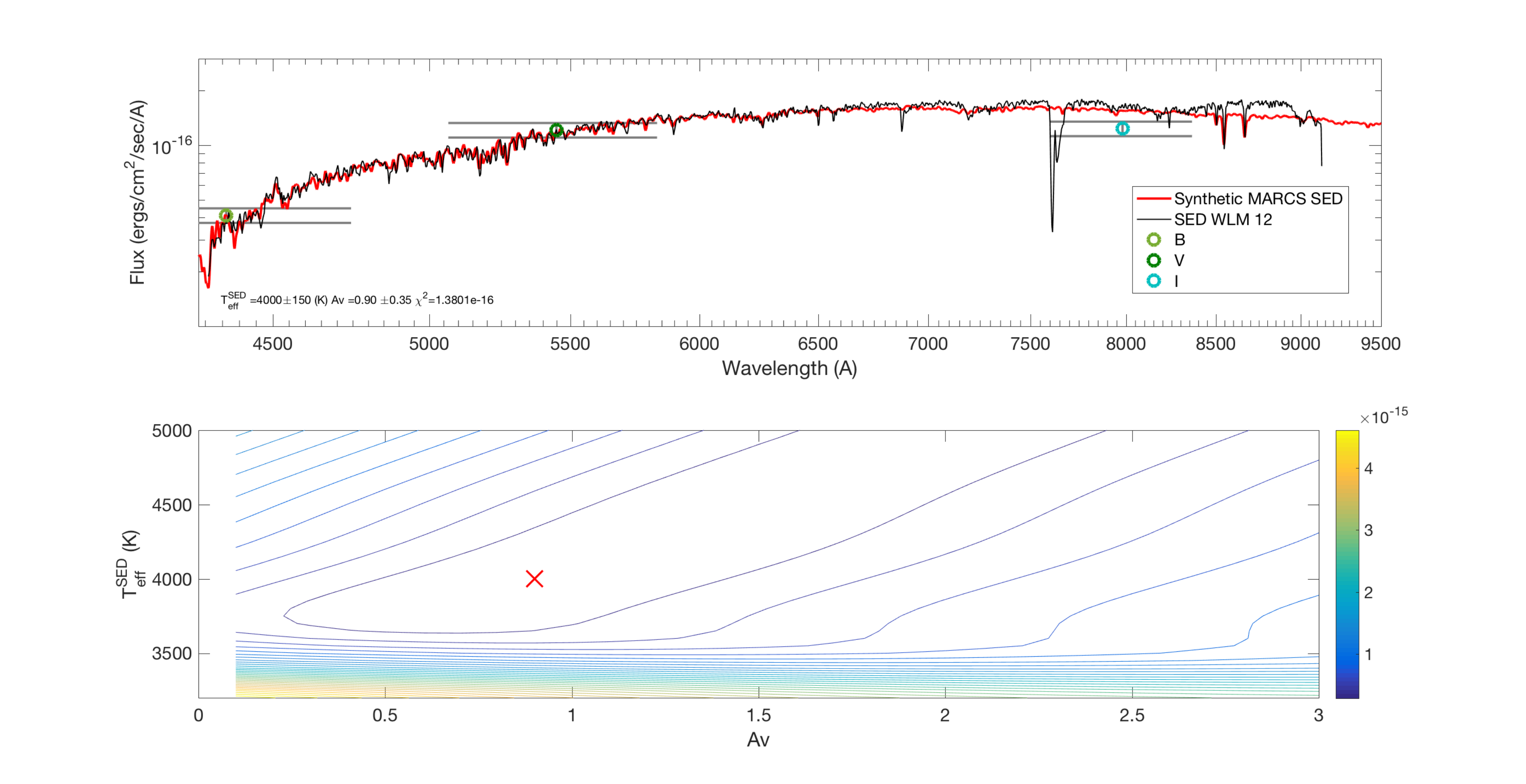}
\includegraphics[width=0.5\linewidth]{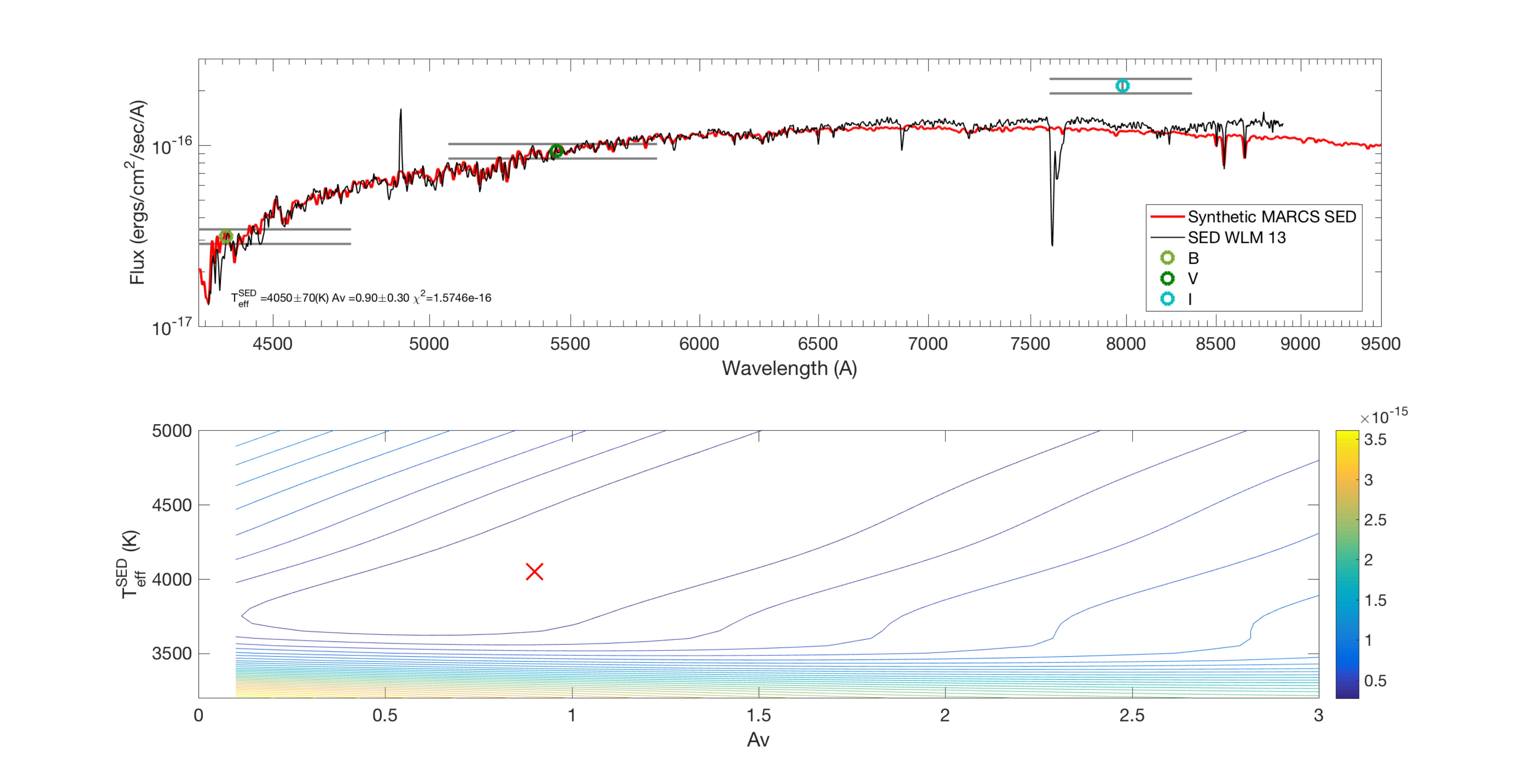}
\includegraphics[width=0.5\linewidth]{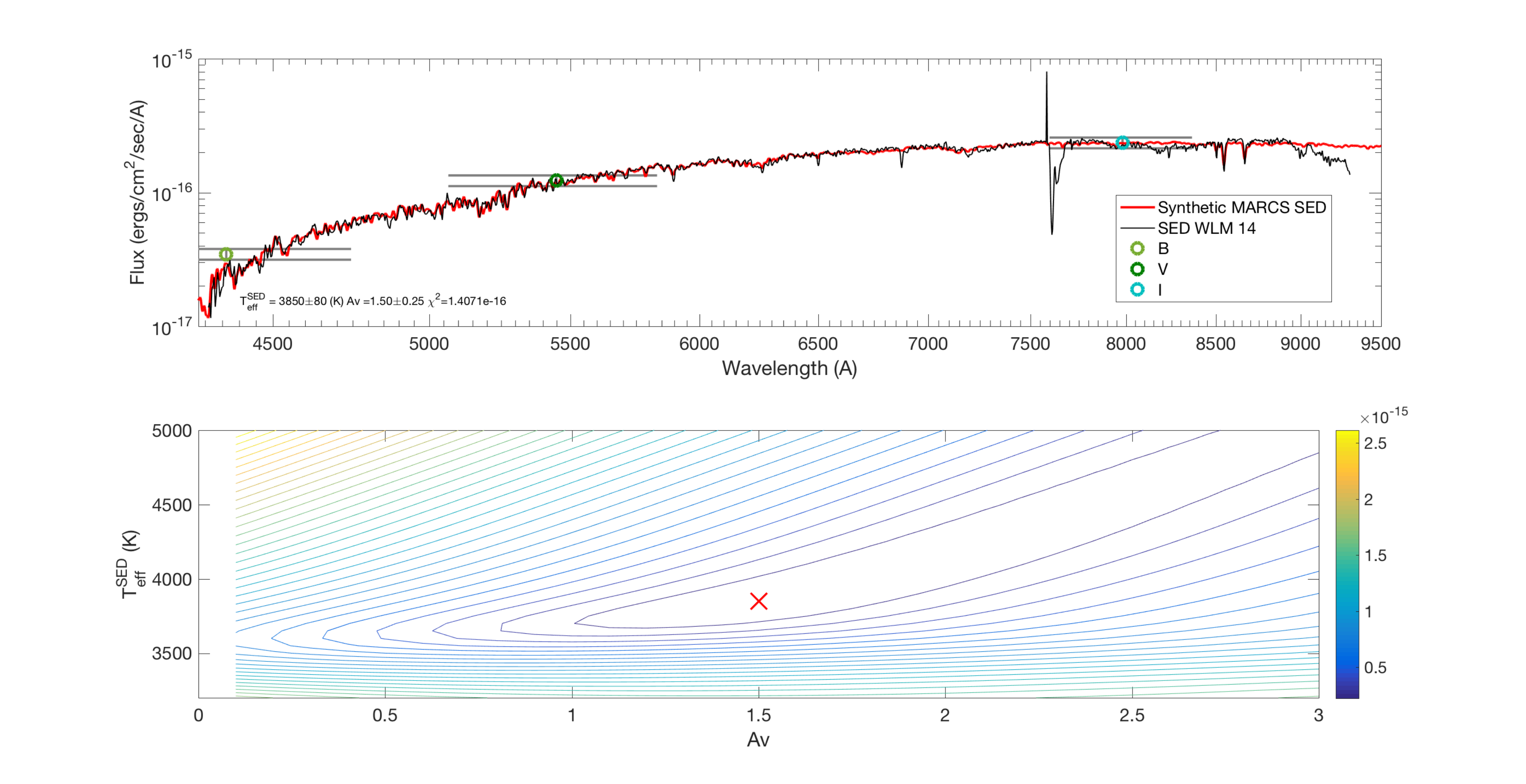}
\caption[h]{Best-fit MARCS model SED for the spectra of RSGs in the WLM galaxy. The bottom panel, the legend and the labels are the same as in Figure \ref{Fig_peg1}.}
\label{Fig_wlm1}
\end{figure*}

\begin{figure*}
\includegraphics[width=0.5\linewidth]{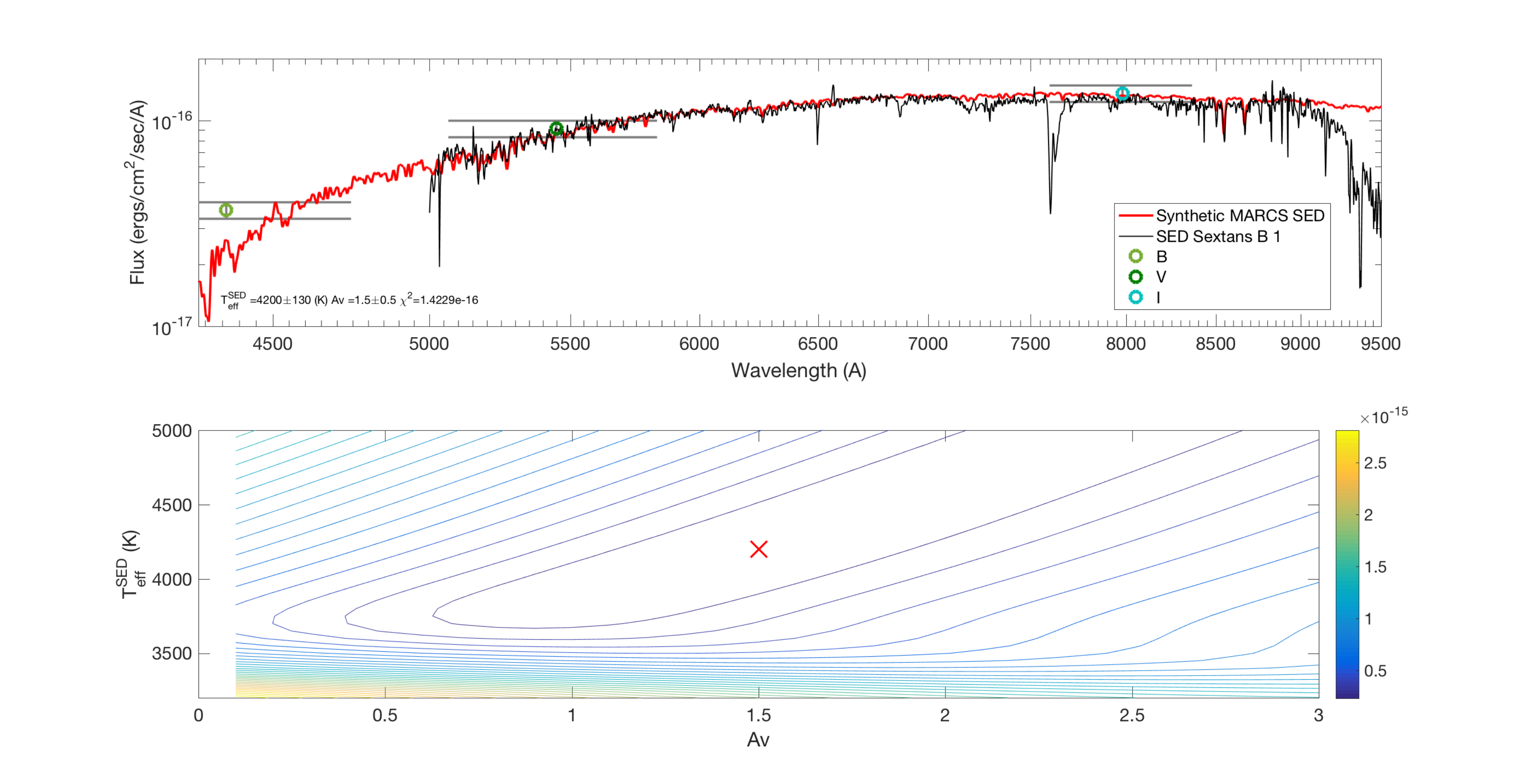}
\includegraphics[width=0.5\linewidth]{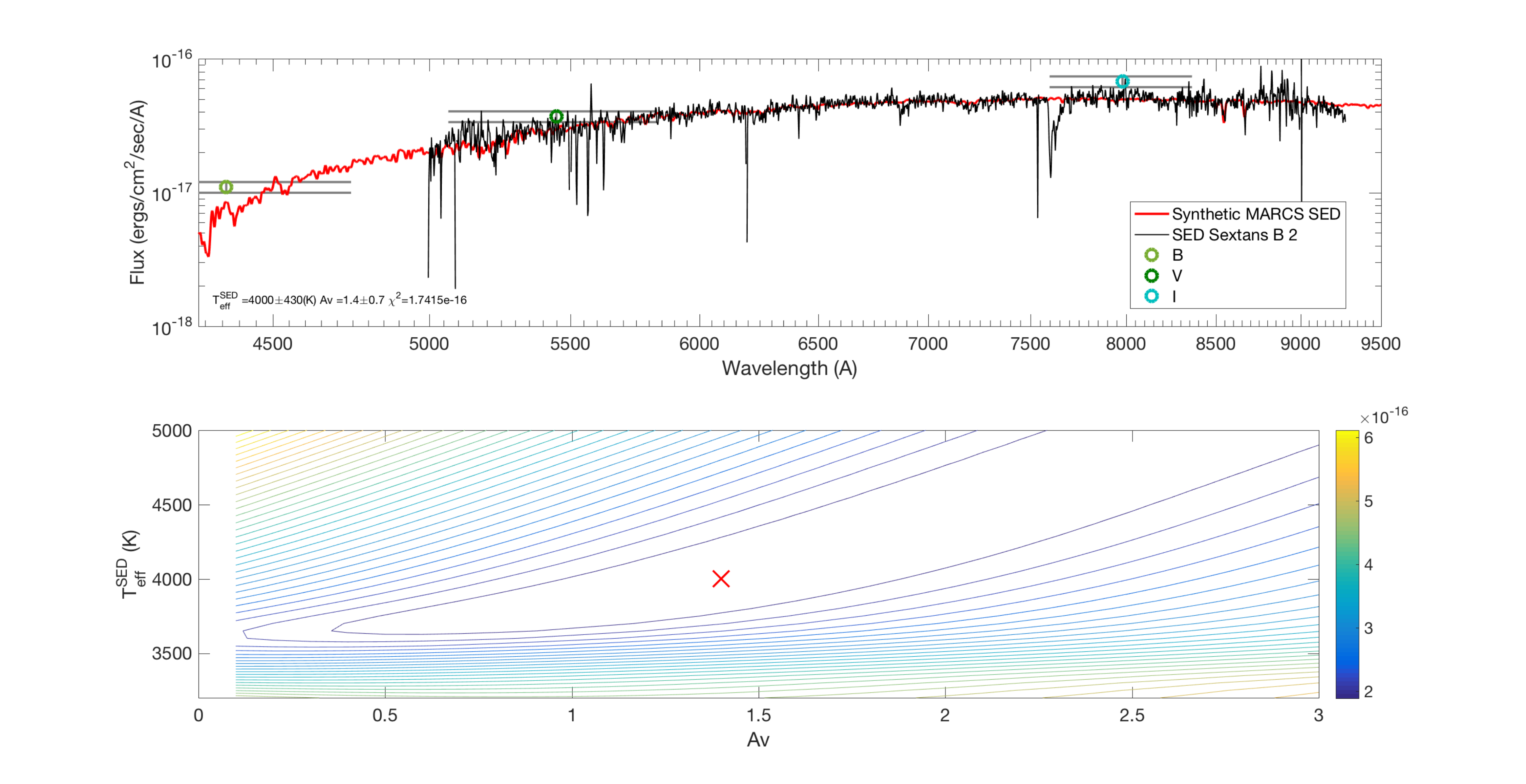}
\caption[h]{Best-fit MARCS model SED for the spectra of RSGs in the Sextans B. The bottom panel, the legend and the labels are the same as in Figure \ref{Fig_peg1}.}
\label{Fig_sexb1}
\end{figure*}

\begin{figure*}
\includegraphics[width=0.5\linewidth]{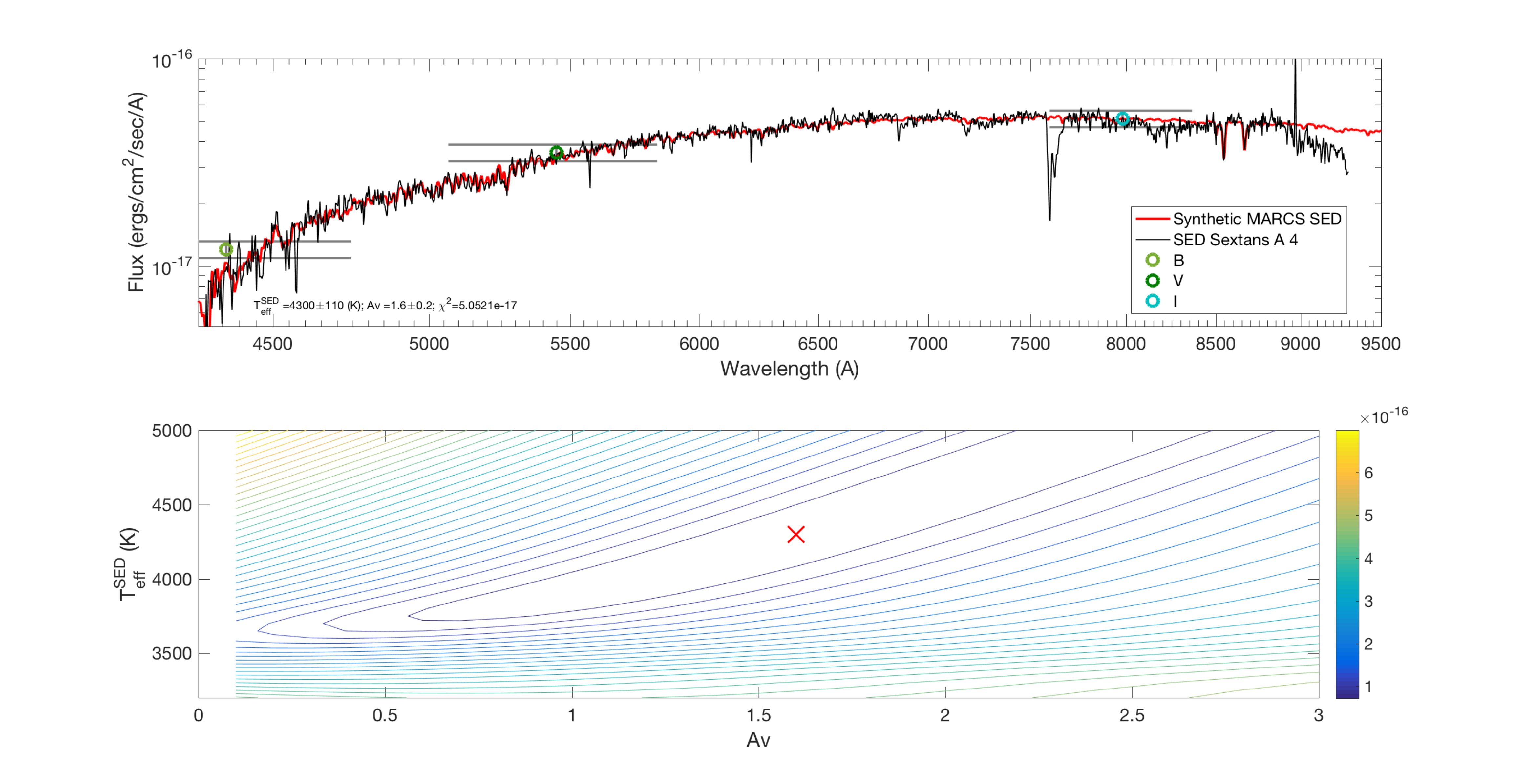}
\includegraphics[width=0.5\linewidth]{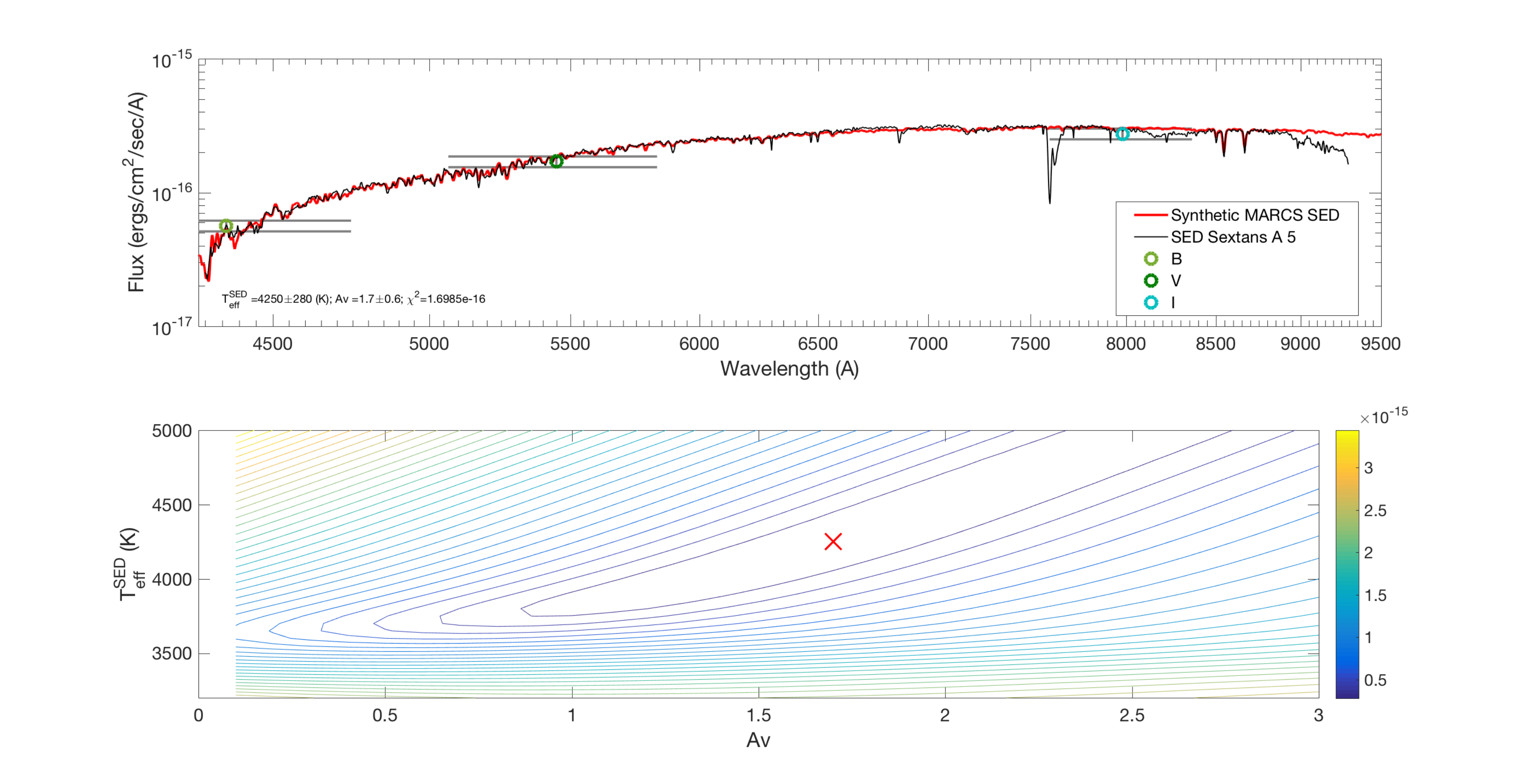}
\includegraphics[width=0.5\linewidth]{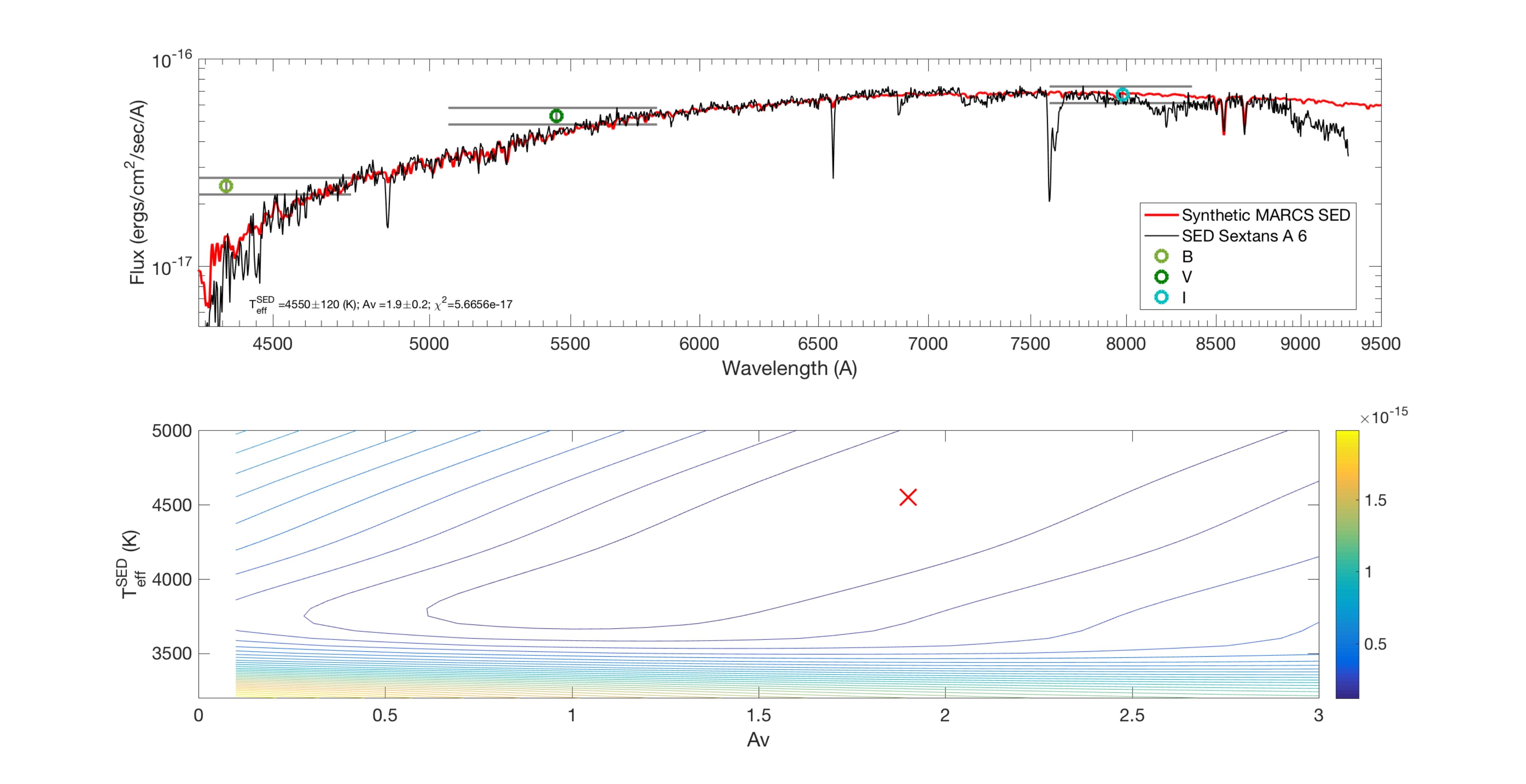}
\includegraphics[width=0.5\linewidth]{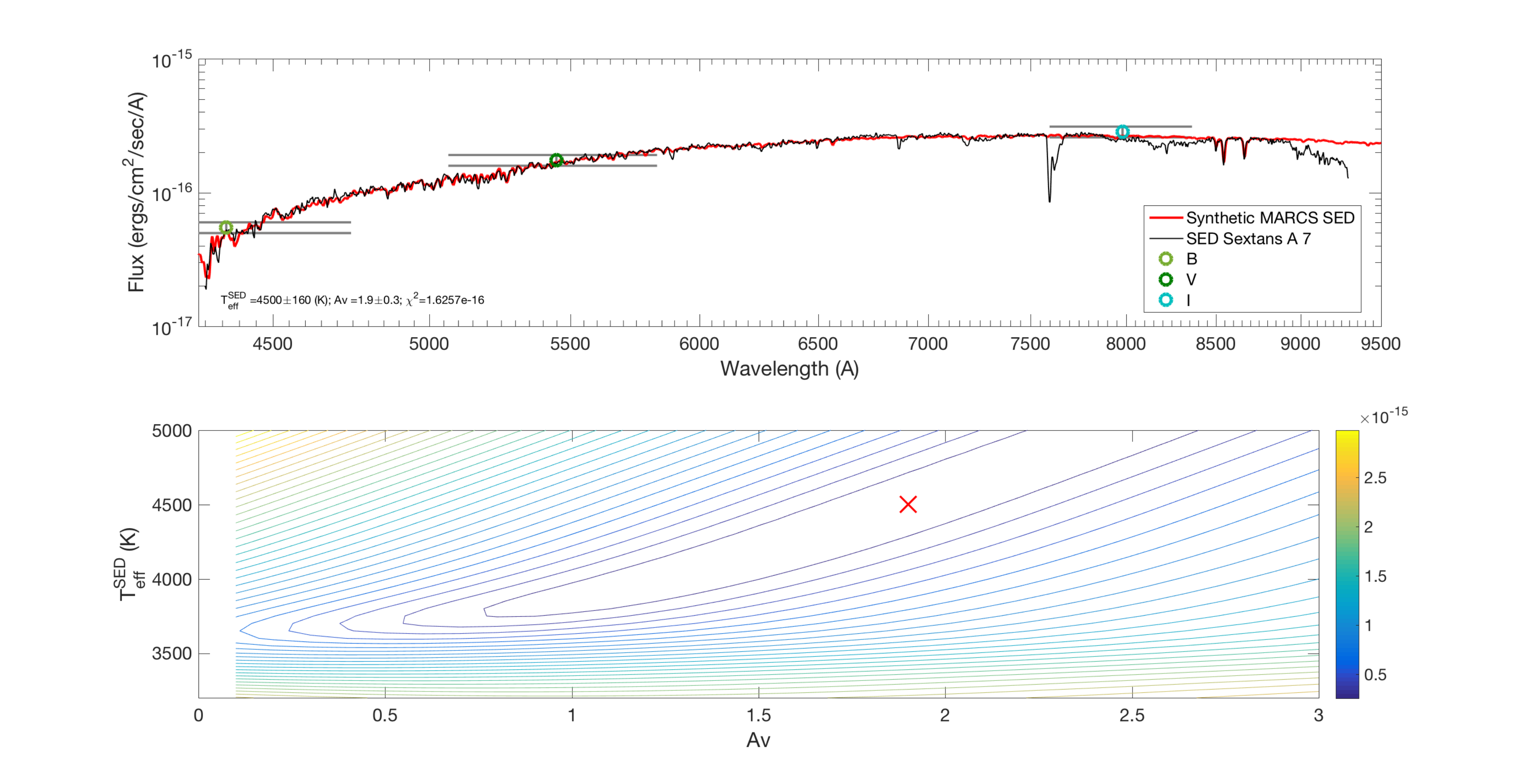}
\includegraphics[width=0.5\linewidth]{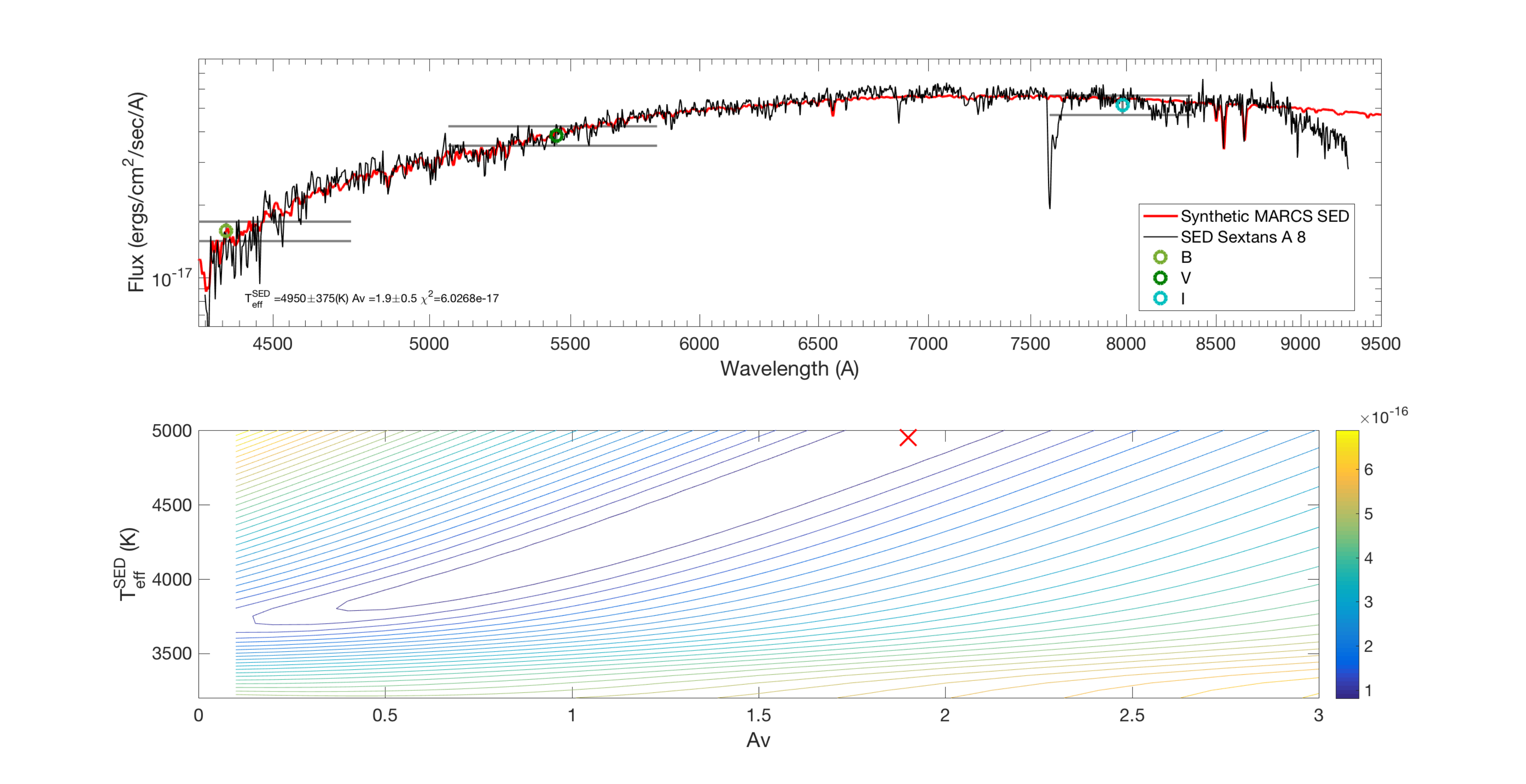}
\includegraphics[width=0.5\linewidth]{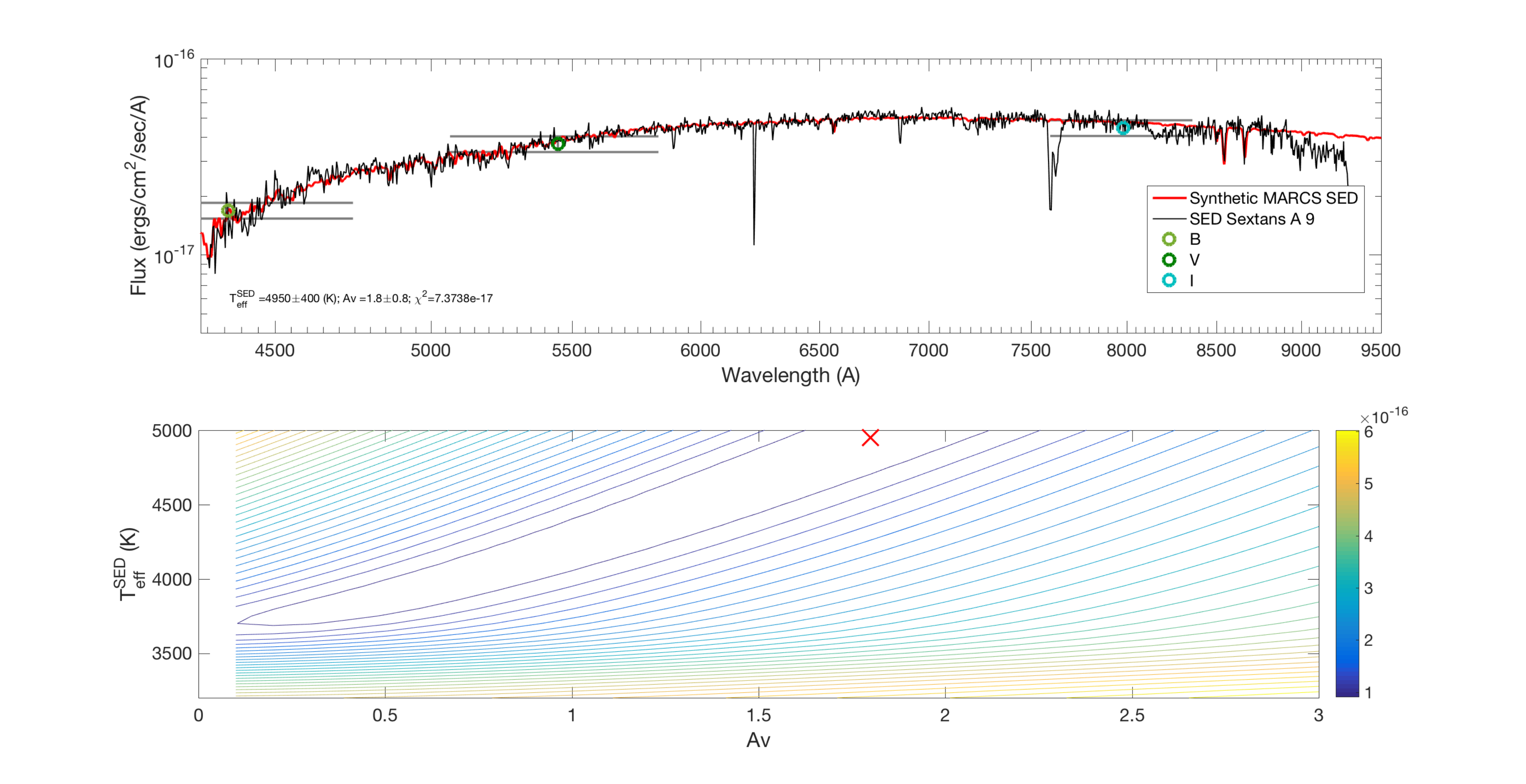}
\includegraphics[width=0.5\linewidth]{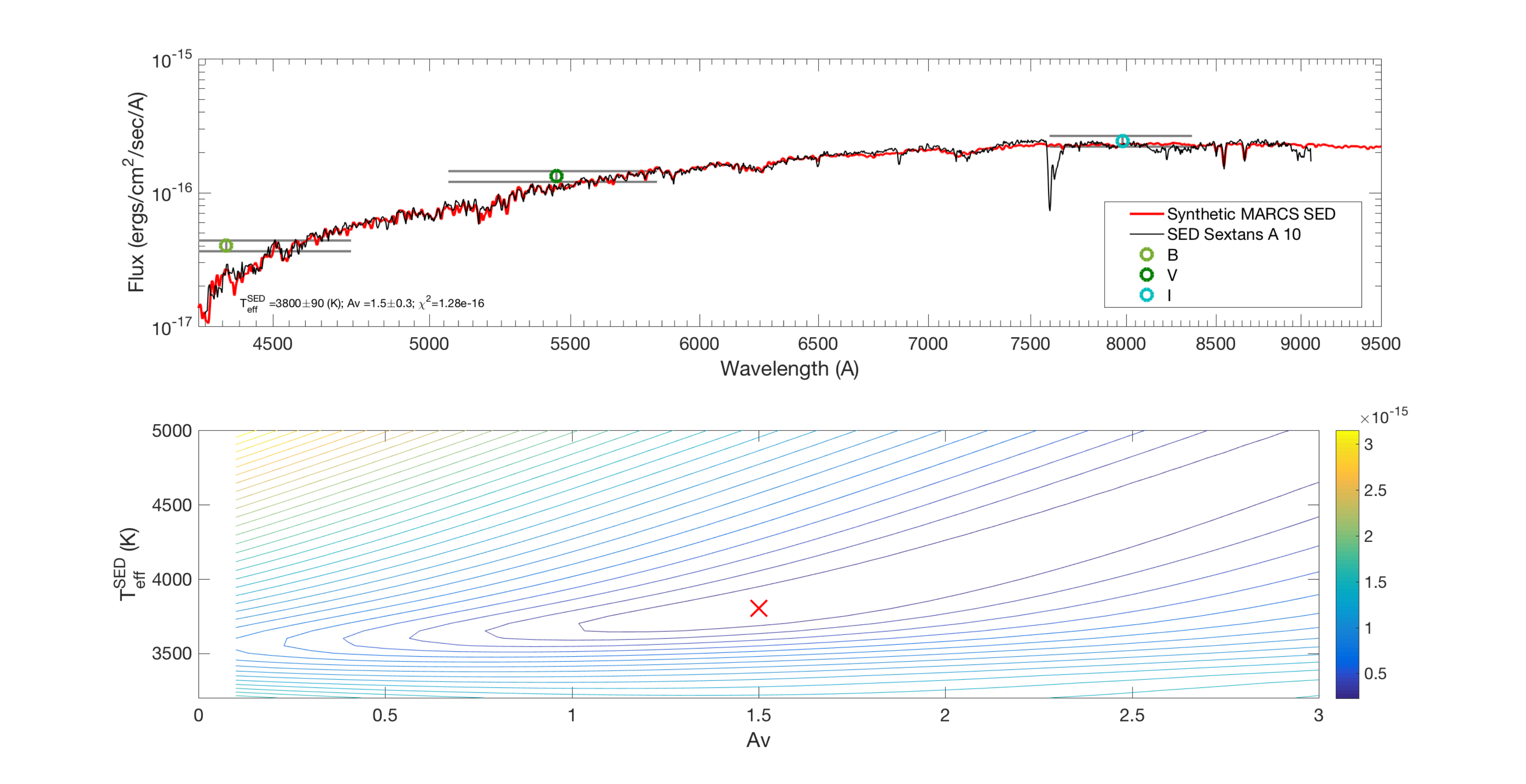}
\caption[h]{Top panel: Best-fit MARCS model SED for the spectra of RSGs in the Sextans A galaxy. The bottom panel, the legend and the labels are the same as in Figure \ref{Fig_peg1}.}
\label{Fig_sexa1}
\end{figure*}


%

\begin{figure*}
\includegraphics[width=0.5\linewidth]{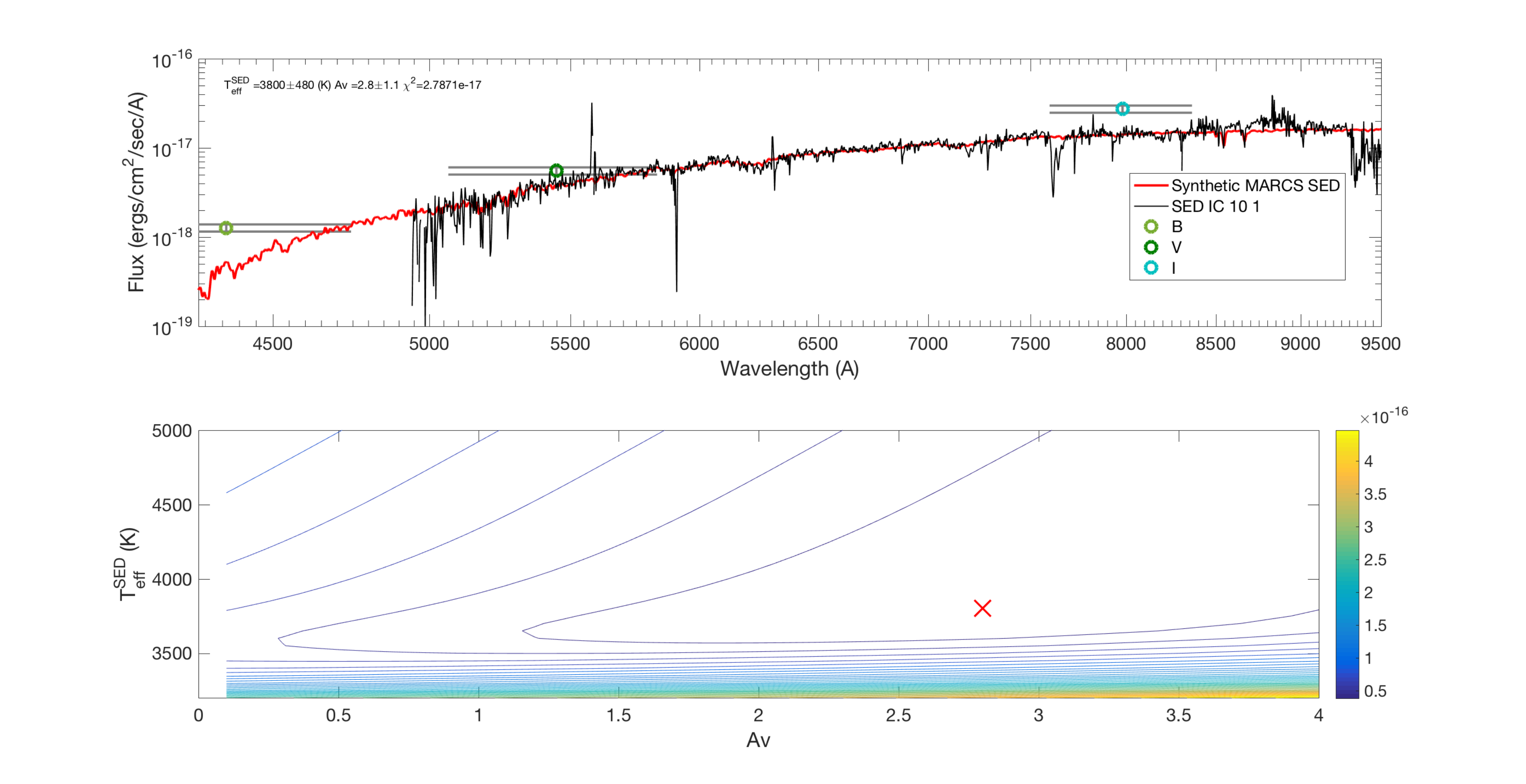}
\includegraphics[width=0.5\linewidth]{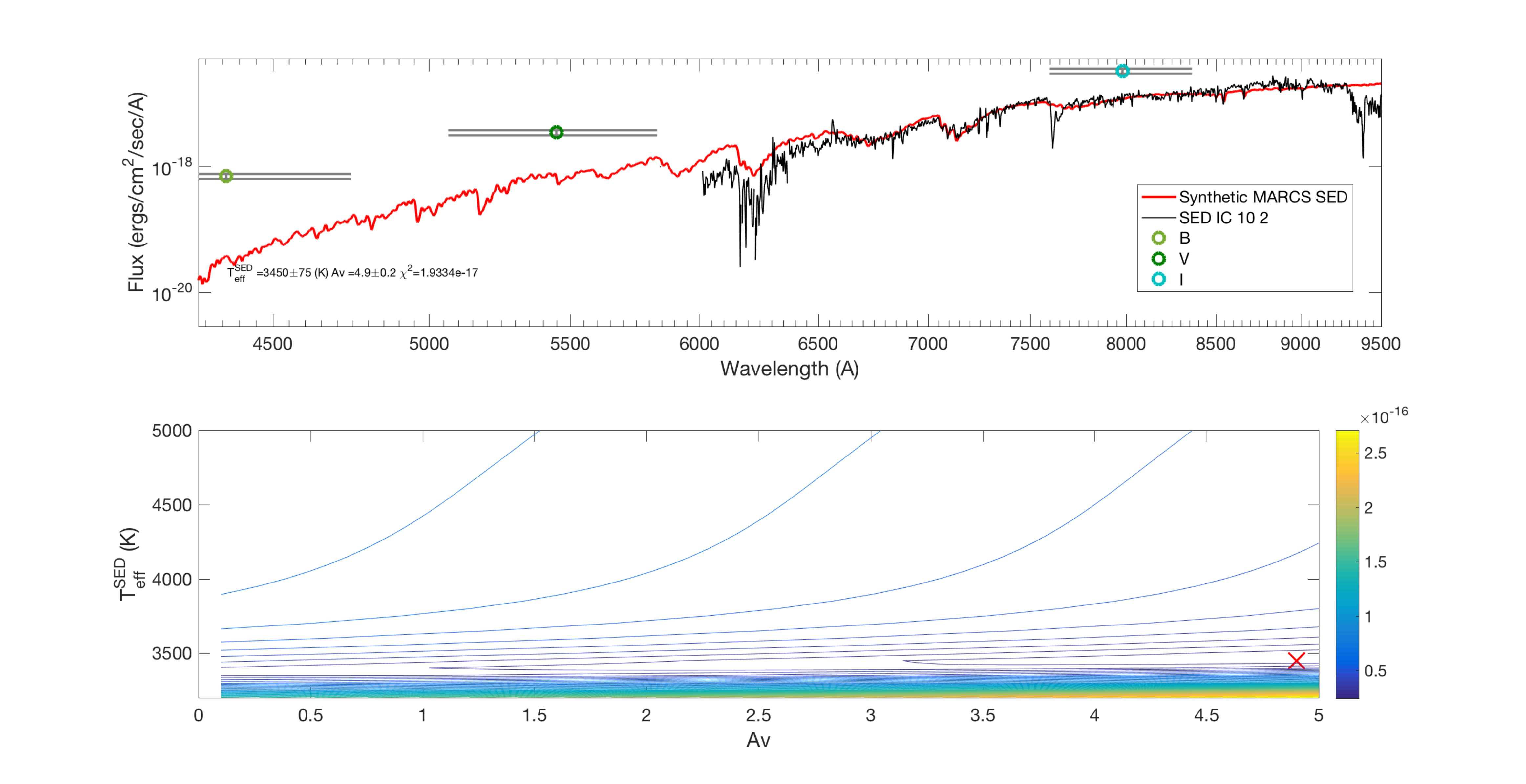}
\includegraphics[width=0.5\linewidth]{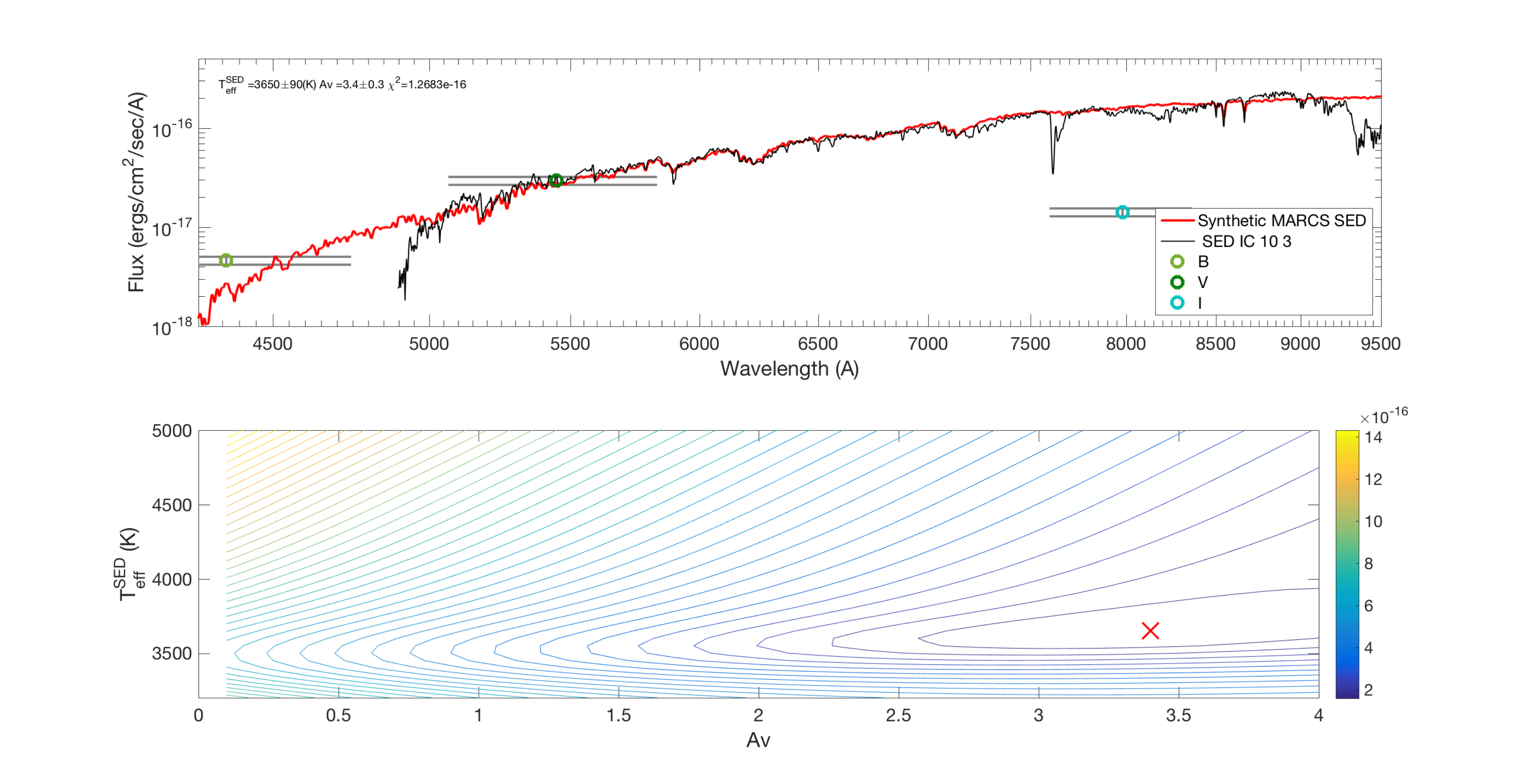}
\includegraphics[width=0.5\linewidth]{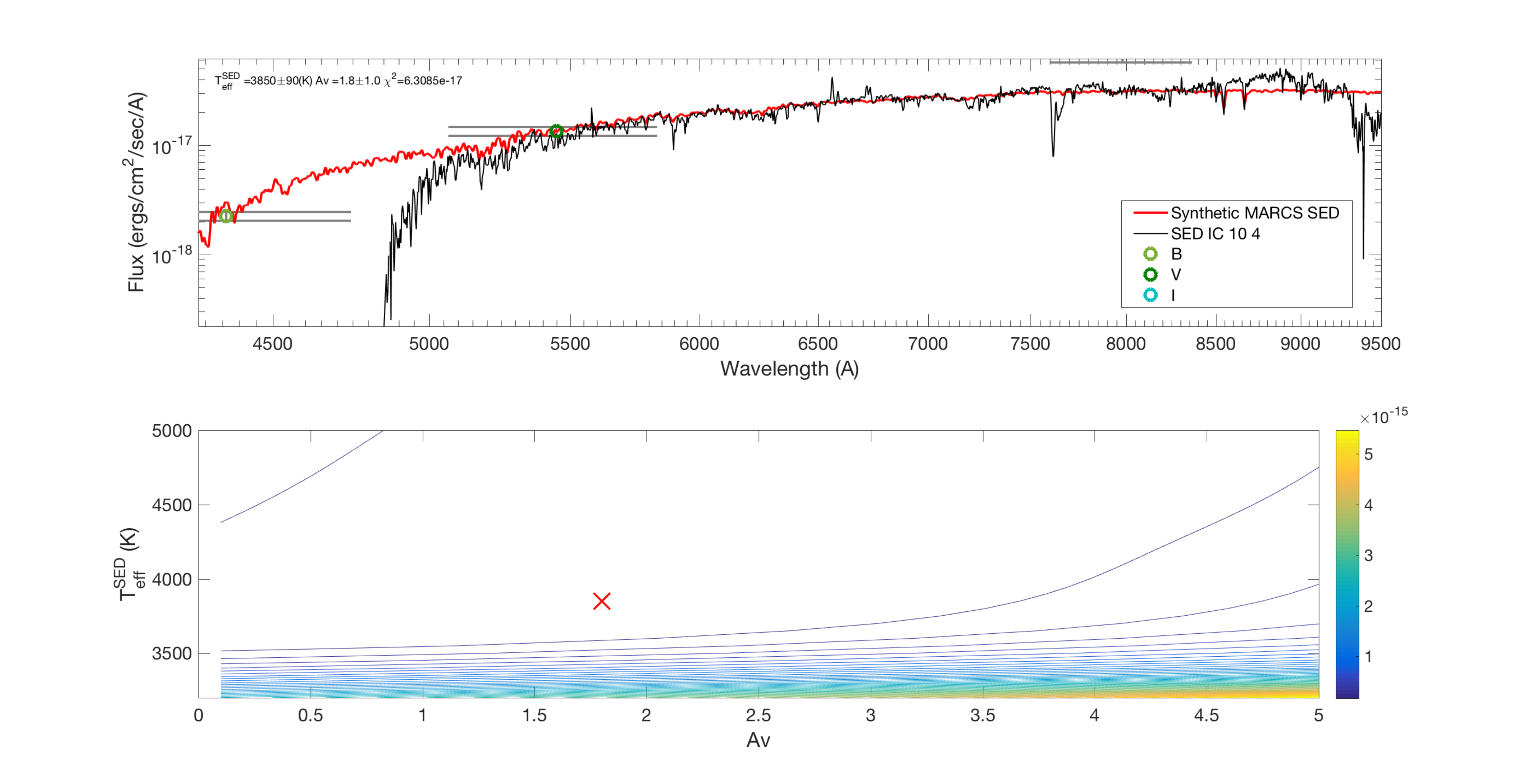}
\includegraphics[width=0.5\linewidth]{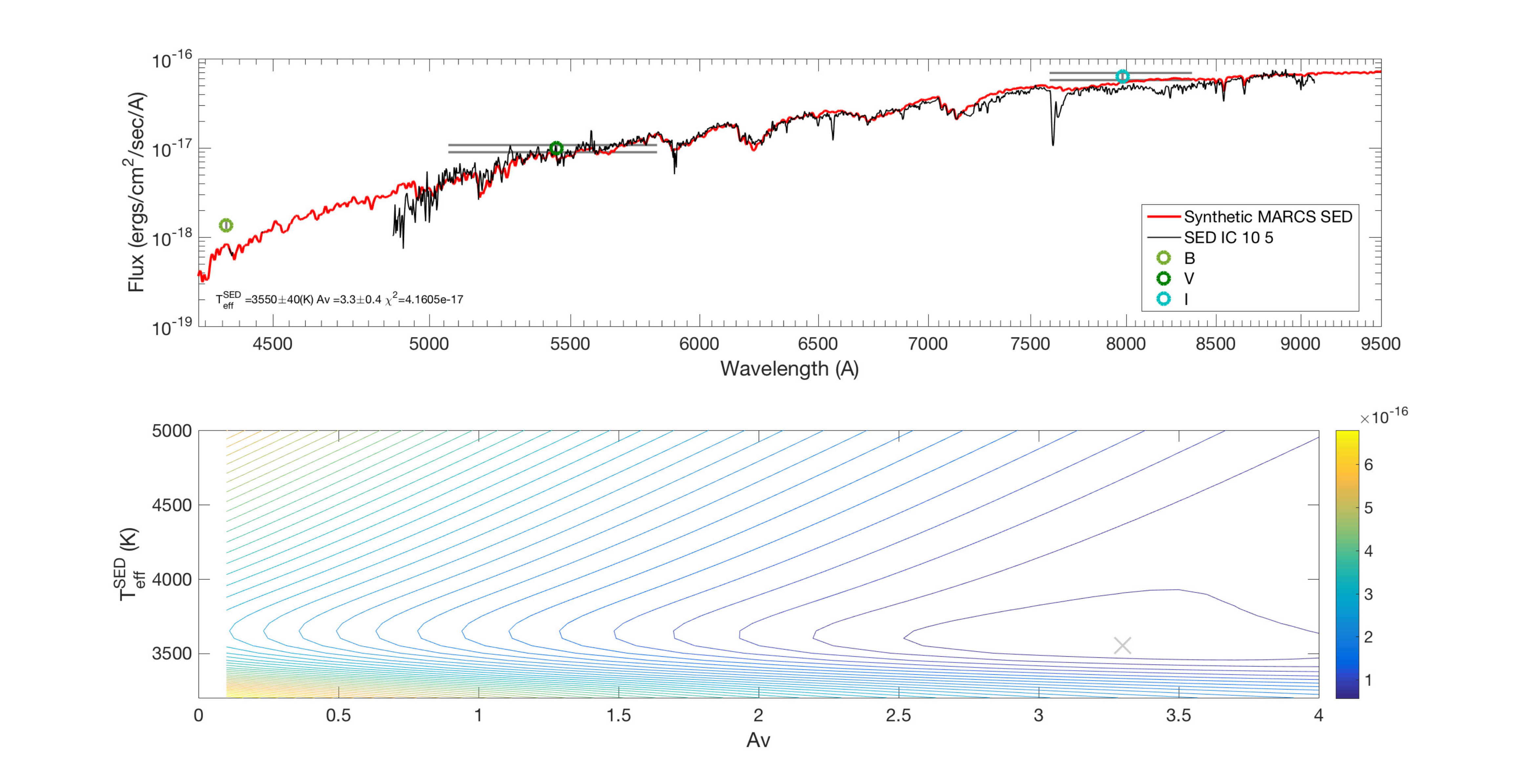}
\includegraphics[width=0.5\linewidth]{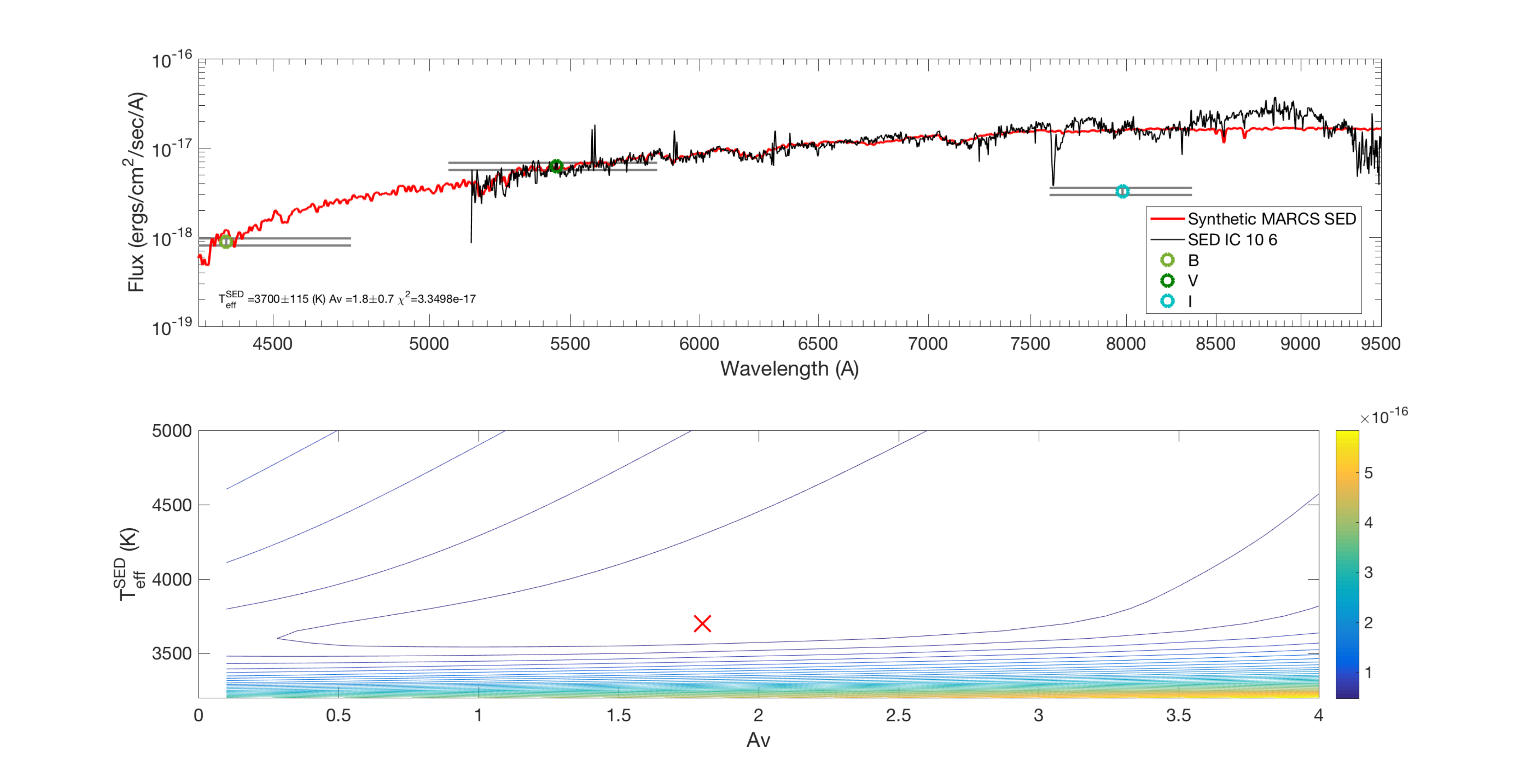}
\caption[h]{Best-fit MARCS model SED for the spectra of RSGs in the IC 10 galaxy. The bottom panel, the legend and the labels are the same as in Figure \ref{Fig_peg1}.}
\label{Fig_ic101}
\end{figure*}

\end{appendix}

\end{document}